\documentclass[useAMS,usenatbib]{mn2e}

\usepackage{epsfig, natbib, graphicx, color,amsmath, cancel, icomma, bm, amssymb}

\input epsf 

\usepackage{color}

\newcommand{\ANNZ}{{\tt ANNZ}}

\newcommand{\Mpc}{\mbox{Mpc}}

\newcommand{\bmm}[1]{\mathbf{#1}}

\newcommand{\ba}{{\bmm a}}

\newcommand{\avg}[1]{\left\langle #1 \right\rangle}

\newcommand{\zmin}{z_{min}}
\newcommand{\zmax}{z_{max}}

\newcommand{\lmin}{l_{{\rm min}}}

\newcommand{\lkhd}{\mbox{$\cal{L}$}}

\newcommand{\bp}{\bmm{p}}

\newcommand{\bCint}{\bmm{C}_{\rm int}}
\newcommand{\bCobs}{\bmm{C}_{\rm obs}}
\newcommand{\bCtot}{\bmm{C}_{\rm tot}}

\newcommand{\be}{\begin{equation}}
\newcommand{\ee}{\end{equation}}
\newcommand{\bea}{\begin{eqnarray}}
\newcommand{\eea}{\end{eqnarray}}

\newcommand{\obs}{{\mathrm{obs}}}

\newcommand{\chired}{{\chi_{\rm red}^2}}

\newcommand{\redmapper}{redMaPPer}
\newcommand{\redmagic}{redMaGiC}
\newcommand{\pmem}{p_{\mathrm{mem}}}
\newcommand{\bc}{\bm{c}}

\newcommand{\zspec}{z_{\rm spec}}

\newcommand{\zred}{z_{\rm red}}
\newcommand{\zphoto}{z_{\rm photo}}

\newcommand{\photoz}{photo-$z$}
\newcommand{\photozs}{photo-$z$s}

\newcommand{\Lmin}{L_{\rm min}}
\newcommand{\chimax}{\chi_{\rm max}^2}
\newcommand{\mz}{m_z}
\newcommand{\mi}{m_i}
\newcommand{\mref}{m_{\rm ref}}
\newcommand{\bq}{\bm{q}}

\newcommand{\zrm}{z_{\rm rm}}

\newcommand{\nmax}{n_{\rm max}}

\newcommand{\ang}{\mathrm{\AA}}

\newcommand{\zsAB}{z_{\rm sAB}}

\newcommand{\OII}{[OII]}
\newcommand{\Halpha}{$\mathrm{H}\alpha$}

\newcommand{\ngmix}{\textsc{ngmix}}
\newcommand{\skynet}{\textsc{SkyNet}}

\title[redMaGiC on DES SV Data]{redMaGiC: Selecting Luminous Red Galaxies from the DES
  Science Verification Data}
\author[E. Rozo, et. al]{E.~Rozo$^{1}$, E.~S.~Rykoff$^{2,3}$, A.~Abate$^{1}$, C.~Bonnett$^{4}$, M.~Crocce$^{5}$, C.~Davis$^{2,6,7}$, 
\newauthor B.~Hoyle$^{8}$, B.~Leistedt$^{9}$, H.V.~Peiris$^{9}$, R.~H.~Wechsler$^{2,3,10}$, T.~Abbott$^{11}$, 
\newauthor F.~B.~Abdalla$^{9}$, M.~Banerji$^{12,13}$, A.~H.~Bauer$^{5}$, A.~Benoit-L{\'e}vy$^{9}$, 
\newauthor G.~M.~Bernstein$^{14}$, E.~Bertin$^{15,16}$, D.~Brooks$^{9}$, E.~Buckley-Geer$^{17}$, D.~L.~Burke$^{2,3}$, 
\newauthor D.~Capozzi$^{18}$, A.~Carnero~Rosell$^{19,20}$, D.~Carollo$^{21}$, M.~Carrasco~Kind$^{22,23}$, 
\newauthor J.~Carretero$^{4,5}$, F.~J.~Castander$^{5}$, M.~J.~Childress$^{24}$, C.~E.~Cunha$^{2}$, 
\newauthor C.~B.~D'Andrea$^{18}$, T.~Davis$^{2,6,7}$, D.~L.~DePoy$^{25}$, S.~Desai$^{26,27}$, H.~T.~Diehl$^{17}$, 
\newauthor J.~P.~Dietrich$^{8,26}$, P.~Doel$^{9}$, T.~F.~Eifler$^{14,28}$, A.~E.~Evrard$^{29,30}$, A.~Fausti Neto$^{19}$, 
\newauthor B.~Flaugher$^{17}$, P.~Fosalba$^{5}$, J.~Frieman$^{17,31}$, E.~Gaztanaga$^{5}$, D.~W.~Gerdes$^{30}$, 
\newauthor K.~Glazebrook$^{32}$, D.~Gruen$^{8,33}$, R.~A.~Gruendl$^{22,23}$, K.~Honscheid$^{34,35}$, 
\newauthor D.~J.~James$^{11}$, M.~Jarvis$^{14}$, A.~G.~Kim$^{36}$, K.~Kuehn$^{37}$, N.~Kuropatkin$^{17}$, 
\newauthor O.~Lahav$^{9}$, C.~Lidman$^{6,37}$, M.~Lima$^{19,38}$, M.~A.~G.~Maia$^{19,20}$, M.~March$^{14}$, 
\newauthor P.~Martini$^{34,39}$, P.~Melchior$^{34,35}$, C.~J.~Miller$^{29,30}$, R.~Miquel$^{4,40}$, 
\newauthor J.~J.~Mohr$^{26,27,33}$, R.~C.~Nichol$^{18}$, B.~Nord$^{17}$, C.~R.~O'Neill$^{7}$, R.~Ogando$^{19,20}$, 
\newauthor A.~A.~Plazas$^{28}$, A.~K.~Romer$^{41}$, A.~Roodman$^{2,3}$, M.~Sako$^{14}$, E.~Sanchez$^{42}$, 
\newauthor B.~Santiago$^{19,43}$, M.~Schubnell$^{30}$, I.~Sevilla-Noarbe$^{22,42}$, R.~C.~Smith$^{11}$, 
\newauthor M.~Soares-Santos$^{17}$, F.~Sobreira$^{17,19}$, E.~Suchyta$^{34,35}$, M.~E.~C.~Swanson$^{23}$, 
\newauthor J.~Thaler$^{44}$, D.~Thomas$^{18}$, S.~Uddin$^{6,32}$, V.~Vikram$^{45}$, A.~R.~Walker$^{11}$, 
\newauthor W.~Wester$^{17}$, Y.~Zhang$^{30}$, L.~N.~da Costa$^{19,20}$ 
 \\ \textit{Affiliations are listed at the end of the paper} \vspace{-250pt}}
  
\begin{document}

\maketitle 

\label{firstpage}

\begin{abstract}
We introduce \redmagic, an automated algorithm for selecting Luminous Red Galaxies (LRGs).
The algorithm was specifically developed to minimize photometric redshift uncertainties 
in photometric large-scale structure studies.  \redmagic\ achieves this by self-training
the color-cuts necessary to produce a luminosity-thresholded LRG sample of constant
comoving density.  We demonstrate that \redmagic\ \photozs\ are 
very nearly as accurate as the best machine-learning based methods, 
yet they require minimal spectroscopic training, do not suffer from extrapolation biases,
and are very nearly Gaussian.
We apply our algorithm to Dark Energy Survey (DES) Science Verification (SV) data to
produce a \redmagic\ catalog sampling the redshift range $z\in[0.2,0.8]$.  Our fiducial sample
has a comoving space density of $10^{-3}\ (h^{-1} \Mpc)^{-3}$, and a 
median \photoz\ bias ($\zspec-\zphoto$) and scatter ($\sigma_z/(1+z)$) of 
0.005 and 0.017 respectively.  The corresponding $5\sigma$ outlier fraction is 1.4\%.
We also test our algorithm with Sloan Digital Sky Survey (SDSS) Data Release 8 (DR8) and 
Stripe 82 data, and discuss how spectroscopic training can be used 
to control \photoz\ biases at the 0.1\% level.
\end{abstract}


\section{Introduction}

Since the beginning of the Sloan Digital Sky
Survey~\citep[SDSS;][]{yorketal00}, it has been recognized that luminous
red galaxies (LRGs) are an ideal probe of large-scale structure\citep[][]{stoughtonetal02}.  
Being luminous, they can be observed to high redshift with relatively shallow
exposures. In addition, the $4000\,\ang$ break in the spectra of these galaxies
enables robust photometric redshift estimates (\photozs) when the break is photometrically
sampled.  To date, red galaxy selection algorithms have been fairly crude:
one typically defines a
color box that isolates LRGs in color--color space, with the specific cuts
being selected in a relatively ad-hoc
manner~\citep[e.g.][]{eisensteinetal01,eisensteinetal05}.  This relative lack
of attention is driven by the fact that spectroscopic follow-up renders high
precision selection of LRGs unnecessary.  With the advent of photometric
surveys with no spectroscopic component like the DES~\citep[][]{desetal05} and the Large Synoptic Survey
Telescope~\citep[LSST;][]{lsst09}, it is now important to develop selection
algorithms designed to minimize photometric redshift uncertainties.

To this end, we have developed \redmagic, a new red-galaxy selection algorithm.
Specifically, our primary motivation is to select galaxies with robust, exquisitely
controlled photometric redshifts. A secondary and
complementary, goal is to develop a new photometric redshift estimator for
these galaxies that is well understood, and has spectroscopic requirements that
are either easily met with existing facilities.
The algorithm relies heavily on the infrastructure built
for red sequence cluster finding with \redmapper\ \citep[][henceforth
  RM1]{rykoffetal14}.  Specifically, \redmapper\ combines sparse spectroscopy
of galaxy clusters with photometric data to calibrate the red sequence of
galaxies as a function of redshift.  We use the resulting calibration as a
photometric template, and select a galaxy as red if this empirical template
provides a good description of the galaxy's color.  We refer to the resulting
galaxy catalog as the {\bf red}-sequence {\bf Ma}tched-filter {\bf G}alaxy 
{\bf C}atalog, or \redmagic\ for short.

We implement our algorithm in the DES Science Verification (SV) data~(Rykoff
et al., in prep) and characterize the \photoz\ properties of
the resulting catalog.  To provide further \photoz\ testing, we have also applied
\redmagic\ to SDSS DR8 and SDSS Stripe 82 data.

The layout of the paper is as follows.  Section~\ref{sec:data} briefly summarizes the data sets used in
this work. Section~\ref{sec:algorithm} described the \redmagic\ selection algorithm and the \redmagic\
\photoz{} estimator.  Section~\ref{sec:performance} evaluates the performance of \redmagic\
in each of the three data sets considered in this work, while section~\ref{sec:spec}
compares the \redmagic\ \photoz\ performance to several other \photoz\ methods.
Section~\ref{sec:selection} demonstrates that \redmagic\ succeeds at selecting galaxies
with clean \photozs\ by comparing \redmagic\ galaxies to the SDSS ``constant
mass'' CMASS sample, which was specifically tailored for spectroscopic
follow-up of galaxies at $z \geq 0.45$~\citep{dawsonetal13}.
Section~\ref{sec:ab} discusses how \redmagic\ can be improved upon if representative
spectroscopic subsamples of \redmagic\ galaxies become available.  
Section~\ref{sec:outliers} characterizes \redmagic\ catastrophic failures,
which we take to mean $5\sigma$ outliers.
A discussion and summary of our conclusions is presented in 
Section~\ref{sec:conclusions}.

{\bf Fiducial cosmology and conventions:} The construction of the
\redmagic\ galaxy samples requires one specify a cosmology for computing the comoving density of galaxies, and for 
estimating luminosity distances.
To do this, we assume a flat $\Lambda$CDM cosmology with $\Omega_m=0.3$ and $h=1.0$ (i.e. distances
are in $h^{-1} \Mpc$).  This is the convention used by \redmapper.

Finally, this work references both $z$-band magnitudes and galaxy redshifts.
To avoid confusion, we
denote $z$-band magnitudes via $\mz$, and reserve the symbol $z$ to signify
redshift.  Similarly, we refer to $i$-band magnitudes via $\mi$ to distinguish
from the counting index $i$.


\section{Data}
\label{sec:data}

\subsection{DES Science Verification Data}
\label{sec:SV}

DES is a wide-field photometric survey in the $grizY$ bands performed with the
Dark Energy Camera~\citep[DECam,][]{diehletal12,flaugheretal15}.  The DECam is installed at the
prime focus of the 4-meter Blanco Telescope at Cerro Tololo Inter-American
Observatory (CTIO).  The full DES survey is scheduled for 525 nights
distributed over five years, covering $5000\,\mathrm{deg}^2$ of the southern
sky, approximately half of which overlaps the South Pole
Telescope~\citep[SPT,][]{carlstrometal11} Sunyaev-Zel'dovich cluster survey.

Prior to the commencement of regular survey operations in August 2013, from
November 2012 to March 2013 DES conducted a $\sim300\,\mathrm{deg}^2$ ``Science
Verification'' (SV) survey.  The main portion of the SV footprint, used in this
paper, covers the $\sim150\ \mathrm{deg}^2$ Eastern SPT (``SPTE'') region, in the range
$65<\mathrm{R.A.}<93$ and $-60<\mathrm{Decl.}<-42$.  SPTE was observed between 2
and 10 tilings in each of the $griz$ filters.  In
addition, DES surveys 10 Supernova fields every 5-7 days, each of which covers
a single DECam 2.2 degree-wide field-of-view.  The median depth of the SV survey
(defined as $10\sigma$ detections for extended sources) are $g=24.0$, $r=23.9$
$i=23.0$, $z=22.3$, and $Y=20.8$.

The DES SV data was processed by the DES Data Management (DESDM)
infrastructure~(Gruendl et al, in prep).  This processing
performs image deblending, astrometric registration, global calibration, image
coaddition, and object catalog creation.  Details of the DES single-epoch and
coadd processing can be found in \citet{sevillaetal11} and
\citet{desaietal12}.  We use {\tt SExtractor} to create object catalogs from
the single-epoch and coadded images~\citep{bertinarnouts96, bertin11}.  Object
detection was performed on a ``chi-squared'' coadd of the $r$+$i$+$z$ image
with {\tt SWarp}~\citep{bertin10}, and object measurement was performed in
dual-image mode with each individual $griz$ image (here we ignore
the shallow $Y$-band imaging).  

After production of these early data, several problems were detected and corrected
for in post-processing, leading to the creation of the ``SVA1 Gold''
catalog~(Rykoff et al., in prep).  First, unmasked satellite trails 
were masked.
Second, 
calibration was improved using
a modified version of the {\tt big-macs} stellar-locus fitting
code~\citep{kellyetal14}\footnote{https://code.google.com/p/big-macs-calibrate/}.
We recomputed coadd zero-points over the full SV footprint on a
HEALPix~\citep{gorskietal05} grid of NSIDE=256.  These zero-points were then
interpolated with a bi-linear scheme to correct the magnitudes of all objects in
the catalog.  Finally, regions around bright stars ($J<13$) from the Two Micron
All Sky Survey~\citep[2MASS;][]{skrutskieetal06} were masked.

Galaxy magnitudes and colors are computed via the
{\tt SExtractor} {\tt MAG\_AUTO} quantity.  These colors
are significantly noisier than those obtained through model fitting.
However, for SV coadd images {\tt MAG\_AUTO} colors are
considerably more stable due to PSF discontinuities in the coadded images
sourced by coadding different exposures.
This is expected to have a negative impact on our results, 
and future work will make use of full galaxy multi-epoch
multi-band color measurements. 

Star-galaxy separation is a particularly challenging issue for red galaxy
selection at high redshift.  In particular, at $z\sim0.7$ the red end of the
stellar locus approaches the red sequence galaxy locus when using purely optical
($griz$) photometry.  Therefore, we have made use of the \ngmix{} multi-band
multi-epoch image processing~(Sheldon et al., in prep; Jarvis et al., in prep)
to select a relatively pure and complete galaxy selection.  Details are presented in
Appendix~\ref{app:stargal}.  As \ngmix{} is primarily used for shape measurements
on DES data, the tolerance for input image quality is relatively tight,
so our footprint is smaller than that of SVA1
Gold~(see Jarvis et al., in prep).
Finally, we only consider regions where the $z$-band $10\sigma$
depth in {\tt MAG\_AUTO} has $\mz >22$~(Rykoff et al., in prep).  In total, we use
$148\,\mathrm{deg}^2$ of DES SV imaging in this paper, and the angular mask
is described in Appendix~\ref{app:catalogs}.

We note that \redmagic{} relies on the red sequence calibration by the
\redmapper{} algorithm, as detailed in RM1.  The DES SV \redmapper{} cluster
catalog is described in Rykoff et al. (in prep).  We refer the reader to that
work for a detailed description of the catalog.  Here, we simply note that the
\redmapper{} calibration of the red sequence requires spectroscopic training
data for galaxy clusters.  This spectroscopic data set is primarily comprised
of existing external spectroscopic surveys, including the Galaxy and Mass
Assembly survey~\citep[GAMA,][]{gamadr1}, the VIMOS VLT Deep
Survey~\citep[VVDS,][]{garillietal08}, the 2dF Galaxy Redshift
Survey~\citep[2dFGRS,][]{collessetal01}, the Sloan Digital Sky
Survey~\citep[SDSS,][]{dr10}, the VIMOS Public Extragalactic
Survey~\citep[VIPERS,][]{garillietal14}, 
the UKIDSS Ultra-Deep Survey~\citep[UDSz,]{bradshawetal13,mclureetal13},
and the Arizona CDFS Environment
Survey~\citep[ACES,][]{cooperetal12}. In addition, we have a small sample of
cluster redshifts from SPT used in the cluster
validation~\citep{bleemetal15}. These data sets have been further supplemented
by galaxy spectra acquired as part of the OzDES spectroscopic survey, which is
performing spectroscopic follow-up on the AAOmega instrument at the
Anglo-Australian Telescope (AAT) in the DES supernova fields \citep{ozdes15}.
The total number of spectroscopic cluster redshifts used in our calibration is
625, most of which are low richness.
By point of comparison, current DES machine learning methods rely on over
46,000 spectra.


\begin{figure*}
\includegraphics[width=140mm]{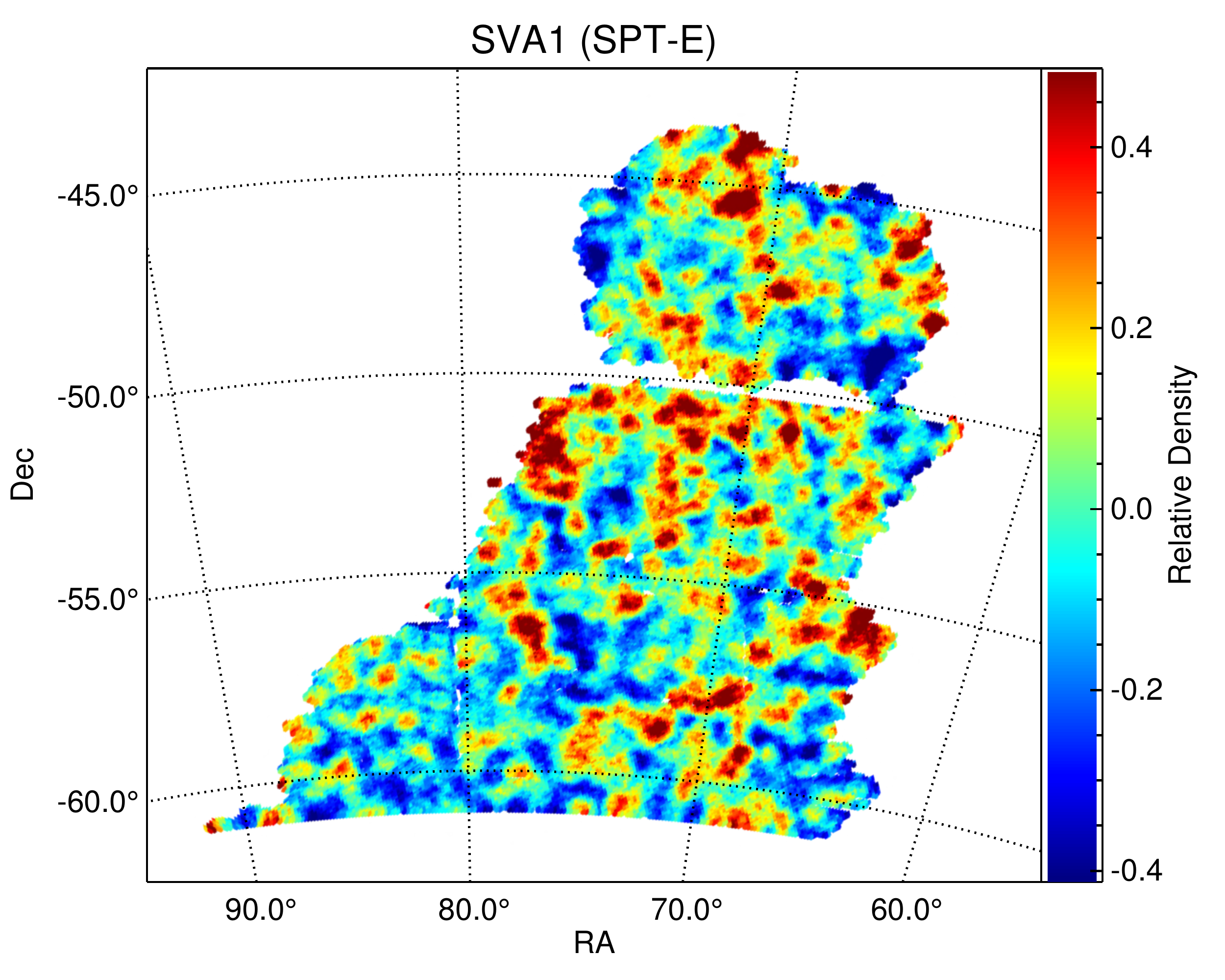}
\caption{Angular galaxy density contrast $\delta=(\rho-\bar \rho)/\bar \rho$ for DES SV \redmagic\
galaxies in the redshift range $[0.2,0.8]$, averaged on a $15'$ scale.  
This plot uses our fiducial \redmagic\ sample (see text).  
}
\label{fig:spte}
\end{figure*}


Figure~\ref{fig:spte} shows the angular density contrast of our fiducial \redmagic\
galaxy sample in the so called DES SV SPTE region.  The full DES SV catalog also includes
the DES supernovae fields, which are disconnected from the SPTE field. 
We note that very nearly all 
the spectroscopic training data sets reside in the DES
supernovae field, which places 
significant limitations in our ability
to validate the performance of \redmagic{} on the DES SV data set.

We note that the survey depth varies significantly over the footprint.
In some regions we can comfortably reach high redshifts
($z\lesssim1$), while in other regions the depth is insufficient.  
To obtain a homogeneous catalog across the full footprint we restrict
ourselves to \redmagic\ galaxies over the redshift range $z\in[0.2,0.8]$.


\subsection{SDSS DR8 Data}

We apply the \redmagic\ algorithm to SDSS DR8 photometric data \citep{dr8}.  
The DR8 galaxy catalog contains $\approx 14,000\ \deg^2$ of imaging, which we
reduce to $\approx 10,000\ \deg^2$ of contiguous high quality
observations using the mask from the Baryon Acoustic
Oscillation Survey (BOSS) \citep{dawsonetal13}.  The mask
is further extended to include all stars in the Yale Bright
Star Catalog~\citep{hoffleitjaschek91}, as well as the area 
around objects in the New General Catalog \citep[NGC][]{sinnott88}.
The resulting mask is that used by \citet{rykoffetal14} to generate the SDSS
DR8 \redmapper\ catalog. We refer the reader to that work for further discussion on the mask.

Galaxies are selected using the default SDSS star/galaxy separator.  We filter
all galaxies with any of the following flags in the $g$, $r$, or $i$ bands:
{\tt SATUR CENTER}, {\tt BRIGHT, TOO MANY PEAKS}, and {\tt(NOT BLENDED OR
  NODEBLEND)}. Unlike the BOSS target selection, we keep objects
flagged with {\tt SATURATED}, {\tt NOTCHECKED}, and {\tt PEAKCENTER}.  
A discussion of these choices can be found in RM1.
Total magnitudes are determined from $i$-band {\tt CMODEL\_MAG} and colors from
$ugriz$ {\tt MODEL\_MAG}. 

The red sequence model is that of the SDSS DR8 \redmapper\ v6.3 cluster
catalog~(Rykoff et al., in prep).
This catalog is an updated version of the \redmapper\ catalog in RM1 (v5.2), and supersedes
both it and the update in \citet[][v5.10]{rozoetal14e}.  Spectroscopic
training data are drawn from the SDSS DR10 spectroscopic data set \citep{dr10}.


\subsection{SDSS Stripe 82 Data}

We apply the \redmagic\ algorithm on SDSS Stripe 82 (S82) coadd
data~\citep{annisetal11}.  The S82 catalog consists of $275\,\mathrm{deg}^2$ of
$ugriz$ coadded imaging over the equatorial stripe. The coadd is roughly 2 magnitudes
deeper than the single-pass SDSS data.  We use the
same flag cuts as those used for the DR8 catalog.  In addition, we clean all
galaxies with extremely large magnitude errors.  Total magnitudes are determined from
$i$-band {\tt CMODEL\_MAG} and colors from $griz$ {\tt MODEL\_MAG}.  Most
modest to high redshift ($z\gtrsim0.3$) red galaxies in S82 are $u$-band
dropouts, so we opted to rely exclusively on $griz$ photometry for S82
runs.  However, in Section~\ref{sec:outliers} we demonstrate the utility of the
$u$-band imaging at low redshift.

We have run the \redmapper\ algorithm in this photometric data set, using SDSS
DR10 spectroscopy as the spectroscopic training data set.  In addition, for
high redshift performance validation we make use of VIPERS~\citep[VIPERS,][]{franzettietal14}. 
During our testing
and validation of the \redmapper\ catalog on these data, we discovered
that $\approx 15\%$ of red cluster member galaxies in the S82 data
set have reported magnitudes that are clearly incorrect in one or more bands.
We do not know the origin of this failure, nor
whether it extends to other galaxies (blue cluster galaxies
or field galaxies).   
These errors inevitably bias the resulting cluster richness estimates.
Consequently, we
have opted not to release the S82 \redmapper\ and \redmagic\ catalogs.  
Nevertheless, we include a discussion of
these data because the \photoz{} performance of \redmagic\ in this
data set provides a valuable baseline to compare against the
DES SV \redmagic\ sample.


\section{The \redmagic\ Selection Algorithm}
\label{sec:algorithm}

The \redmagic\ algorithm can be summarized very simply:
\begin{enumerate}
\item Fit {\it every} galaxy to a red sequence template.  Compute the corresponding best
fit redshift $\zphoto$, and the goodness-of-fit $\chi^2$ of the template fit.
\item Given $\zphoto$, compute the galaxy luminosity $L$.
\item If the galaxy is bright ($L\geq \Lmin$), and it is a good fit to the red sequence template
	($\chi^2 \leq \chimax$), include it in the \redmagic\ catalog.  Otherwise, drop it.
\end{enumerate}
As long as $\chimax$ is sufficiently aggressive,
the resulting catalog will be very nearly comprised of red sequence galaxies exclusively.
In addition, if the red sequence photometric template is accurate, then the resulting 
redshifts should be of excellent quality.
In what follows, we describe how we construct our red sequence template,
and how the maximum goodness-of-fit value $\chimax$ is selected so as to ensure that the resulting
\redmagic\ galaxy sample has a constant comoving space density.  
It should be note that our template is {\it not} a spectroscopic template.  Rather, we model the colors
as a function redshift and magnitude directly, without ever going through a spectrum.  
When we refer to \redmagic\ template,
we always mean our model colors.


\subsection{The \redmagic\ Template}

The \redmagic\ algorithm relies on the \redmapper\ calibration of the red-sequence,
so we begin our discussion by reviewing how the
\redmapper\ template is constructed.
Let $\bc$ be the color vector of a galaxy, and $m$ denote the galaxy's
magnitude in some reference band.  When possible, the reference
band should lie redwards of the 4000\,\AA{} break at all redshifts,
which leads us to select $\mz$ as the reference magnitude
for the DES \redmagic\ sample.  The lower redshift range of the SDSS catalogs
allows us to use $\mi$ in those data sets.  One could in principle use $m_z$ in
SDSS as well, but since SDSS $\mi$ is much less noisy than $\mz$, we rely on
$i$-band for the SDSS data.

Red sequence galaxies populate a narrow ridgeline in color magnitude
space, though with some intrinsic scatter, which we model as Gaussian.
In this case, the ridgeline corresponds to the mean color of red sequence
galaxies.  We write
\be
\avg{\bc|m,z} = \ba(z) + \balpha(z) (m - \mref(z)).
\ee
Here $\ba(z)$ and $\balpha(z)$ are the unknown redshift-dependent amplitude and slope of the
red sequence.
The magnitude $\mref(z)$ defines the pivot point of
the color--magnitude relation.  Its value is arbitrary and can be freely chosen by the experimenter.  \redmapper\ selects $\mref(z)$ 
so that it traces the median magnitude of the cluster member galaxies.
The unknown functions $\ba(z)$ and $\balpha(z)$ are parameterized via spline interpolation, with the model parameters
being the value of the functions at a grid of redshifts.

The covariance matrix $\bCint$ characterizing the intrinsic width of the red sequence in multi-dimensional color space is assumed to 
be independent of magnitude.  The covariance matrix is, however, assumed to vary as a function of redshift.  As with the functions
$\ba(z)$ and $\balpha(z)$, the matrix $\bCint(z)$ is parameterized via spline interpolation, with the model parameters
being the values of each independent matrix element along a grid of redshifts.
Together with the parameters for $\ba(z)$ and $\balpha(z)$, this set of model parameters $\bp$ fully specifies the 
color distribution of red sequence galaxies $P(\bc|\bp;m,z)$.

The parameters $\bp$ specifying our color model are fit using an iterative maximum likelihood approach.  Briefly,
given a cluster galaxy with a spectroscopic redshift $\zspec$, and
a rough estimate for the parameters $\bp$, one can photometrically select cluster galaxies using a matched-filter
approach.  Given these initial photometric cluster members, one then defines the likelihood
\be
\lkhd(\bp) = \prod P(\bc_i|\bp;\mi,z_{\rm cluster})
\ee
where the product is over all the selected cluster members.  
In practice, the likelihood is modified to allow for contamination by interlopers \citep{rykoffetal14}.
A new set of parameters $\bp$ is estimated by maximizing the above likelihood, and the whole procedure
is iterated until convergence.  For further details, see RM1. 
The end result of the above procedure is a strictly empirical calibration of the red sequence of cluster galaxies as a function
of redshift.  


\subsection{\redmagic\ Photometric Redshfits}
\label{sec:photoz}

We want to estimate the photometric redshift of a galaxy of magnitude $m$ and
and color $\bc$.  We use an updated version of the photometric
redshift estimator $\zred$ introduced in RM1.  The probability that a red
galaxy selected from a constant comoving density sample have redshift $z$,
magnitude $m$, and color $\bc$ is denoted via $P(\bc,m,z)$.  One has
\be
P(\bc,m,z) = P(\bc|m,z) P(m|z) P(z).
\ee
We are interested in the redshift probability distribution
\bea
P(z|\bc,m) & = & \frac{P(\bc,m,z)}{P(\bc,m)} \\
	& = & \frac{ P(\bc|m,z) P(m|z) P(z) }{ P(\bc,m) }.
\eea
Since the denominator is redshift independent, we can ignore it.  The corresponding likelihood is
\be
\lkhd(z) = P(\bc|m,z) P(m|z) P(z).
\ee

For a constant comoving density sample $P(z)\propto |dV/dz|$.  
$P(m|z)$ is modeled assuming the galaxies follow a Schechter luminosity function,
\be
P(m|z) \propto 10^{-0.4(m-m_*)(\alpha+1)} \exp\left[ -10^{-0.4(m-m_*)} \right].
\ee
The value $m_*(z)$ is set to $\mi=17.85$ at $z=0.2$ to match \redmapper.
The evolution of $m_*(z)$ is computed using the
\citet[][BC03]{bruzualcharlot03} stellar population synthesis code 
as implemented in the {\tt EzGal} Python package\footnote{http://www.baryons.org/ezgal}.  
We model $m_*(z)$ using a single star formation burst at $z=3$, and we have 
confirmed this evolution matches that in RM1 at $z<0.5$.
The normalization condition for $\mz$ for DES is then derived
from the BC03 model using the DECam passband.  
Finally, $P(\bc|m,z)$ of our red sequence model, so that
\be
P(\bc|m,z) \propto \exp\left( - \frac{1}{2} \chi^2(z) \right)
\ee
where
\be
\chi^2(z) = (\bc - \avg{\bc|m,z}) \bCtot^{-1}(\bc-\avg{\bc|m,z}) \label{eq:chisq}
\ee
and 
\be
\bCtot = \bCint+\bCobs
\ee
is the total scatter about the red sequence color.  Here, $\bCobs$ is the covariance
matrix describing the photometric errors in the galaxy colors.
Our final expression for the redshift likelihood is therefore
\be
\ln \lkhd(z) = -\frac{1}{2}\chi^2(z) + \ln P(m|z) + \ln \left \vert \frac{dV}{dz} \right \vert.
\ee

The photometric redshift $\zred$ is the redshift at which this log-likelihood function is maximized,
and the corresponding $\chi^2$ value is denoted $\chired$.   In addition, the galaxy is also assigned
a luminosity $l=L/L_*(\zred)$,
\be
l(m,\zred) = \frac{L}{L_*} = 10^{-0.4(m-m_*(\zred))}.
\ee
The photometric redshift error $\sigma_z$ is estimated using the variance of the posterior,
\be
\sigma_z^2 = \avg{z^2}-\avg{z}^2
\ee
where
\be
\avg{z^n} = \frac{ \int dz\ \lkhd(z)z^n }{ \int dz\ \lkhd(z) }.
\ee


\subsection{Selection Cuts}
\label{sec:chicut}

We wish to select luminous red galaxies.  Consequently, we demand that all
galaxies have a luminosity $l\geq \lmin$, where $\lmin = L_\mathrm{min}/L_*$ is
a selection parameter that is to be determined by the experimenter.  To ensure
that our final galaxy sample is comprised of red sequence galaxies, we further
demand that our red sequence template be a good fit by applying the selection
cut
\be
\chired \leq \chimax(\zred).
\ee
Note the $\chi^2$ cut $\chimax(z)$ can be redshift dependent.
The simplest possible model is $\chimax(z)=k$
for some constant $k$, but this is rather arbitrary.
What we really want is to be able to select the ``same'' sample
of galaxies at all redshifts.
In the absence of merging, red sequence galaxies evolve passively, resulting
in a constant comoving density sample.  Of course, galaxies do merge, so this
approximation cannot be exactly correct, but this can nevertheless be a useful
approximation for comparing galaxies across relatively narrow redshift intervals.
Thus, rather than applying a constant $\chi^2$
cut, we construct the selection
threshold $\chimax(z)$ such that the resulting galaxy sample has a constant comoving galaxy
density.  This selection also justifies our assumption that $P(z)\propto |dV/dz|$ in the construction
of the redshift likelihood. 

To ensure a constant comoving space density of \redmagic\ galaxies, we parameterize $\chimax(z)$ using 
spline parameterization.  The model parameters $\bq$ are the values of $\chimax$ along a grid
of redshifts, and the value of $\chimax(z)$ everywhere else is defined via spline interpolation.
We will come back to how the parameters $\bq$ are chosen momentarily.  Before we do so, however, we 
need to describe an additional
calibration step we take in order to improve the photometric redshift performance of the \redmagic\
algorithm.


\subsection{Photo-$z$ Afterburner}
\label{sec:afterburner}

The \redmagic\ selection cuts are fully specified by the parameter $\lmin$
and the parameters $\bq$ defining the function $\chimax(z)$.   
If a random fraction of the selected galaxies have spectroscopic
redshifts $\zspec$, we can use these galaxies to remove
any biases in our \photozs.
For instance, given the \redmagic\ selection specified by $\lmin$ and $\bq$,
we could split the spectroscopic galaxies in two, a training sample and a validation
sample.  We can then use the training sample to compute the median redshift offset
$\zspec-\zred$ in bins of $\zred$.  We denote this quantity as $\Delta z(\zred)$. 
Our new photometric redshift estimator is
\be
\zrm = \zred + \Delta z(\zred),
\ee
which we can validate with the validation data set.

In practice, $\Delta z(\zred)$ is defined using spline interpolation, with the
spline parameters being determined by minimizing the cost function
\be
E_\Delta = \sum_j |z_{{\rm spec},j} - z_{{\rm rm},j}|
\ee
where the sum is over all spectroscopic \redmagic\ galaxies.
We add the absolute values rather than the squares to reduce the impact of possible catastrophic outliers.

Of course, in general one is hardly assured spectroscopic redshifts for a large
{\it representative} sample of \redmagic\ galaxies. We overcome this problem by
relying instead on \redmagic\ galaxies that are members of \redmapper\ clusters
(membership probability $\pmem \geq 0.9$), using the \redmapper\ photometric cluster
redshift $z_\lambda$ as the ``spectroscopic'' redshift of the calibration
galaxies.  Roughly, the redshift $z_\lambda$ is obtained by simultaneously fitting
the ensemble of cluster galaxies with a single photometric redshift.
It has already been shown that \redmapper\ redshifts 
are unbiased and much more accurate than the photometric redshifts of individual
galaxies.  We emphasize that by making use of photometric cluster
members our calibration sample is not restricted to the brightest \redmagic{}
galaxies, as would be the case of a typical spectroscopic calibration sample.

In addition to modifying the photometric redshift estimate $\zrm$, we also modify the photometric redshift errors.  
Imagine again binning the galaxy calibration sample by $\zrm$.  For each bin, we could compute the Median
Absolute Deviation $MAD=\rm{median}\{|\zred-z_\lambda|\}$.  For a Gaussian distribution, $\avg{MAD}=\sigma_z$/1.4826,
where $\sigma_z$ is the standard deviation.  Thus, the quantity $1.4826|\zrm-z_\lambda|$ is an estimator for $\sigma_z$.
Let then $\sigma_0$ be our original photometric redshift error estimate as per Section~\ref{sec:photoz}.  We assume
that the corrected photometric redshift error $\sigma_1$ for each galaxy is given by $\sigma_1=r(\zrm)\sigma_0$,
where $r(\zrm) = \sigma_z/\sigma_0$.  Rather than doing this in bins, we parameterize $r(z)$ via spline
interpolation, with the best fit parameters being those which minimize the cost function
\be
E_\sigma = \sum_j \big| \ 1.4826|z_{{\rm rm},j}-z_{\lambda,j}| - r(z_{{\rm rm},j})\sigma_{0,j}\  \big |.
\ee
The sum is over all calibration galaxies, and we again use absolute values to reduce the impact
of possible catastrophic outliers.
We note that the afterburner perturbations to the photometric redshifts are small, but do improve
photometric redshift performance.

With the new estimator $\zrm$ in hand and its improved error estimate, 
we can recompute the luminosity $l$ and $\chi^2$ of every galaxy in the survey, 
and reapply our selection cuts to arrive at an improved \redmagic\ sample.


\subsection{$\chimax$ Calibration}
\label{sec:chimax_calib}

We have seen how to select \redmagic\ galaxies given the selection parameters
$\bq$, but we have yet to specify how the parameters $\bq$ are selected.  To do
so, we first define a series
of redshift bins $z_j$ going from the minimum redshift of interest $\zmin$ to
the maximum redshift $\zmax$.  Given a set of selection parameters $\bq$, we
construct the \redmagic\ sample by applying the luminosity and $\chi^2$ cuts as
above.  Next, we compute the \photoz{} afterburner parameters for the sample
derived from the parameters $\bq$, which allows us to compute $\zrm$ for every
galaxy.  We then measure the comoving space density $n_j(\bq)$ in each redshift
bin $j$.  Since we want to enforce a constant comoving density
$\bar n$, we define the cost function $E(\bq)$ via 
\be
E(\bq) = \sum_{j} \frac{ (n_j(\bq) - \bar n)^2 }{ \bar n V_j^{-1} }
\ee
where the sum is over all redshift bins, $V_j$ is the comoving volume of
redshift bin $j$, $n_j$ is the empirical \redmagic\ galaxy density in redshift
bin $j$.  The denominator is the expected Poisson error for a galaxy density
$\bar n$.  The spline parameters $\bq$ are obtained by minimizing the cost
function $E(\bq)$ using the downhill-simplex method of \citet{nelder65}.  We
always use redshift bins that are significantly narrower than the spacing
between spline nodes, and we take care to ensure that the number of galaxies
$n_jV_j \gg 1$ in every redshift bin.  We emphasize that the \photoz{}
afterburner parameters are re-estimated at every iteration in the minimization,
to ensure that we have a consistent sample selection given the updated galaxy
redshifts.  Finally, with the spline parameters determined, we apply the
corresponding $\chired \leq \chimax(\zrm)$ cut to arrive at the final
\redmagic\ galaxy sample.


\begin{figure*}
\hspace{-12pt} \includegraphics[width=90mm]{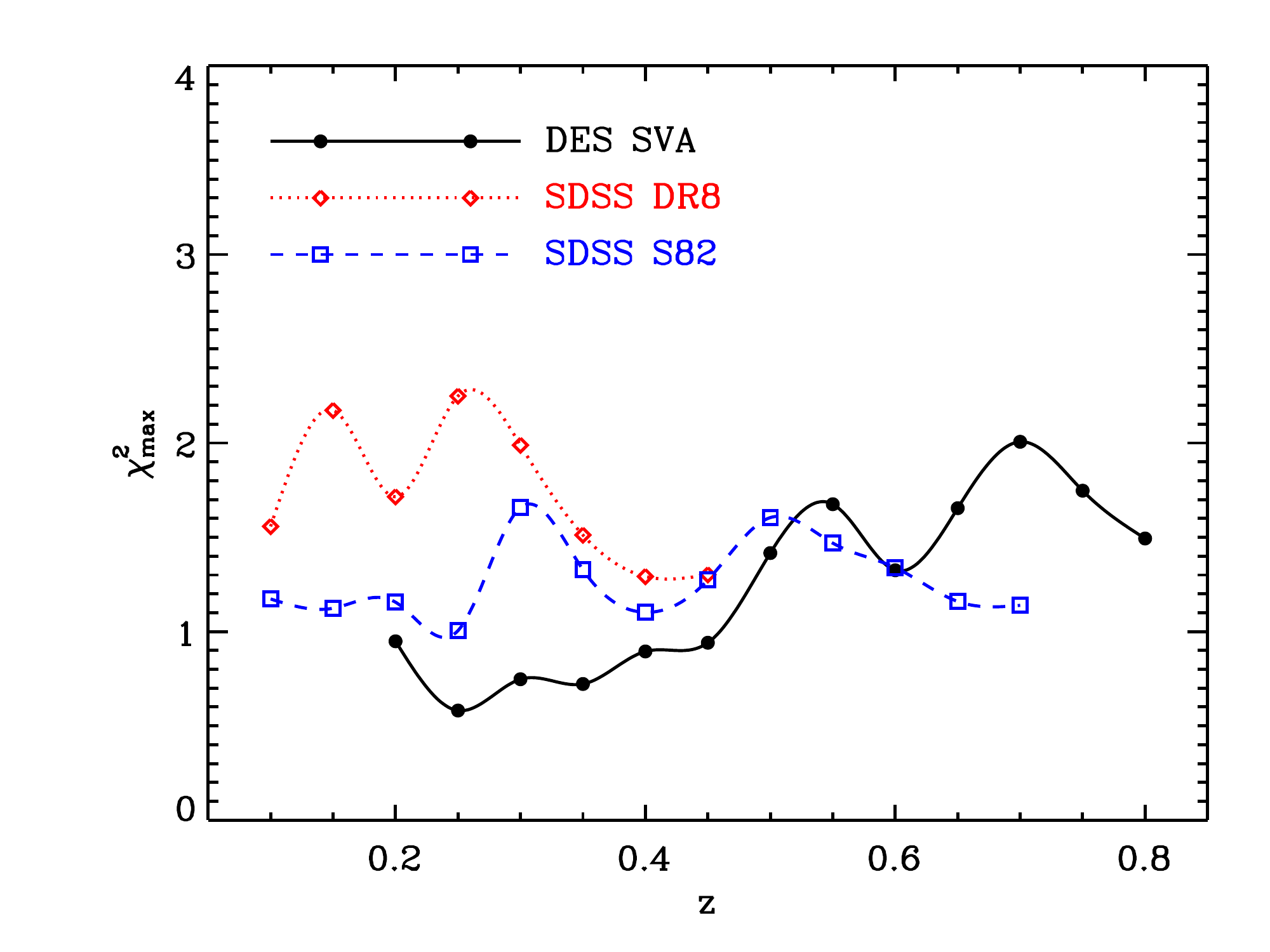}
\hspace{-12pt} \includegraphics[width=90mm]{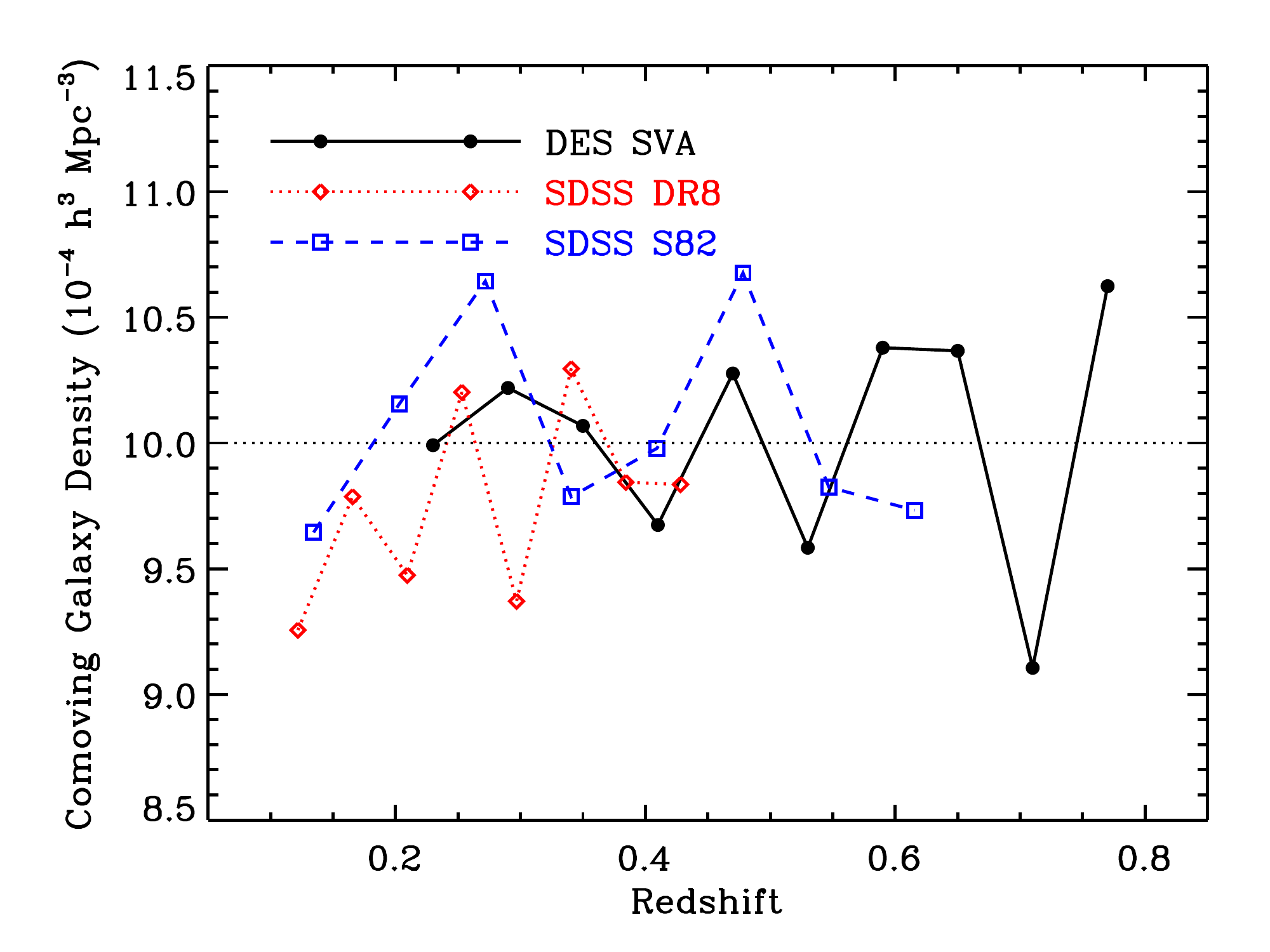}
\caption{{\bf Left:} Selection cut $\chi^2_{\rm{max}}$ as a function of redshift defining each
of the \redmagic\ galaxy samples, as labelled.  The symbols mark the spline nodes
defining the function $\chimax(z)$, while the lines show the corresponding
spline interpolation at every point.
{\bf Right:} \redmagic\ comoving galaxy density as a function of redshift for 
each of the three data sets employed in this work, as labelled.  The target comoving
space density was $10^{-3}\ h^3\ \Mpc^{-3}$ (horizontal dotted line).
}
\label{fig:sel}
\end{figure*}


\subsection{Selection Summary}
\label{sec:selection_summary}

Despite the computational complexity of the above selection, it is worth
emphasizing that our selection algorithm contains only two free
parameters, both of which have clear physical interpretations: the luminosity cut
$\lmin$, and the desired space density $\bar n$ of the resulting galaxy sample.
Importantly, the ``color cuts'' that select red-galaxies are self-trained from the
data.  By comparison, the SDSS CMASS galaxy selection involves 12 parameters
hand-picked {\it a priori} to produce an approximately stellar-mass limited sample at $z\geq 0.45$~\citep{dawsonetal13}.

It is also important to note that our selection makes it very easy to test different selection 
thresholds, allowing one to optimize galaxy selection for scientific purposes.  
Some patterns emerge: $\lmin$ must always be low enough for
the corresponding $\chimax$ threshold to be reasonable (i.e. $\chi^2/\mathrm{dof} \lesssim
2$).  If $\lmin$ is too large, \redmagic\ will start pulling in galaxies with
large $\chi^2$ values in order to attempt to reach the desired space density,
which will result in a large number of \photoz{} outliers.  We find that when this
happens it becomes difficult to construct a truly flat $n(z)$ sample, so
checking the comoving space density of the \redmagic\ catalog is a quick
an easy way to test whether the
\redmagic\ algorithm is performing as desired.

We illustrate the performance of our algorithm in Figure~\ref{fig:sel} for a set
of fiducial cuts $\lmin=0.5$ and $\bar n = 10^{-3}\ h^3\ \Mpc^{-3}$. 
The left panel shows the $\chi^2(z)$ threshold for each of our three \redmagic\
samples, while the right panel shows the resulting galaxy comoving densities as a function
of redshift. We see that in all cases the observed space density is
close to flat, and that the $\chi^2$ thresholds are low, as desired.


\section{Photo-$z$ Performance}
\label{sec:performance}

We consider two sets of \redmagic\ galaxies.  The first is our fiducial sample,
selected to be galaxies brighter than $0.5L_*$ and with a space density $\bar
n=10^{-3}\ h^3\ \Mpc^{-3}$.  Unless otherwise stated, all of the results noted
below correspond to these fiducial selection parameters.  The second sample is a high
luminosity, low space density \redmagic\ sample, comprised of galaxies brighter
than $L_*$ with a space density of $2\times 10^{-4}\ h^3\ \Mpc^{-3}$.  
This high luminosity sample will be useful for comparing against other
commonly used galaxy samples, particularly CMASS.

\begin{figure*}

\hspace{-12pt} \includegraphics[width=90mm]{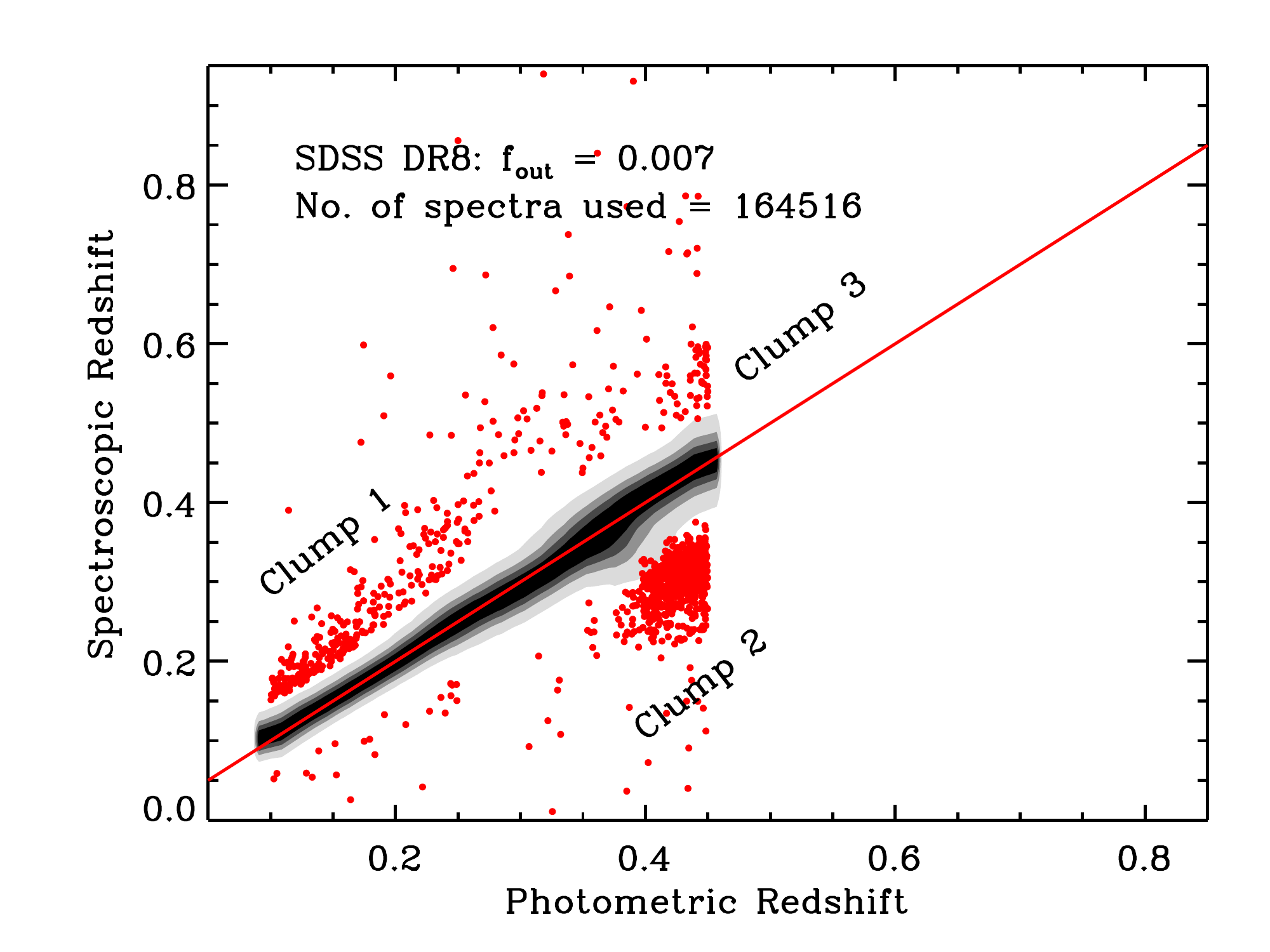}
\hspace{-12pt} \includegraphics[width=90mm]{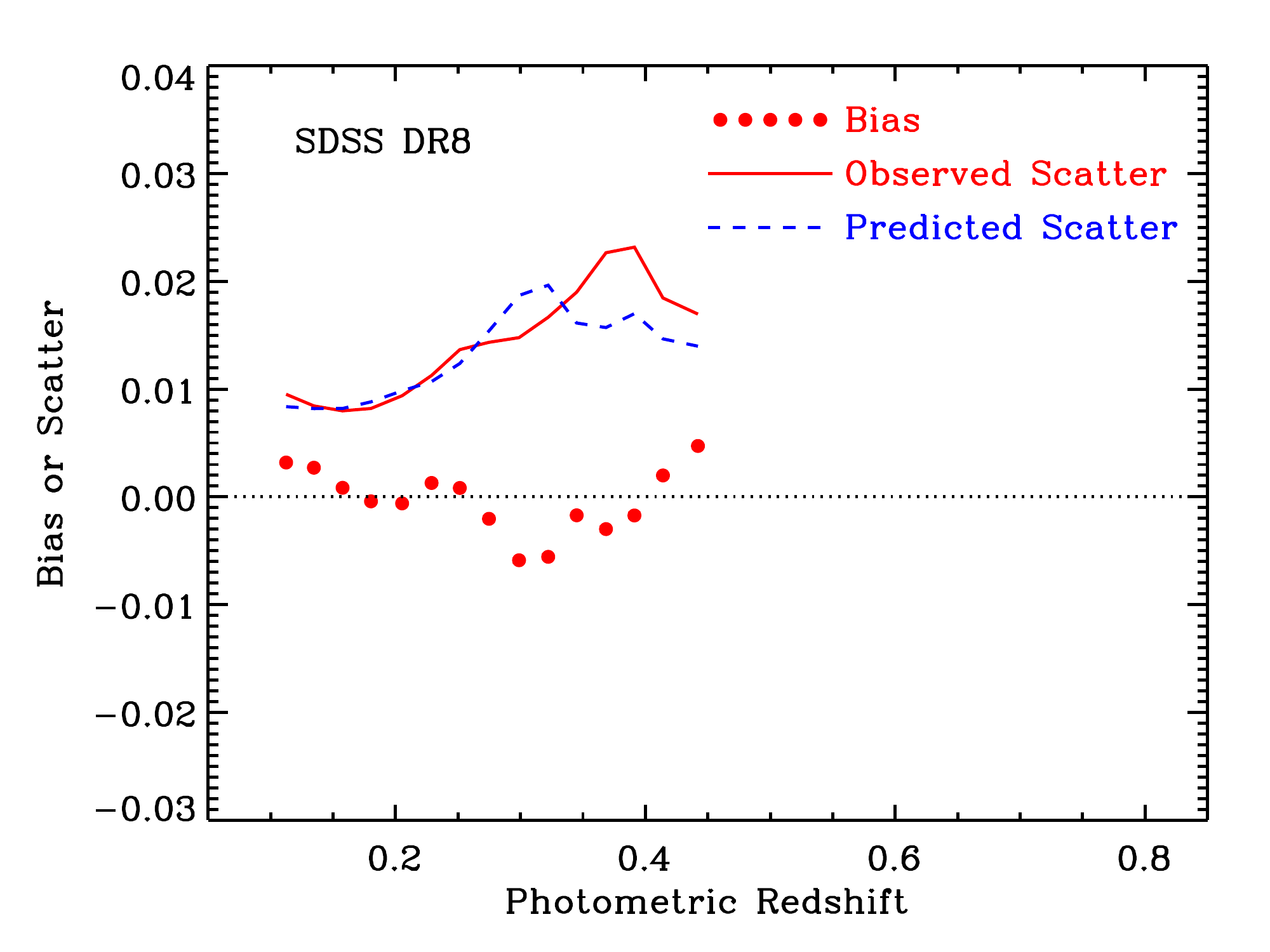}
\vspace{-5pt}

\hspace{-12pt} \includegraphics[width=90mm]{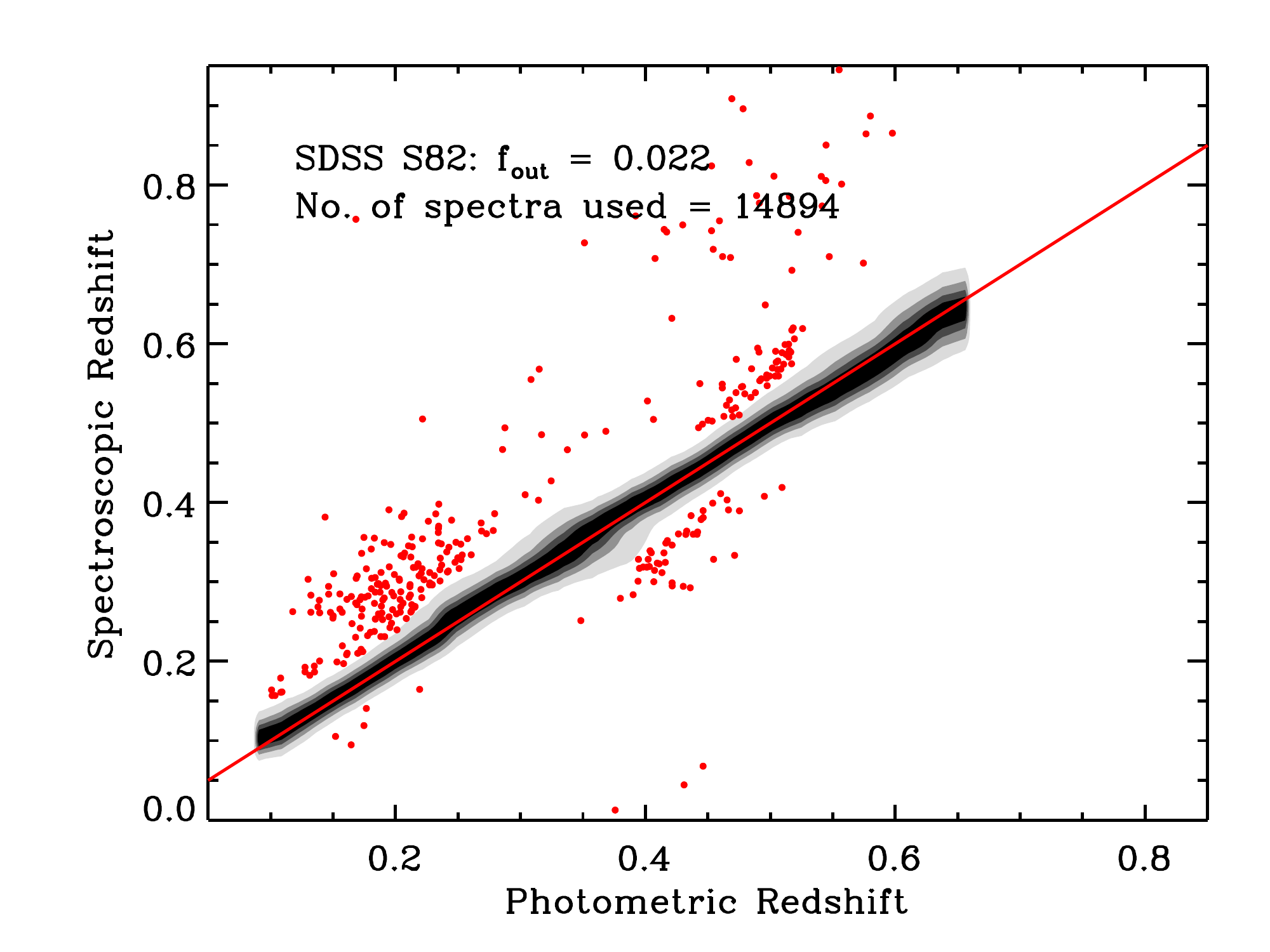}
\hspace{-12pt} \includegraphics[width=90mm]{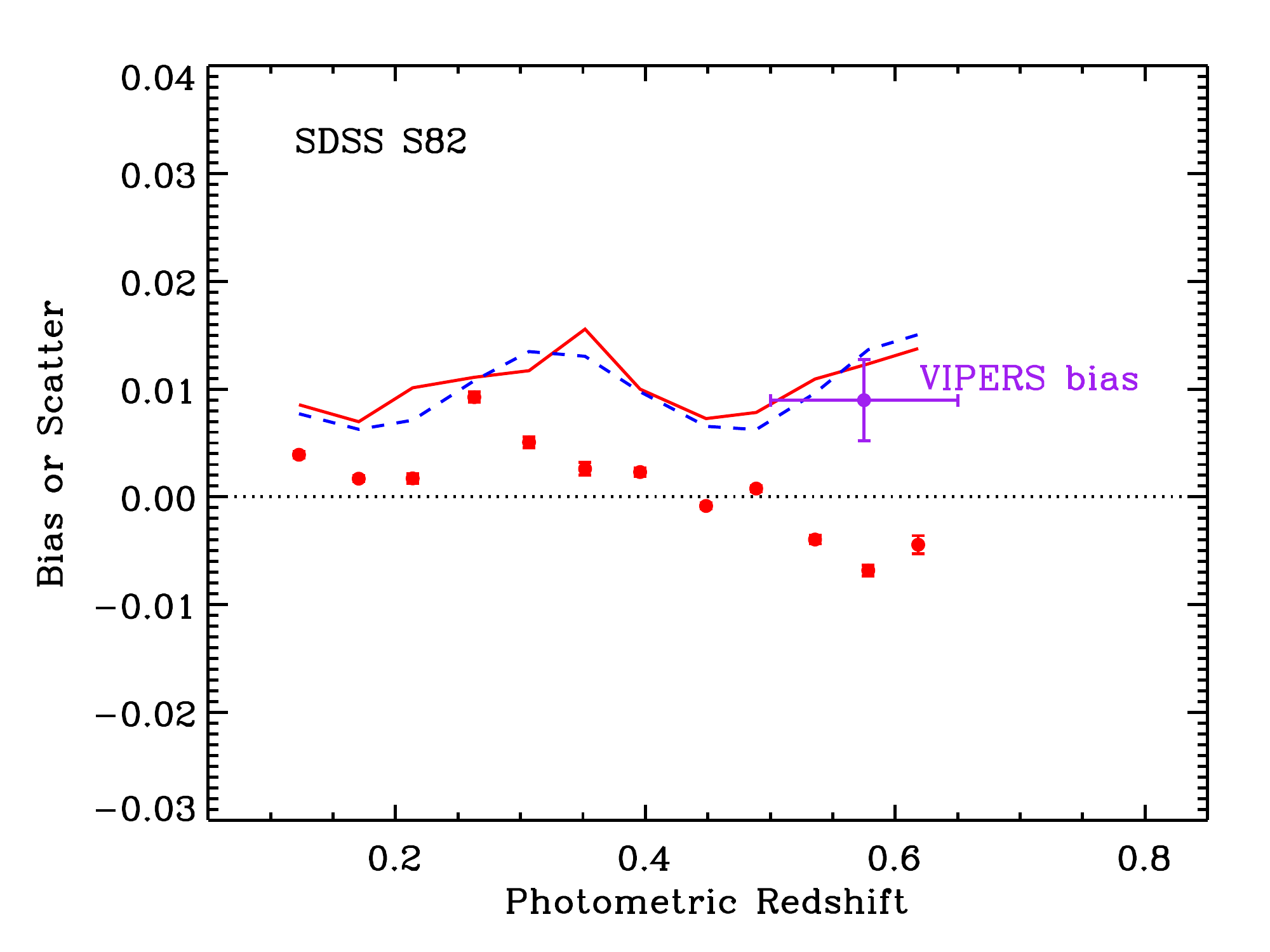}
\vspace{-5pt}

\hspace{-12pt} \includegraphics[width=90mm]{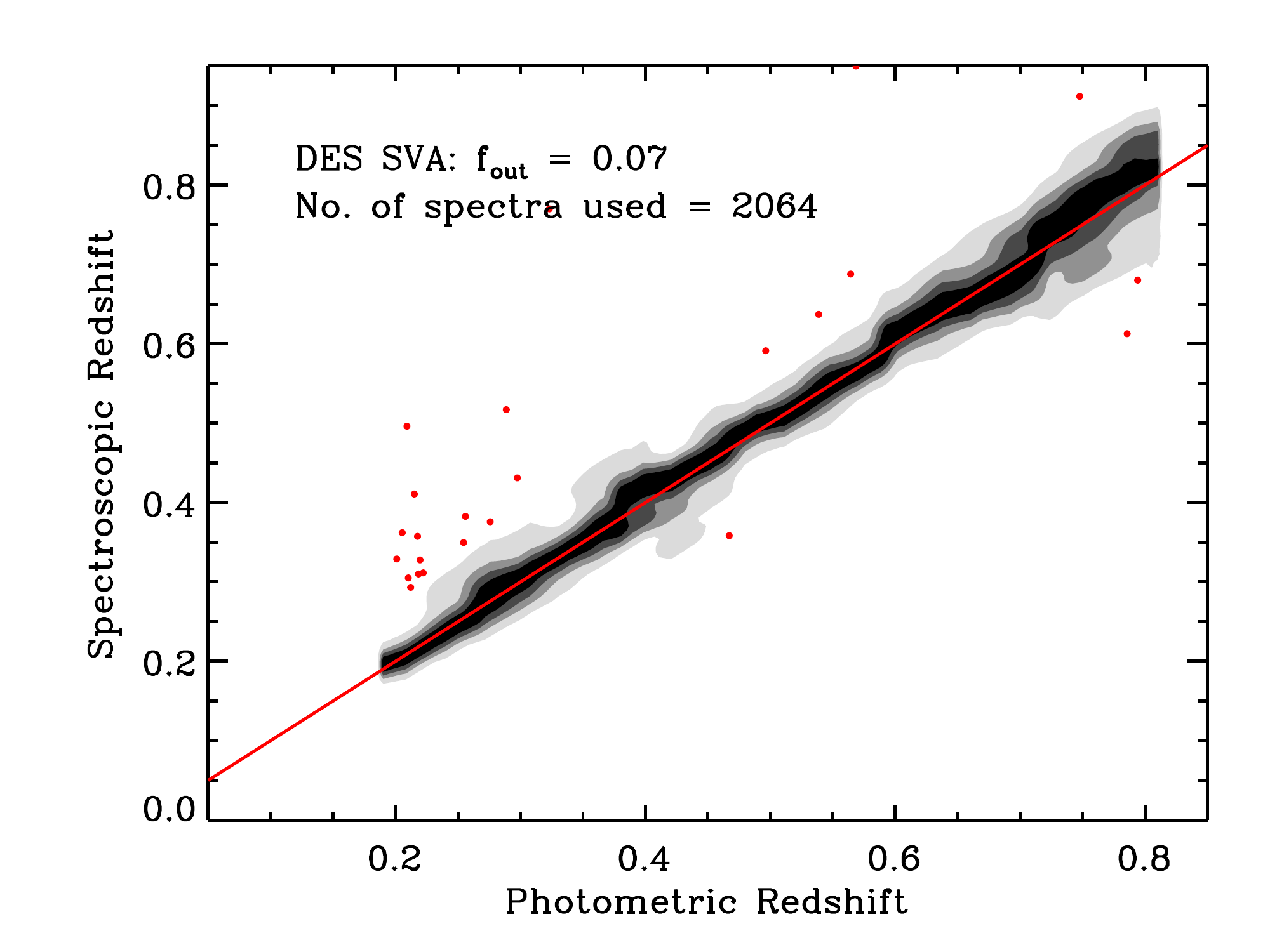}
\hspace{-5pt} \includegraphics[width=90mm]{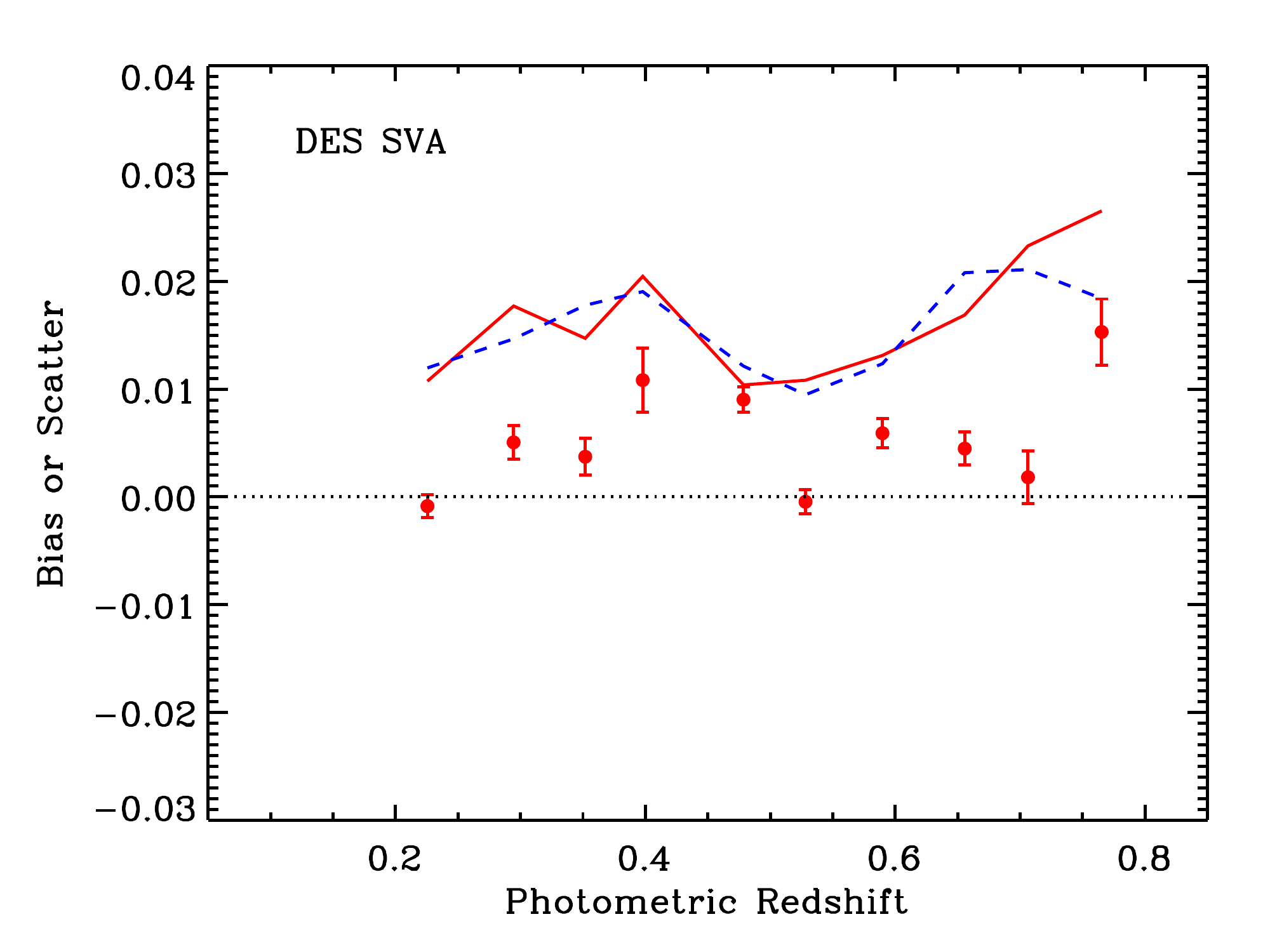}

\caption{{\bf Left:} Spectroscopic redshift vs photometric redshift for the fiducial \redmagic\
galaxy sample in each of the various data sets considered in this work.  Red points are $5\sigma$ outliers, while the
red line corresponds to $\zspec=\zphoto$.  
{\bf Right:} Photometric redshift performance statistics.  Red points with error bars are the photometric redshift bias,
defined as the median value of $\zspec-\zphoto$.  All statistics for the SDSS data sets are computed using SDSS spectroscopy,
except for the purple VIPERS point for S82.
The red curve is the observed scatter of $(\zphoto-\zspec)/(1+\zspec)$, while the dashed blue curve is the predicted scatter 
based on the available photometry. 
The horizontal error bar for the S82 plot shows the width of the redshift bin used in the VIPERS measurement.
}
\label{fig:photoz}
\end{figure*}

Figure~\ref{fig:photoz} shows the photometric redshift performance for our fiducial selection
in the SV, DR8, and S82 data sets.
The spectroscopic data used to characterize the photometric
redshift performance were described in Section~\ref{sec:data}.
The photometric redshift bias $\overline{\Delta z}$ is defined
as the median offset of $\Delta z = \zspec-\zrm$.  The scatter is defined as $1.4826\times MAD$,
where $MAD$ is the median absolute deviation, i.e. the median of $| \Delta z - \overline{\Delta z}|$.  
For Gaussianly distributed data, $\overline{\Delta z}$ and $.4826\times MAD$
are unbiased estimators of the mean and standard deviation of these offsets.  In using
median statistics, our results are robust to a small fraction of gross outliers. 

The most obvious features in the left-hand plots of Figure~\ref{fig:photoz}
are the three clumps of outlier points.  These are obvious for both DR8 and
S82 data, but not apparent in the DES SV data.  
We are confident this reflects the paucity of spectra in the DES data
rather than a sudden and unexpected improvement in the \redmagic\ performance.
We discuss each of these clumps in Section~\ref{sec:outliers}.


Turning to the bias and scatter plots in the right column of
Figure~\ref{fig:photoz}, we see that for all data sets there is excellent
agreement between the observed redshift scatter (red solid line) and the
predicted \photoz{} uncertainty (dashed blue line).  The latter is simply the
median \photoz{} error in each bin. Note that the predicted redshift errors
in the SDSS S82 and DES SV data sets are clearly double-humped.  This is
expected: photometric redshift uncertainties increase whenever the
$4000\,\ang$ break feature in the spectra of these galaxies falls in between
filters.  At
$z\approx 0.35$ there is a peak associated with the $g$ to $r$ filter
transition, and at $z\approx 0.65$ we see a second peak associated with the
$r$ to $i$ filter transition.


\begin{figure*}
\hspace{-12pt} \includegraphics[width=90mm]{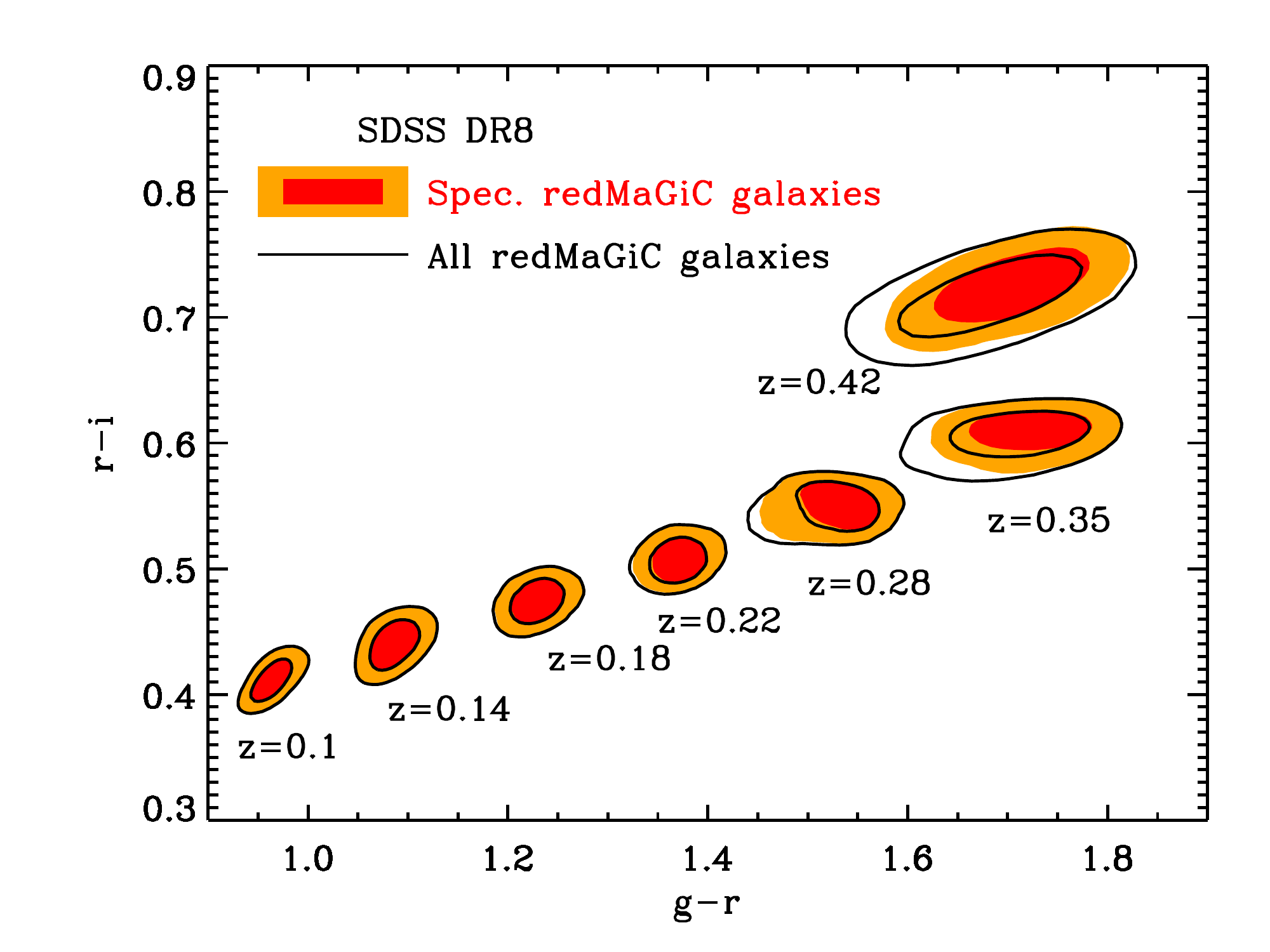}
\hspace{-12pt} \includegraphics[width=90mm]{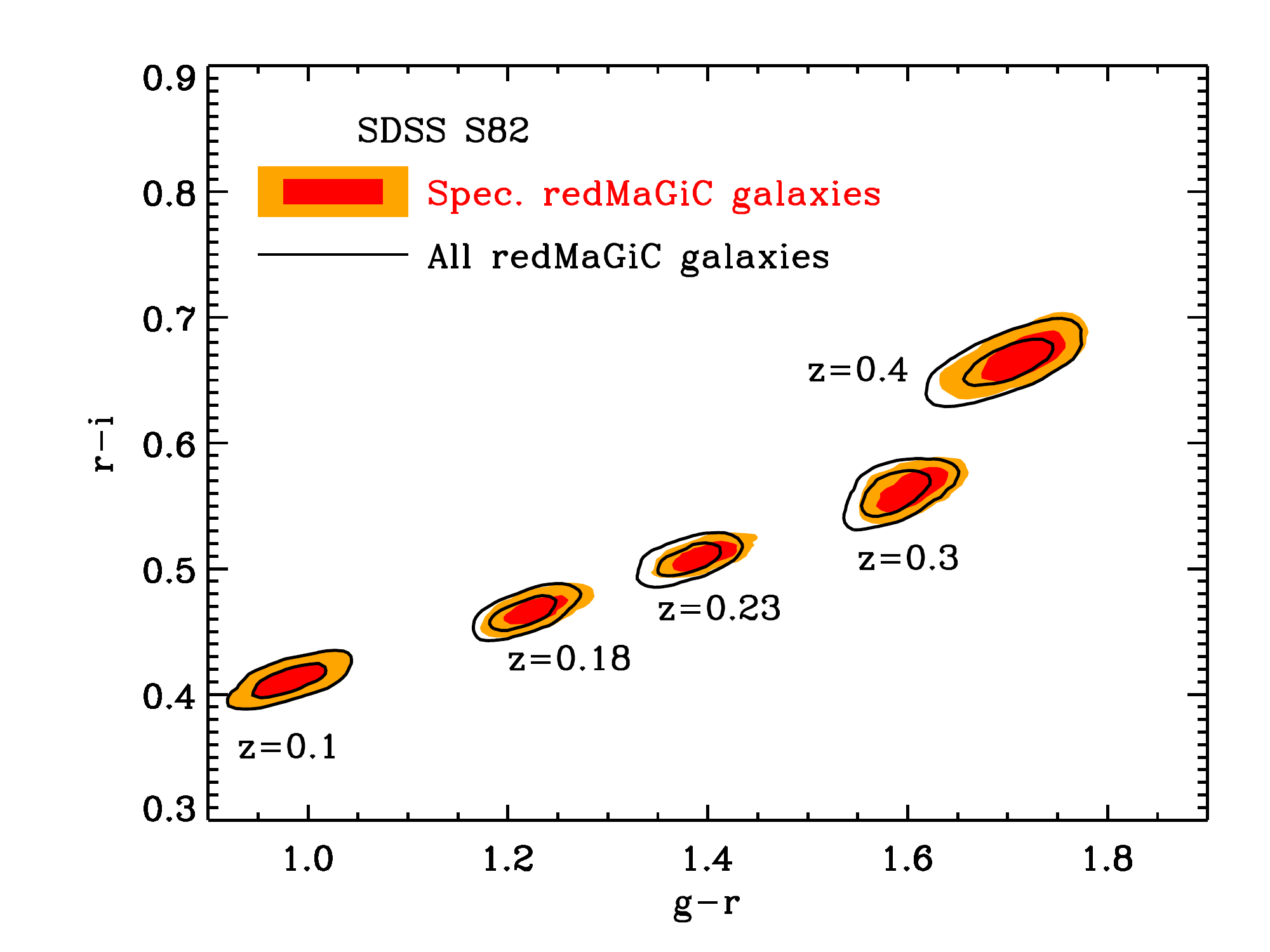}
\caption{68\% and 95\% galaxy density contours in $g-r$ vs $r-i$ space for DR8 and S82 \redmagic\
galaxies for a variety of redshift bins, as labelled. Red/orange contours
correspond to \redmagic\ galaxies with spectroscopic redshifts,
while the solid black curves show the contours for the full \redmagic\ sample.  A mismatch between the colored and black 
ellipses implies biased spectroscopic sampling of \redmagic\ galaxies.
}
\label{fig:biasing}
\end{figure*}


Comparing the three data sets,
we see DR8 and S82  have nearly identical
photometric redshift errors at low redshifts, which demonstrates that the redshift
errors are set by the intrinsic width of the red sequence.   By contrast, 
at $z\gtrsim 0.4$ the photometric errors
in DR8 are clearly important, and so its \photoz\ errors are larger than those in S82.
Notably, DES has larger photometric
redshift scatter than the SDSS data sets.  There are several contributors
to this result.  First, the
spectroscopic training set for \redmapper{} training is still quite
sparse, and so the \redmapper{} calibration is expected to be noisier than in
the SDSS data sets.  Second, 
DES SV {\tt MAG\_AUTO} colors are expected to be intrinsically noisier than SDSS 
{\tt MODEL\_MAG} colors, leading to larger uncertainties.

Turning to the bias, we see that the
DR8 \redmagic{} sample appears to have a negative bias at $\zrm \approx 0.3$.
By contrast, the S82 sample exhibits a slight positive bias at the same
photometric redshift.  The situation reverses at $z\approx 0.25$, with S82 galaxies
exhibiting bias while DR8 galaxies do not.
We believe these biases are driven by non-representative spectroscopic sampling
of \redmagic\ galaxies.  Specifically, our photometric redshift tests rely on
the subset of \redmagic\ galaxies that have spectra.  If that subset is biased relative
to the full population, we would in fact expect to see a photometric redshift bias.

Figure~\ref{fig:biasing} shows \redmagic\ galaxy
density contours in the $g-r$ vs $r-i$ plane for several photometric redshift
bins.  The filled red and orange contours show the regions containing 68\% and 95\%
of all \redmagic\ galaxies with spectroscopic redshifts.  The solid ellipses
show the corresponding regions for all \redmagic\ galaxies with a magnitude
threshold set by the spectroscopic \redmagic\ sub-sample. 
Offsets between the
red--orange contours and the solid line contours imply a non-representative spectroscopic
sampling of the \redmagic\ galaxy population.

It is clear from Figure~\ref{fig:biasing} that DR8 spectroscopic sampling
is biased at $z\gtrsim 0.3$, with the reddest galaxies start being
somewhat over-sampled. There is a similar trend of over-sampling the reddest
\redmagic{} galaxies in S82 starting at $z\approx 0.23$.
These differences appear to be correlated with the presence of ``large'' \photoz\
biases in Figure~\ref{fig:photoz}.

The \photoz{} bias at $z\approx0.6$ in the S82 data is rather unusual. It is large and negative
($\approx -0.005$) when using SDSD spectroscopy, but large and positive ($\approx 0.009$)
when using VIPERS.  The difference between the two spectroscopic data sets further highlights
the importance that spectroscopic sampling can have on our conclusions.  


\begin{center}
\begin{table*}
\centering
\caption{Photometric redshift performance of \redmagic\ galaxies. All quantities are first computed in redshift bins, and then
the median of the binned values is reported.  Bias and $\vert$Bias$\vert$ are the median values for $(\zspec-\zphoto)$
and $|\zspec-\zphoto|$ respectively.  The scatter is $1.4826\times MAD$ where $MAD$ is the median
absolute deviation of $|\zspec-\zphoto|/(1+\zspec)$. The predicted scatter is the median
value of $\sigma_z/(1+\zphoto)$ where $\sigma_z$ is the reported \photoz\ error.
}
\begin{tabular}{lllccccc}
Space Density & Redshift Range & Data Set & Bias & $\vert$Bias$\vert$ & Scatter & Predicted Scatter  & $5\sigma$ Outlier Fraction \\
\hline
$10^{-3}\ h^3\ \Mpc^{-3}$ & $z\in[0.2,0.8]$ & DES SV & 0.51\% & 0.51\% & 1.69\% & 1.78\% & 1.4\% \\
& $z\in[0.1,0.65]$ & SDSS S82 & 0.17\% & 0.39\% & 1.10\% & 0.97\% & 2.2\%  \\
& $z\in[0.1,0.45]$ & SDSS DR8  & -0.04\% & 0.20\% & 1.43\% & 1.40\% & 0.8\% \vspace{5pt} \\
\hline
$2\times 10^{-4}\ h^3\ \Mpc^{-3}$ & $z\in[0.2,0.8]$ & DES SV & 0.19\% & 0.37\% & 1.50\% & 1.59\% & 0.9\% \\
& $z\in[0.1,0.65]$ & SDSS S82 & 0.14\% & 0.22\% & 1.04\% & 1.03\% & 1.5\% \\
& $z\in[0.1,0.45]$ & SDSS DR8 & -0.22\% & 0.22\% & 1.40\% & 1.46\% & 1.9\% \vspace{5pt} \\
\end{tabular}
\label{tab:photoz}
\end{table*}
\end{center}

\begin{figure*}
\hspace{-12pt} \includegraphics[width=90mm]{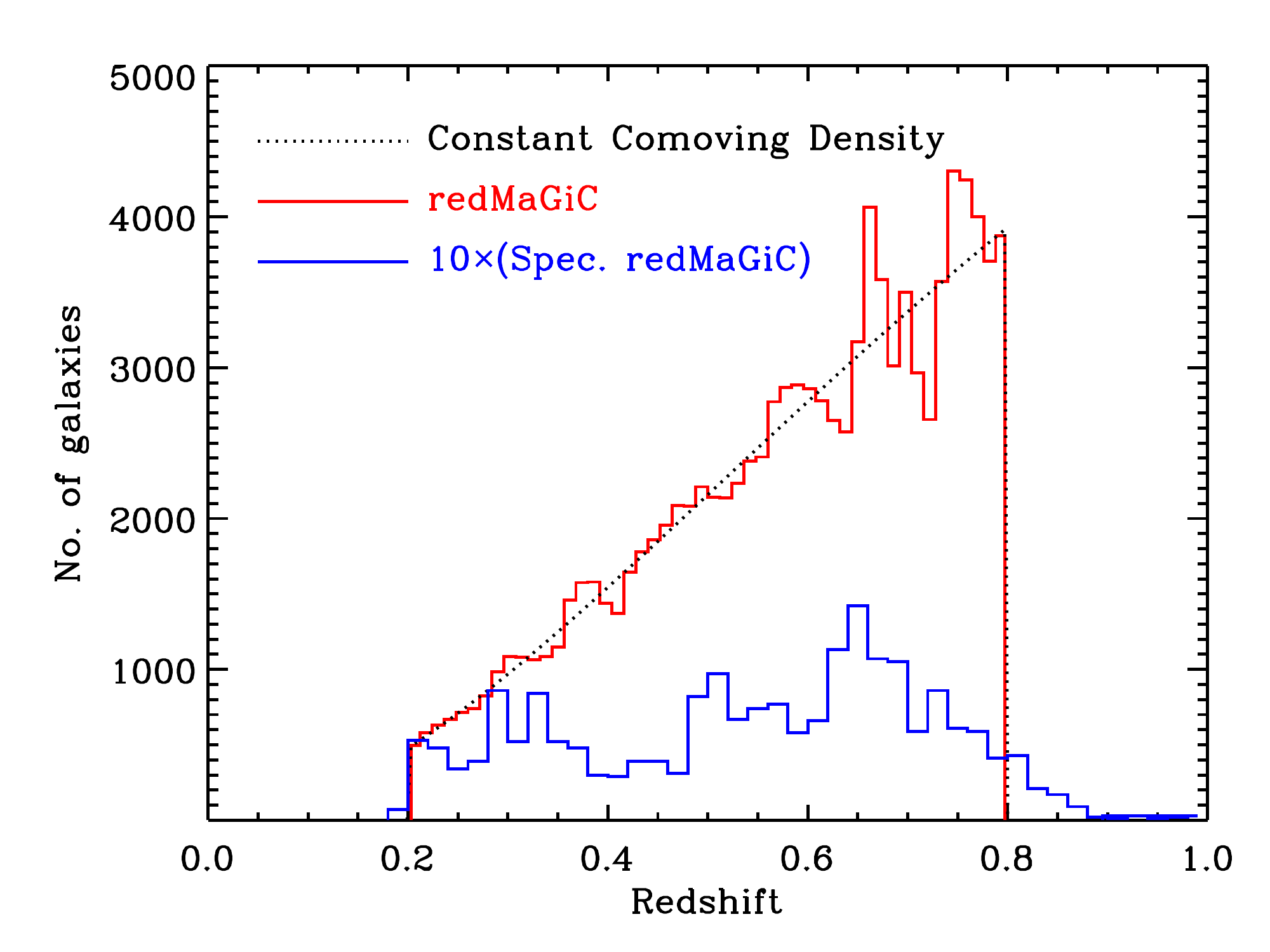}
\hspace{-12pt} \includegraphics[width=90mm]{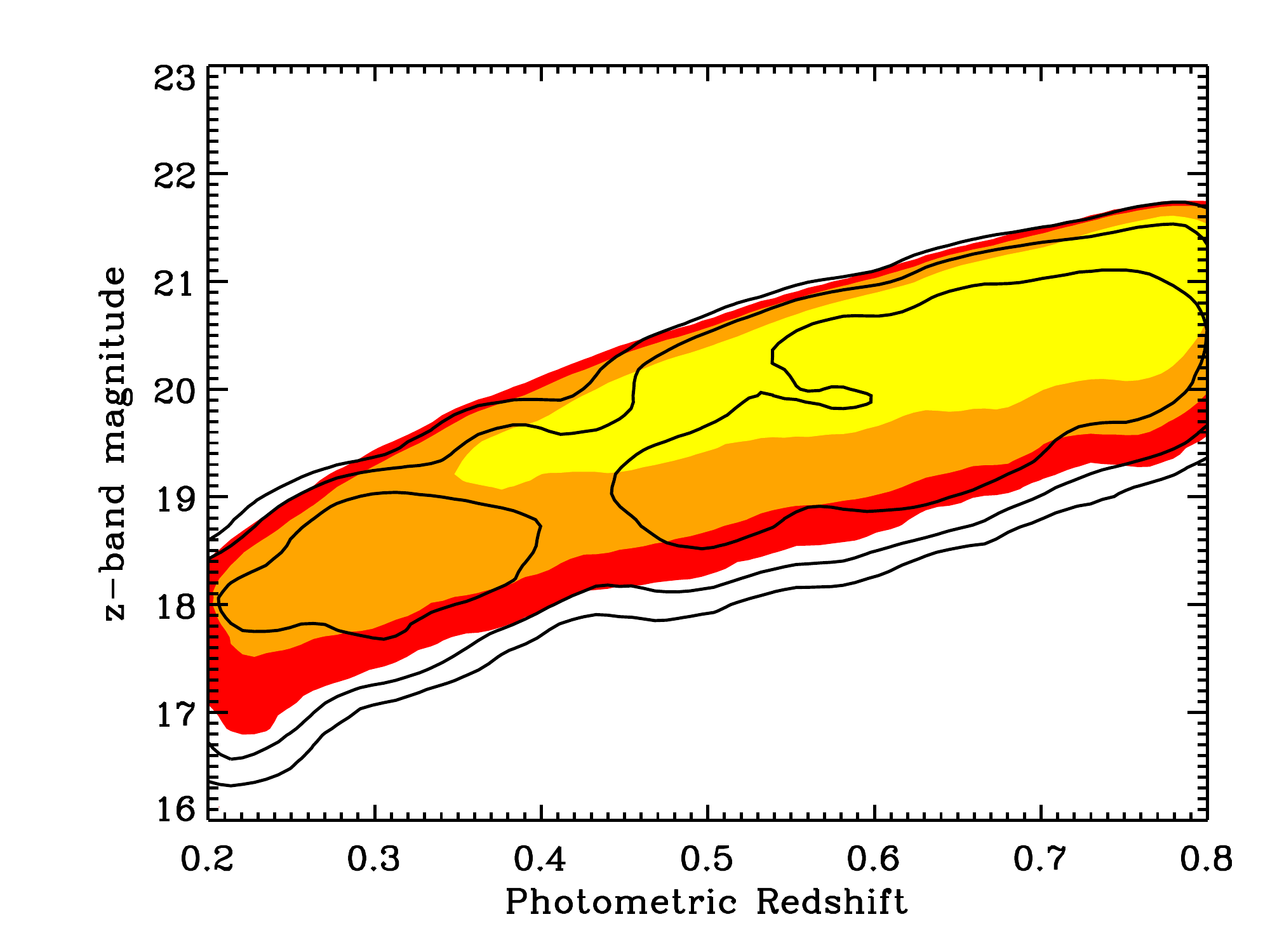}
\caption{{\bf Left:} $dN/dz$ histogram for the fiducial \redmagic\ galaxy sample.  The dotted line is the expected distribution for a constant
comoving density sample.  The red histogram is the \redmagic\ data binned by our photometric redshift estimate.  The blue
histogram shows the number counts for the \redmagic\ sample with spectroscopic redshifts,
boosted by a factor of 10 for clarity. 
{\bf Right:} Contours containing 68\%, 95\%, and 99\% of \redmagic\ galaxies (colored contours)
or \redmagic\ galaxies with spectroscopic redshifts (solid contours).  The dearth of galaxies at $z\approx 0.4$
and the relative excess of bright galaxies in the spectroscopic sample is apparent.
}
\label{fig:dndz}
\end{figure*}

The origin of the redshift biases in the DES SV \redmagic{}
sample are much more difficult to ascertain. First, 
the spectroscopic training set for \redmapper\ is very sparse,
and is most certainly not representative of the sample as a whole.
For instance, there is a dearth of spectroscopic galaxies at $z\approx 0.4$.
A histogram of the number of \redmagic\ galaxies as a function of redshift
is shown in Figure~\ref{fig:dndz}, along with a contour plot showing how
these galaxies populate the redshift--magnitude space.
Second, most of the
redshifts available to us come from training sets in the SN fields,
adding up to $\approx 30\ \deg^2$. The small area results in only  a handful of spectroscopic clusters for 
red sequence calibration.  
Third, our reliance on {\tt MAG\_AUTO} colors in the DES is expected to
adversely affect \photoz\ performance.  
Fortunately, all of these difficulties will be considerably ameliorated if not
entirely removed as the DES images larger areas and updates the data reduction
pipelines.


A summary of the statistical performance of \redmagic\ is presented
in Table~\ref{tab:photoz}.


\section{Comparison to Existing Photo-$z$ Algorithms}
\label{sec:spec}

\subsection{DR8 Comparisons}

\begin{figure*}
\hspace{-12pt} \includegraphics[width=90mm]{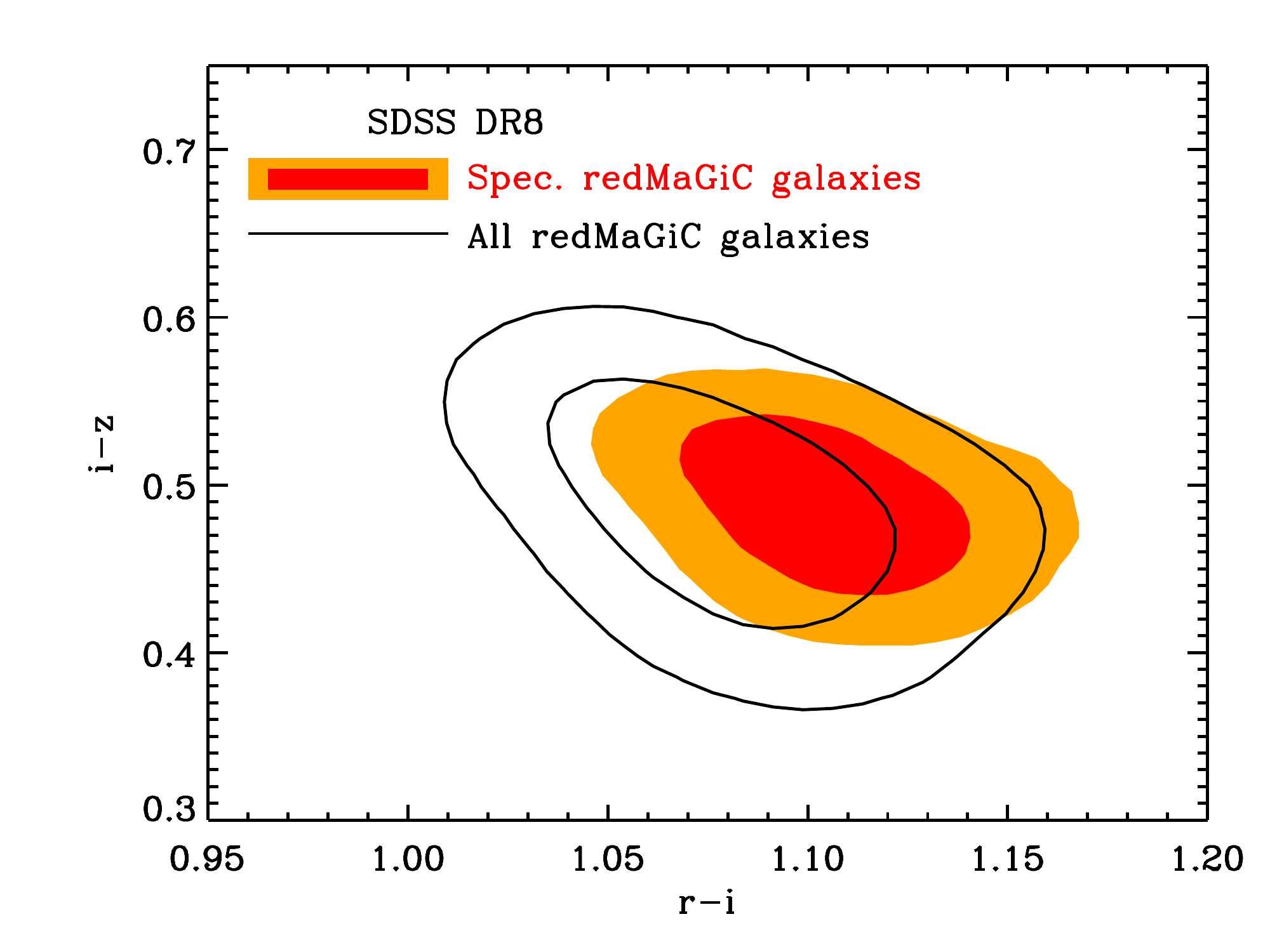}
\hspace{-12pt} \includegraphics[width=90mm]{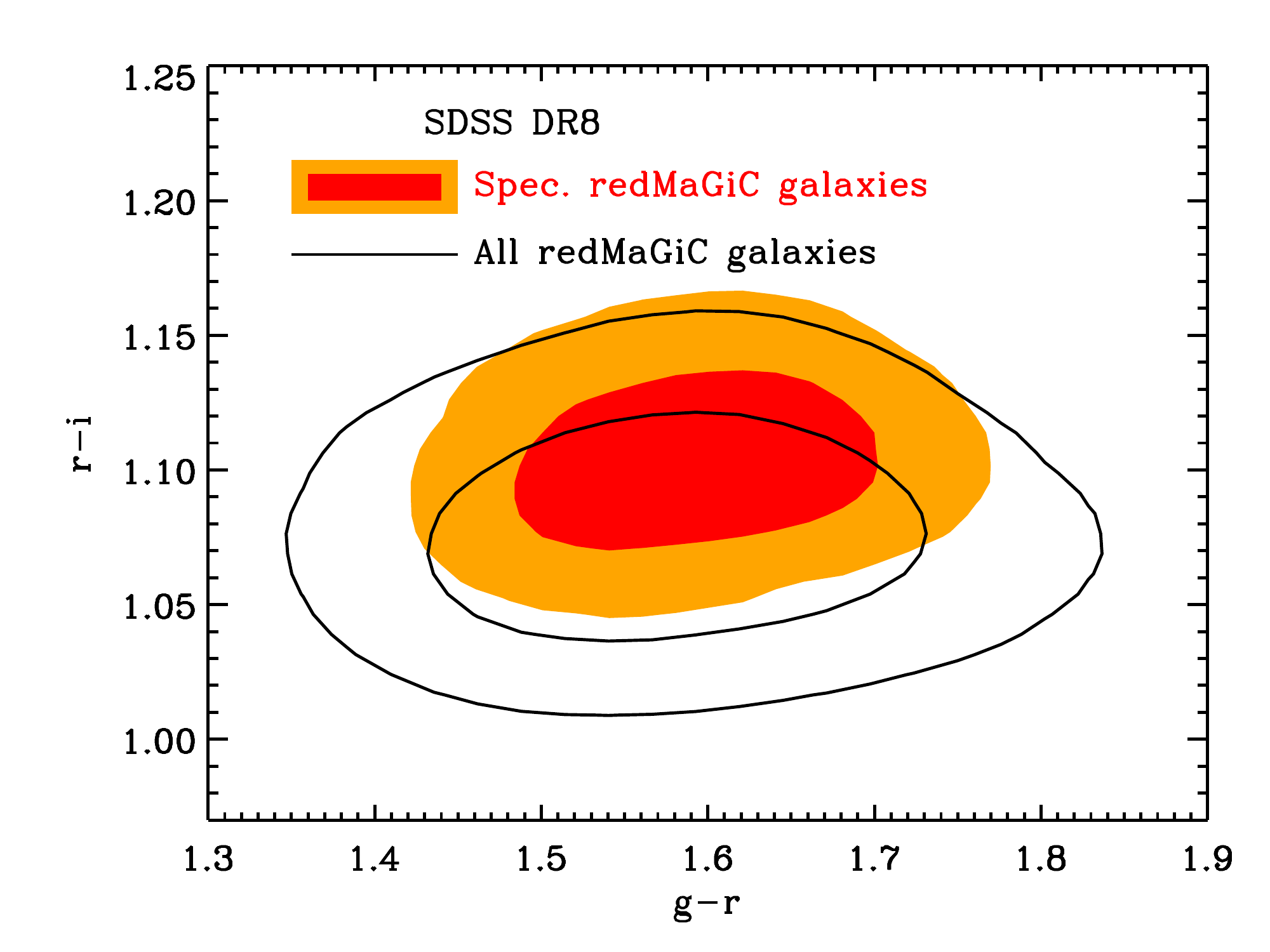}
\caption{Distribution of \redmagic\ galaxies in the photometric redshift bin $\zphoto\in[0.54,0.55]$.
Orange/red contours show the color distribution of
\redmagic\ galaxies with spectroscopic redshifts, while the solid ellipses
show the distribution of all \redmagic\ galaxies.  The large offsets between the two sets of ellipses are due
to biased spectroscopic sampling of the \redmagic\ galaxies.
}
\label{fig:cmass_sampling}
\end{figure*}

As noted in the introduction, \redmagic\ seeks both to select galaxies
with robust photometric redshifts, and to develop a photometric redshift
estimator that can be used on these galaxies with minimal spectroscopic
training data.  For the latter to be useful, however, the performance of our
algorithm must be comparable to that of existing algorithms.
We now test how the \redmagic{} \photozs{} compare with state-of-the-art
photometric redshift codes run on \redmagic\ galaxies.  We start with the SDSS
data set.  To make the comparison as fair as possible we rely on the high
luminosity ($L\geq L_*$), low space density \redmagic\ sample, as the typical
magnitudes of these galaxies are closer to the magnitudes of the galaxies with
spectroscopic redshifts.  Note this high luminosity \redmagic\ sample goes up
to a maximum redshift $z=0.55$ rather than the $z=0.45$ redshift we could achieve
with the low luminosity sample.  However, we restrict our attention to $z\in[0.1,0.5]$
rather than $z\in[0.1,0.55]$.  This is because for $z\geq 0.5$, the
spectroscopic sampling of \redmagic\ galaxies becomes increasingly biased, as
illustrated in Figure~\ref{fig:cmass_sampling}.


We consider three \photoz\ algorithms.
 The first set of \photozs\ are those included with SDSS DR7 \citep{abazajianetal09},
which we shall refer to simply as the SDSS \photozs.  
These were obtained through a hybrid method that combines
the spectral templates of \citet{budavarietal00} with the machine learning
method of \citet{csabaietal07}.
A second set of \photozs\ we compare against are those from \citet{hoyleetal15},
which we will refer to as the RDF \photozs.  
This algorithm uses a combination of decision trees
and feature imporance to derive photometric redshift estimates.
RDF \photozs\ use 85 galaxy features with a 60\%/40\% split for training and validation.
Finally, we utilize the publicly available code
\ANNZ~\citep{collisteretal04} to estimate the redshifts of \redmagic\ galaxies.  
This choice is motivated by the results of
\citet{abdallaetal11}, who performed a detailed comparison of six photometric
redshift algorithms, and found \ANNZ\ performed best in luminous red galaxy
samples.  We train \ANNZ{} with 2/3 of the full spectroscopic training sample,
and test on the remaining 1/3.  The neural net had 5 input nodes (4 {\tt
  MODEL\_MAG} galaxy colors, and a total $\mi$, for which we use
{\tt CMODEL\_MAG}).  We utilized two hidden layers of 10 nodes each, as per the
standard architecture.


\begin{figure*}
\hspace{-12pt} \includegraphics[width=90mm]{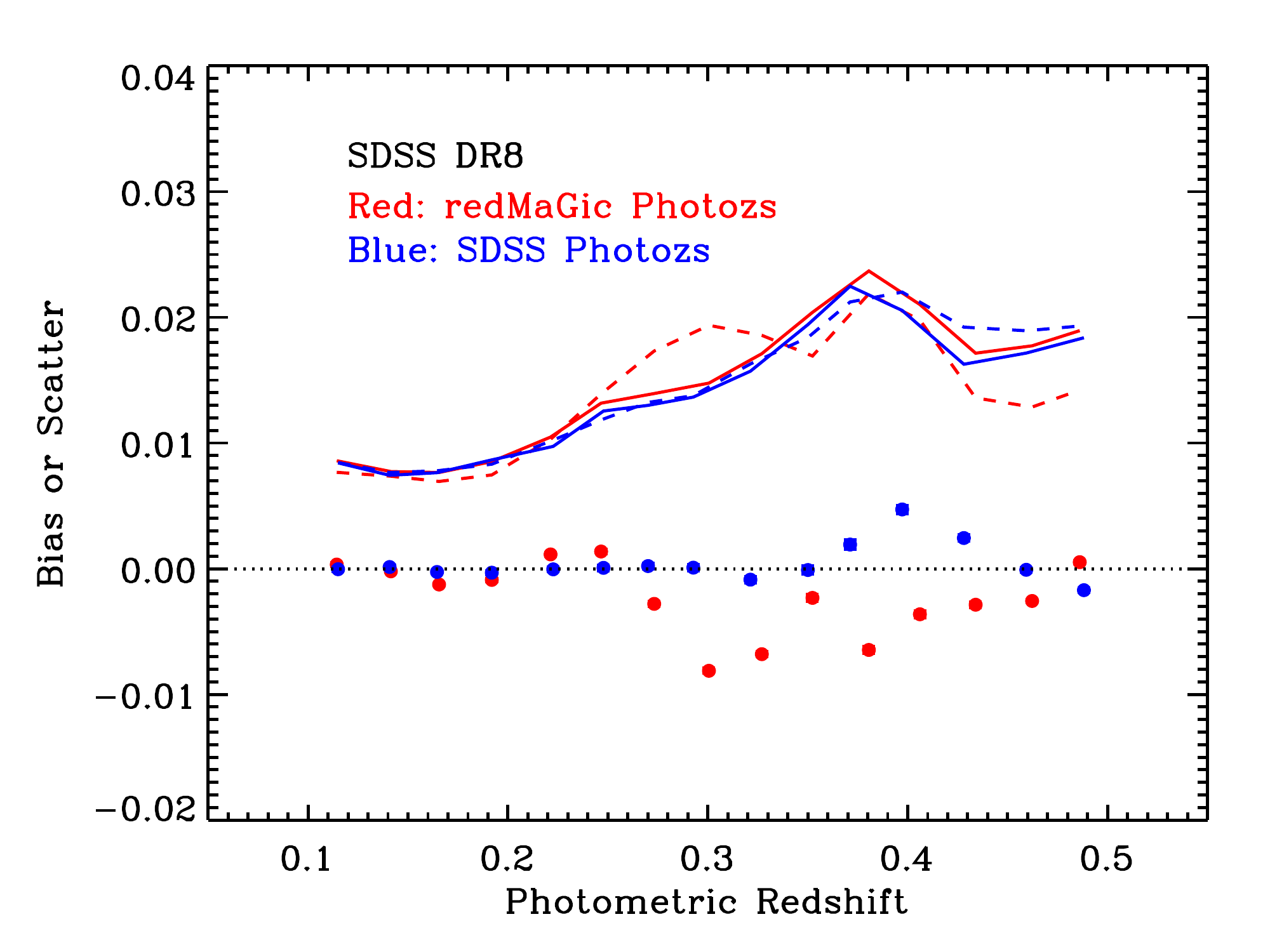}
\hspace{-12pt} \includegraphics[width=90mm]{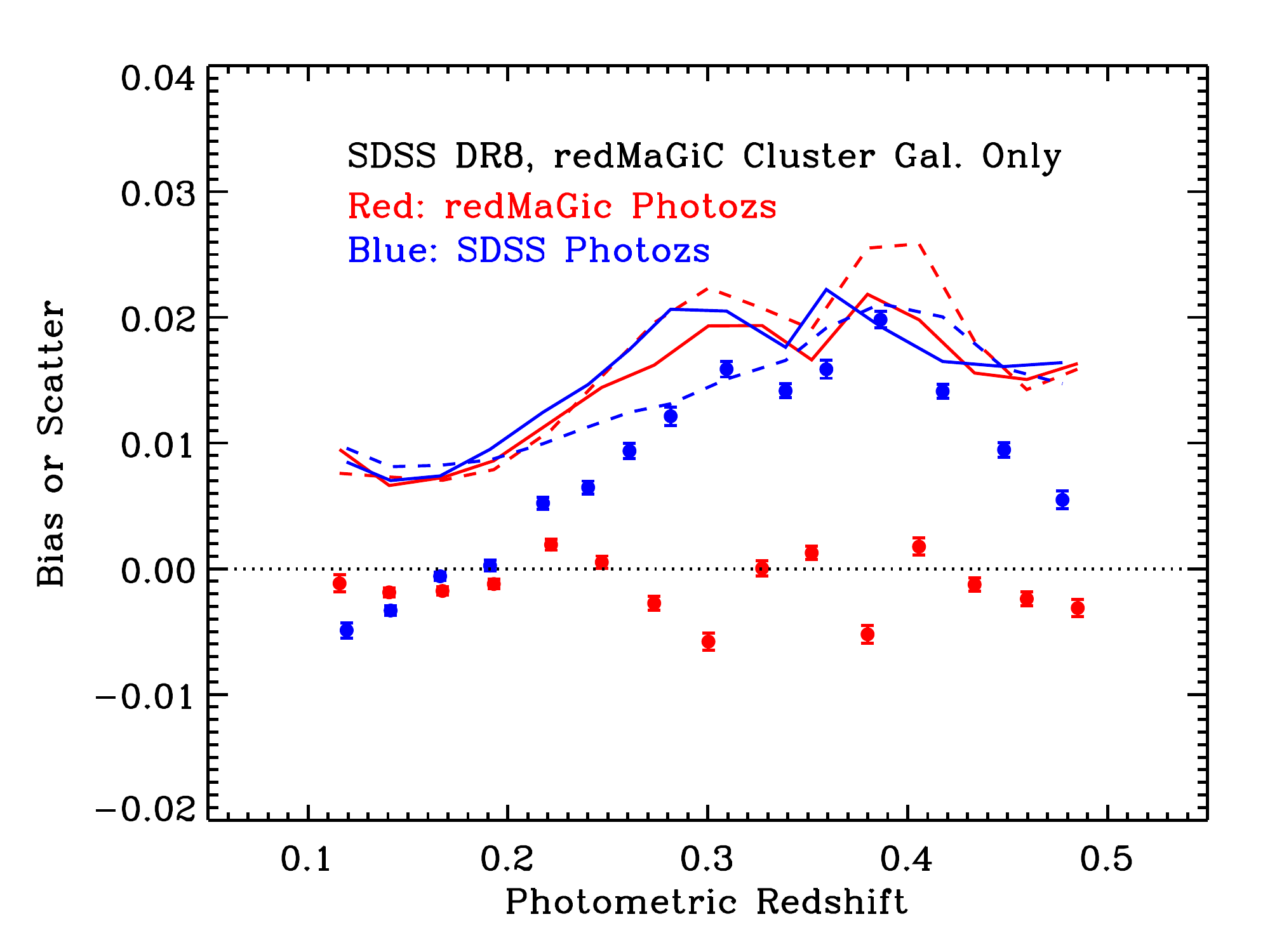}
\caption{{\bf Left:} Comparison of the photometric redshift performance of \redmagic\ (red)
and SDSS \photozs\ for \redmagic\ galaxies (blue).  This plot uses SDSS spectroscopic redshifts
to compute the redshift bias and scatter of the \redmagic\ \photozs, and is therefore limited to the
brightest \redmagic\ galaxies.
Points with error bars show the median redshift bias for each of the two samples.  Solid lines show the observed
\photoz\ scatter, while dashed lines show the predicted scatter. 
{\bf Right:} As for the left panel, only now we test the \photoz\ performance of the sub-sample of \redmagic\
galaxies that are members of \redmapper\ clusters.  For these galaxies, we assign the photometric
redshift of the host \redmapper\ clusters as the ``spectroscopic'' redshift of the \redmagic\ galaxy for the purposes
of computing photometric redshift biases and scatter.  By doing so, we can test the accuracy of the photometric
redshifts of faint \redmagic\ galaxies with no spectroscopic redshift. 
}
\label{fig:comparison}
\end{figure*}


A comparison of the \redmagic\ \photoz{} to the SDSS \photozs{} is shown in
Figure~\ref{fig:comparison}.  We find the SDSS \photozs\ are slightly 
less biased than the \redmagic\ \photozs, but have nearly identical scatters.
The SDSS \photozs\ also do a better job of error characterization,
though the difference is not large.
The picture is much the same for \ANNZ, except that \ANNZ\ grossly underestimates
the photometric redshift scatter (not shown).
RDF redshifts are clearly superior to the SDSS, \ANNZ, and \redmagic\ \photozs,
though the improvement remains modest: the scatter decreases from 1.48\% in
\redmagic\ to 1.28\% in RDF (not shown).
The agreement between the \ANNZ, SDSS, and \redmagic\ redshifts
strongly suggest that the improvement seen with RDF
is primarily due to the large number of features used (85 observables), rather than more optimal
use of the limited information used in \redmagic{} (5 bands).

A quantitative summary of these results is presented in Table~\ref{tab:photozcomp}.
Also reported there are the fraction of galaxies where $|\zphoto-\zspec|/(1+\zspec) \ge 0.07$,
corresponding roughly to $5\sigma$ for \redmagic\ galaxies.  This number characterizes how
large the tails of the \photoz\ errors are.  All methods we consider here have comparable tails.

We caution, however, that these tests represent a best-case scenario for training set methods. 
Specifically, machine learning
methods do not  extrapolate outside their training sets very well.  Consider red galaxies
as a specific example.  Because the red sequence is tilted, a faint red sequence galaxy
will appear bluer than a bright red sequence galaxy.  Consequently, red sequence galaxies
fainter than the training data set of a machine learning algorithm will have $\zphoto \leq \zspec$.

We can indirectly verify this expectation by looking at members of galaxy clusters. Specifically,
we select all \redmapper\ high probability (membership probability $\geq 90\%$) cluster members,
and assign to all such members a ``spectroscopic'' redshift equal to the photometric cluster redshift.
We then compare the \redmagic\ and SDSS \photozs\ of these galaxies to their assigned cluster redshifts.
The redshift bias $z_{\rm cluster}-\zphoto$ and corresponding scatter are shown in the right panel of
Figure~\ref{fig:comparison}.  We see that our expectation that $\zphoto \leq \zspec$ is borne out by the
data, and that the bias can be large, $\approx 0.02$.  At very high redshifts, the luminosity threshold
in \redmapper\ approaches the spectroscopic magnitude limit, and so the bias starts to decrease 
with redshift.

The main take aways from these test are that \redmagic\ \photozs\ perform
as well the best machine learning methods run with the same photometric input.
However, machine learning methods can improve on \redmagic\ by exploiting additional
data.  Critically, however, machine methods do not extrapolate well, and appear to be subject
to large redshift biases for galaxies that are not well represented in the training data sets 
\citep[however, see][]{hoyleetal15b}.
Because of how the \redmagic\ algorithm is structured, this is not a problem for \redmagic\
\photozs.


\subsection{DES Comparisons}


\begin{center}
\begin{table*}
\centering
\caption{As Table~\ref{tab:photoz}, but comparing the redshift performance of different \photoz\ algorithms
on \redmagic\ galaxies.  We only consider the \redmagic\ sample with space density $2\times 10^{-4} h^3\ \Mpc^{-3}$.
The redshift range of consideration is $z\in[0.1,0.5]$ for DR8, and $[0.2,0.8]$ for DES.
``Bad fraction'' is the fraction of galaxies where $|\zphoto-\zspec|/(1+\zspec) \ge 0.07$ (for SDSS)
or $\ge 0.08$ (for DES), corresponding roughly to $5\sigma$ for \redmagic\ \photozs.
DR8 Spec AB data sets correspond to \redmagic\ with a spectroscopic afterburner (see section~\ref{sec:ab}).
}
\begin{tabular}{lccccc}
Data Set & Bias & $\vert$Bias$\vert$ & Scatter & Predicted Scatter & Bad Fraction\\
\hline
SV \redmagic & 0.35\% & 0.35\% & 1.82\% & 1.80\% & 1.4\% \\
SV \skynet & -0.36\% & 0.59\% & 1.58\% & 5.31\% & 1.1\% \\
SV BPZ & 1.48\% & 2.95\% & 1.59\% & 9.821\% & 11.6\% \vspace{5 pt}\\
\hline
DR8 \redmagic & -0.23\% & 0.23\% & 1.48\% & 1.39\% & 1.4\% \\
DR8 SDSS \photoz & -0.00\% & 0.02\% & 1.37\% & 1.38\% & 1.3\% \\
DR8 RDF \photoz & 0.01\% & 0.03\% & 1.25\% & 1.28\% & 1.3\%\\
DR8 ANNZ \photoz & -0.09\% & 0.13\% & 1.33\% & 1.29\% & 1.5\% \\
DR8 Spec AB & 0.01\% & 0.03\% & 1.49\% & 1.47\% & 1.1\%\\
\end{tabular}
\label{tab:photozcomp}
\end{table*}
\end{center}


We compare \redmagic\ photozs to two algorithm currently in use 
within the DES collaboration \citep{sanchezetal14}, specifically \skynet{} and BPZ \photozs.
\skynet{} is a machine learning method that relies on neural networks to
``classify'' galaxies into redshift bins \citep[][]{graffetal14,bonnet15}, 
while BPZ is a popular template based code \citep[][]{bpz00}.
We use BPZ with its default configuration (8 templates, {\tt INTERP=2},
and we do not allow for zero point offsets).
While there are other machine learning methods available in 
DES, they all have comparable performance, so we have arbitrarily chosen 
to focus on \skynet{} to simplify our analysis.  

\begin{figure}
\hspace{-12pt} \includegraphics[width=90mm]{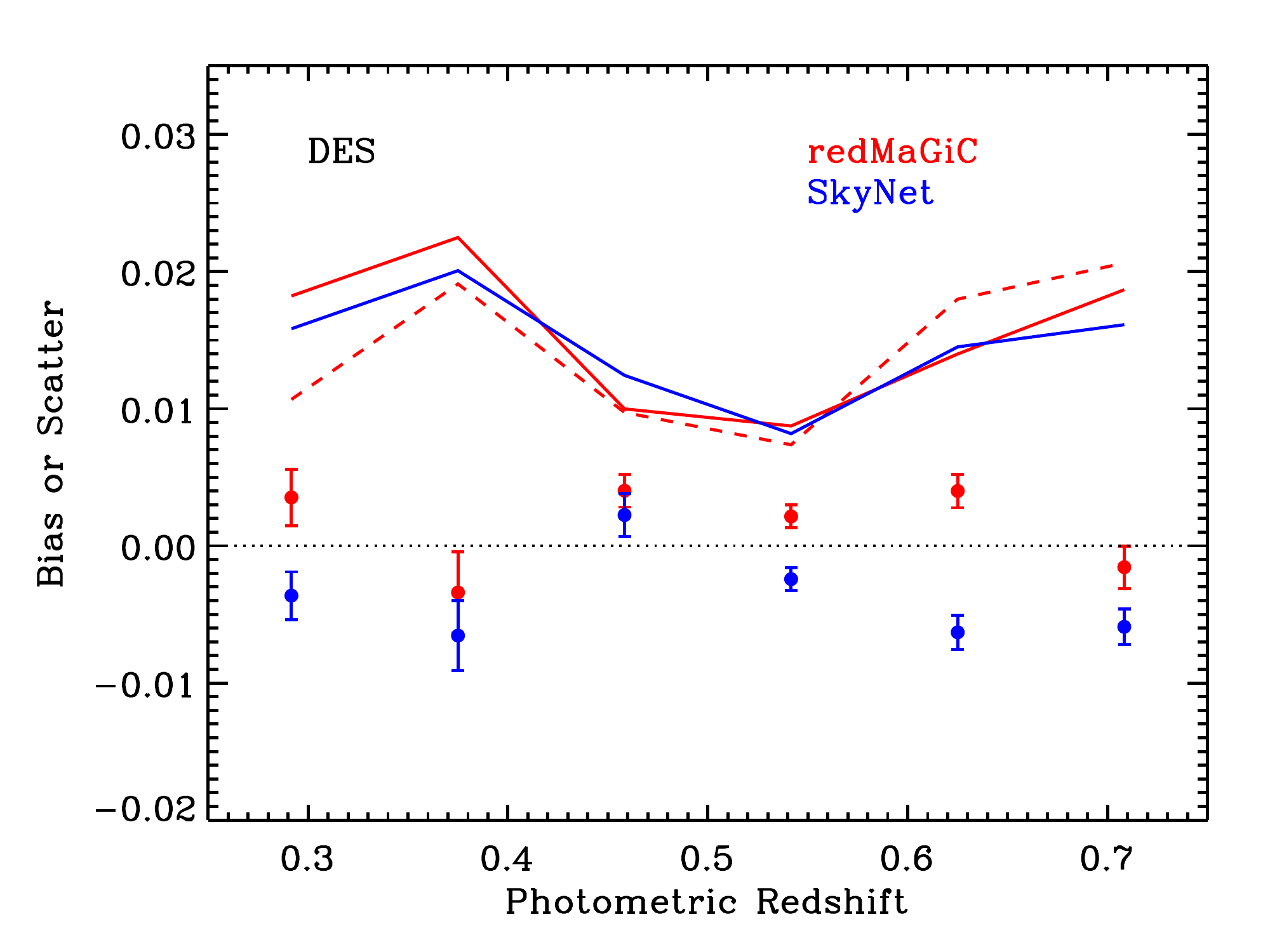}
\caption{As Figure~\ref{fig:comparison}, only now we compare SV \skynet{} \photozs\ (blue)
to SV \redmagic\ \photozs (red).  The predicted \skynet{} scatter is not shown, as the \skynet{}
predicted error are a factor of 3.5 larger than the observed scatter.
}
\label{fig:des_comparison}
\end{figure}

Figure~\ref{fig:des_comparison} compares the performance of \skynet{}
on the \redmagic\ galaxy sample to that of the \redmagic\ \photozs.
The two algorithms perform equally well in terms of \photoz\
biases and scatter.  However, \skynet{} grossly overestimates the photometric redshift
uncertainty, with the \skynet{} predicted uncertainties being a factor of 3.5 times larger than
the observed errors.  This is not unexpected: \skynet\ and the other machine learning codes
used in the DES SV data have their photometric redshifts smoothed and broadened (for details,
see Appendix C in Bonnett et al. in preparation), which improves \photoz\ performance
for lensing sources, but, as evidenced here, has a deleterious effect on the \photoz\ error estimates for
\redmagic galaxies.
\skynet{} and \redmagic\
also exhibit similar tails.

BPZ performs very poorly at low redshifts, exhibiting a redshift bias of $\approx 0.1$.  
The bias decreases to $\approx 0.02$ at higher redshifts, but 
remains well above the \skynet/\redmagic\ biases.
The redshift scatter for BPZ is comparably to that of \skynet/\redmagic,
but the uncertainties are overestimated by a factor of $\approx 6$. 
Nearly 12\% of all galaxies have
$|\zspec-\zphoto|/(1+\zspec) \geq 0.08$ for BPZ, compared with $\approx 1.4\%$
for \redmagic/\skynet.


Our results confirm the basic picture we obtained from the DR8 comparisons:
\redmagic\ performs as well as the best performing machine learning methods,
despite not requiring representative spectroscopic training samples.  
BPZ performance is especially poor.
Importantly, \redmagic\ continues to have extremely well characterized scatter,
whereas \skynet/BPZ do not.


\section{Why Selection Matters}
\label{sec:selection}

The primary motivation of the \redmagic\
algorithm is not to improve upon existing photometric redshift algorithms, but rather
to select a galaxy sample with robust \photozs.   The results in the previous
section clearly demonstrate that \redmagic\ galaxies do, in fact, have photometric
redshifts that are both precise and accurate.  In this section we investigate
whether this feature is unique to the \redmagic{} sample.  In particular, we
look at the current work-horse for large-scale structure measurements in the SDSS, 
the CMASS galaxy sample.  
CMASS galaxies were specifically selected to be roughly stellar mass limited at $z\geq 0.45$.  
Here, we test whether the \redmagic\ selection can lead to improved photometric redshift
performance relative to CMASS.  Note that any gains we make are not of critical important for
spectroscopic experiments, as such experiments are not sensitive to large photometric redshift
scatter and/or catastrophic \photoz\ failures.

A fair comparison of CMASS to \redmagic\ galaxies is difficult.  In particular, we'd like to
compare samples that have comparable space densities (which control the errors in clustering
signal) and luminosities (which set the photometric error uncertainty).
For comparison purposes, Table~\ref{tab:densities} quotes typical densities for a couple of
standard SDSS galaxy samples, namely LRG~\citep{eisensteinetal01}, and LOWZ and CMASS~\citep{dawsonetal13}
Also shown is the minimum luminosity of galaxies in that sample at a typical redshift.
Densities for the standard SDSS
samples are based on Figure 1 of \citet{tojeiroetal14}.
We see that even our bright \redmagic\ sample has a comparable density to CMASS,
but a lower luminosity threshold, reflecting the more stringent color cuts applied in \redmagic.
We will compare CMASS against this sample. 
Note CMASS galaxies are $\approx 0.3$ 
magnitudes brighter than the \redmagic\ galaxies we compare against.
This added noise should degrade the photometric
redshift performance in \redmagic\ galaxies relative to CMASS.  That is, the match-up is purposely
stacked against \redmagic\ for this comparison.


\begin{figure*}
\hspace{-12pt} \includegraphics[width=90mm]{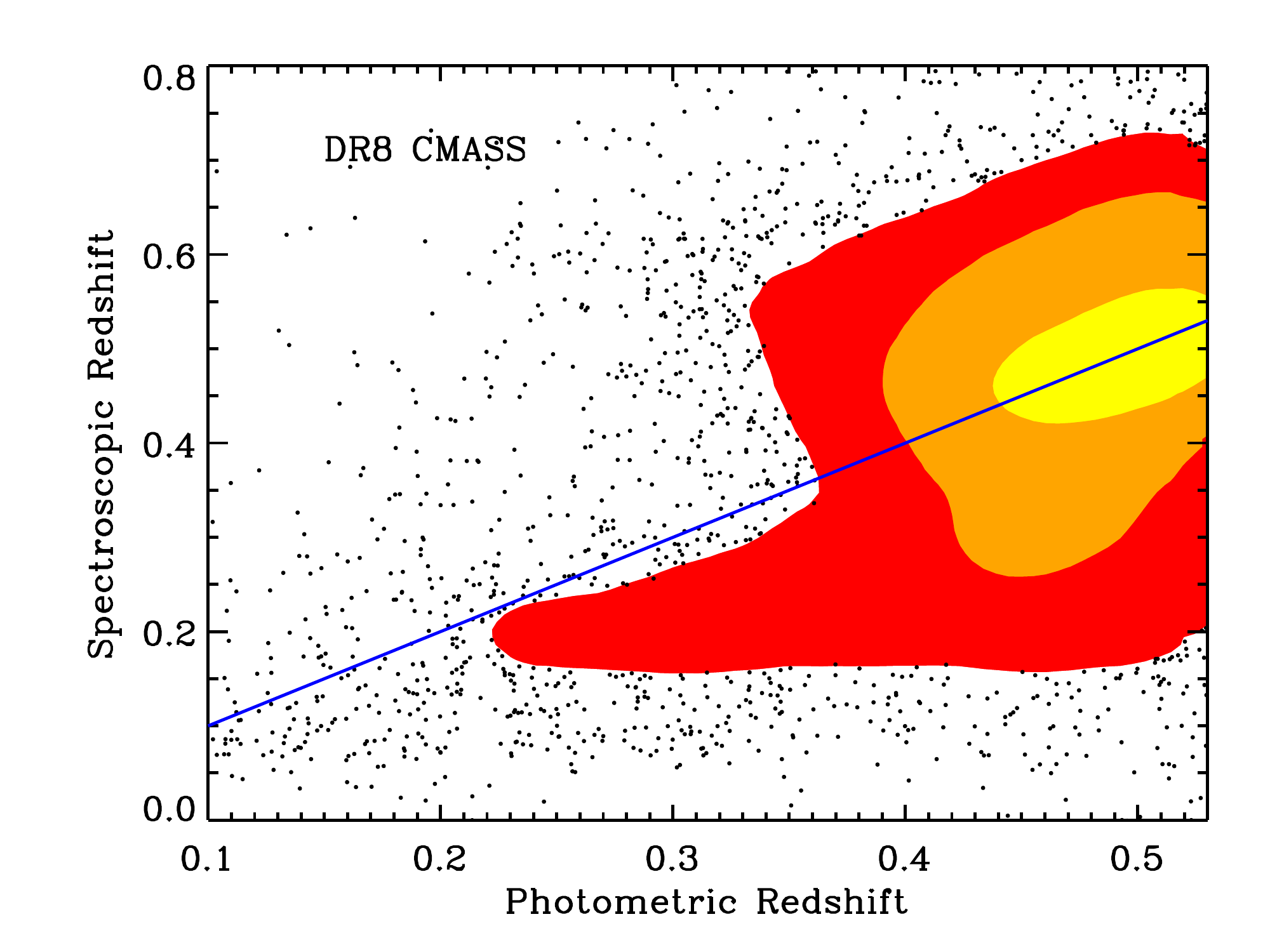}
\hspace{-12pt} \includegraphics[width=90mm]{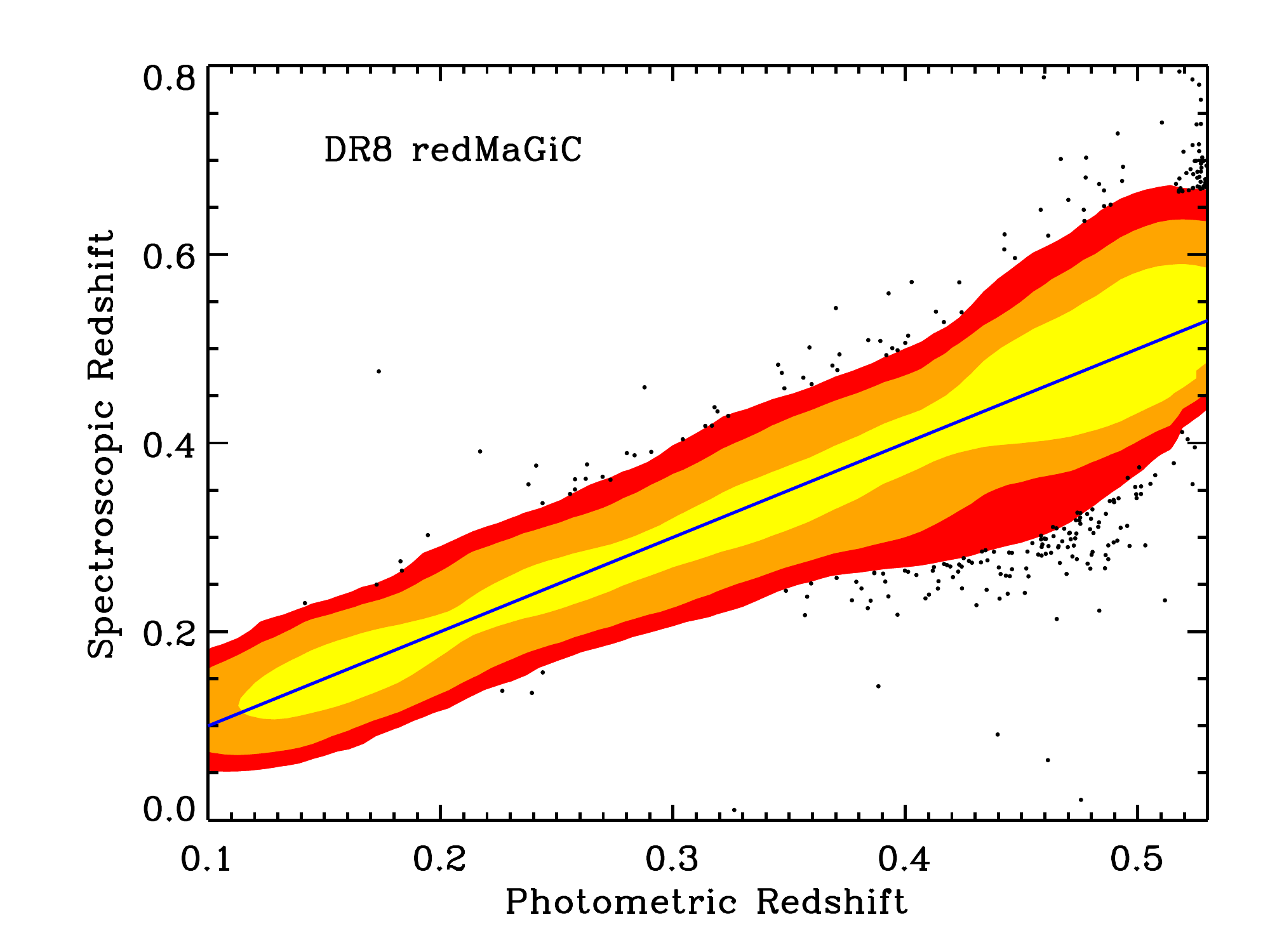}
\caption{{\bf Left:} Spectroscopic vs. photometric redshifts for CMASS galaxies
using SDSS \photozs.  Colored regions contain 68\%, 95\%, and 99\% 
of the points.  The remaining 1\% of galaxies are shown as points.
The blue line is the $y=x$ line.
{\bf Right:} As left panel, but for \redmagic\ galaxies.  
}
\label{fig:selection}
\end{figure*}


Figure~\ref{fig:selection} shows how galaxies fall in the $\zspec$--$\zphoto$ plane
for both CMASS (left panel) and \redmagic\ (right panel).  For the CMASS data set
we rely on SDSS \photozs~\citep[][]{csabaietal07}, while we use \redmagic\ \photozs\ for \redmagic.
Note that \redmagic\ and SDSS \photozs\ had nearly identical performance on
\redmagic{} galaxies, so the performance in the right-hand plot would be much
the same if we replaced \redmagic{} \photozs{} with SDSS \photozs.

The benefit of the \redmagic\ selection is immediately apparent: despite probing fainter
galaxies, the \redmagic\ galaxies have clearly better behaved photometric
redshifts than those of CMASS.
The \photoz\ scatter is 1.5\% for \redmagic, and 2.1\% for CMASS.  
In addition,
the fraction of galaxies with large redshift errors  ($|\Delta z|/(1+z) \geq 0.07$) 
is much larger for CMASS (6.4\%) than for \redmagic\ (1.4\%).
We note that the \photoz\ scatter for CMASS galaxies quoted here is significantly lower than
that reported in \citep{rossetal11}.  This is partly because we define scatter as $\sigma_z/(1+z)$,
while \citet{rossetal11} quote $\sigma_z$, and partly because we estimate $\sigma_z$
using median statistics, while \citet{rossetal11} use $\sigma_z=\sqrt{ \avg{\zspec-\zphoto}^2}$,
which is more sensitive to gross outliers than the MAD-based estimate.

It is also clear from Figure~\ref{fig:selection} that CMASS galaxies with $\zspec \lesssim 0.3$ are
particularly ill-behaved.  This is not particularly problematic for experiments like BOSS, where the
spectroscopic follow-up of the targets ensures that these contaminants don't percolate into 
cluster measurements at $z\approx 0.5$.  By contrast, a photometric survey would end up including
those galaxies in its clustering measurements, leading to systematic errors in the clustering signal.
This further highlights the importance of \redmagic\ selection for photometric large-scale structure studies.

We can also compare the performance of the RDF photometric redshifts in the CMASS
sample to \redmagic.  Relative to the SDSS \photozs, RDF shows clear improvement:
the scatter is reduced to 1.9\%, and the fraction of galaxies with larger errors
goes down to 2.2\%.   This is not surprising: RDF redshifts were trained on CMASS galaxies,
whereas the SDSS \photozs\ were not.  This highlights the importance of training for machine learning
methods, a weakness not shared by \redmagic.  Just as importantly, even RDF redshifts
for CMASS galaxies are worse than \redmagic\ redshifts for \redmagic.

In short, we find \redmagic\ is extremely successful at identifying galaxies with
robust photometric redshift estimates.  Of course, CMASS was designed to be used for a
spectroscopic survey, so the differences highlighted here are much less important in that
case.  For purely photometric surveys, however, our selection algorithm is clearly superior.


\begin{center}
\begin{table}
\centering
\caption{Typical space density and luminosity cuts for a variety of different SDSS galaxy samples.}
\begin{tabular}{lll}
Sample & Space Density & Minimum Luminosity\\
	&$(h^{-1}\ \Mpc^{-3})$ & $(\Lmin/L_*)$\\
\hline
LRG & $1 \times 10^{-4}$ & $2.1$ (at $z=0.35$) \\
LOWZ & $3 \times 10^{-4}$ & $1.6$ (at $z=0.35$) \\
CMASS & $2\times 10^{-4}$ & $1.5$ (at $z=0.5$) \\
\redmagic\ Bright & $2\times 10^{-4}$ & $1.0$ \\
\redmagic\ Faint & $1\times 10^{-3}$ & $0.5$\\
\end{tabular}
\label{tab:densities}
\end{table}
\end{center}



\section{Spectroscopic Training of \redmagic}
\label{sec:ab}

We consider whether $\zrm$ from \redmagic{} can be significantly
improved with further spectroscopic training data.  Specifically, in the
\redmagic{} algorithm, we use a \photoz{} ``afterburner'' that relies on
photometric cluster galaxies to help fine-tune our \photozs.  We now consider
what happens if we apply a further ``afterburner'' using spectroscopic redshift
information for the \redmagic\ sample. As a proof-of-concept, we use the \redmagic{}
galaxies that are in the SDSS DR10 spectroscopic catalog, and split the sample
in half for training and validation.  All results shown are for the validation sample only.

\begin{figure*}
\hspace{-12pt} \includegraphics[width=90mm]{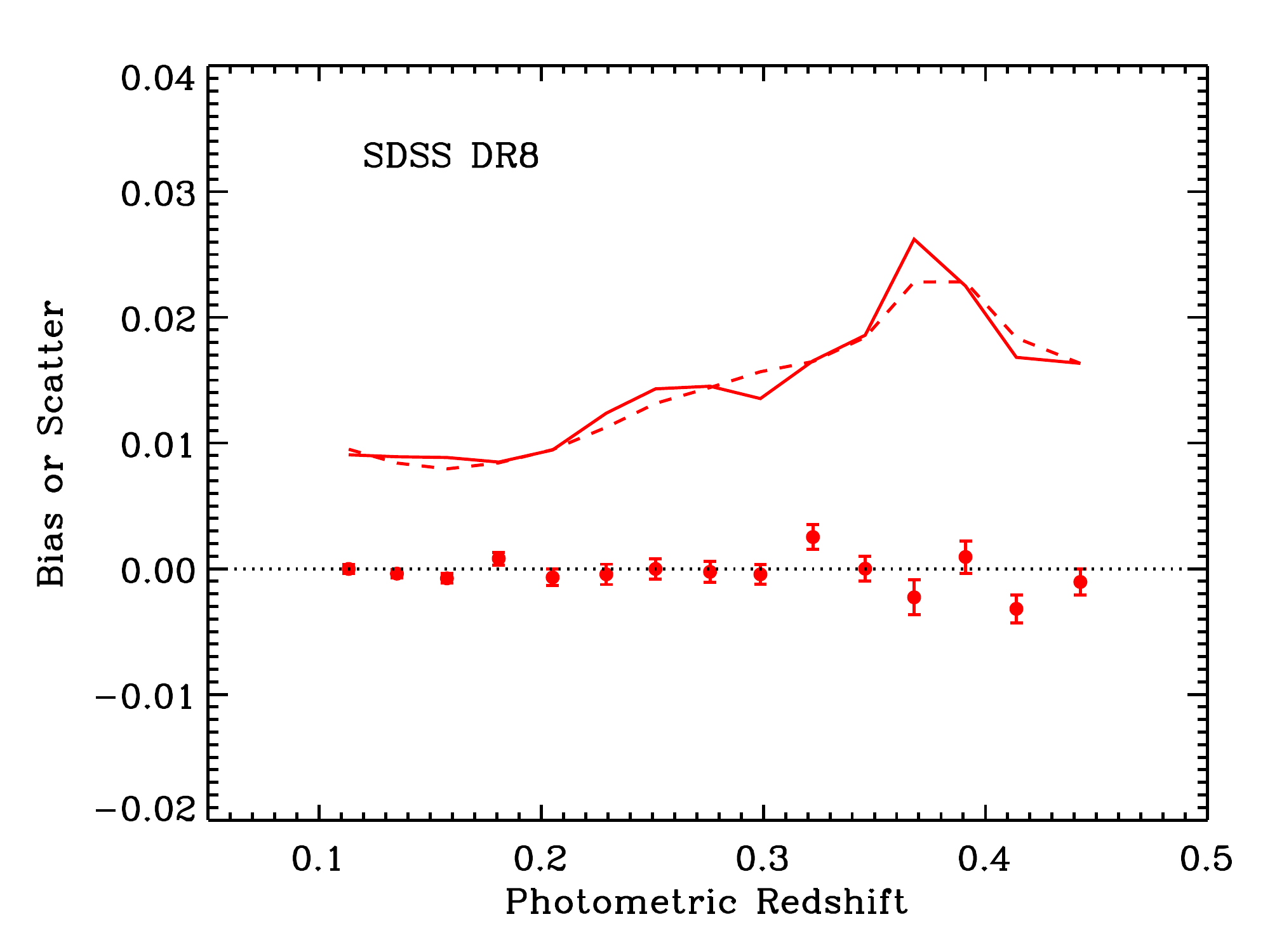}
\hspace{-12pt} \includegraphics[width=90mm]{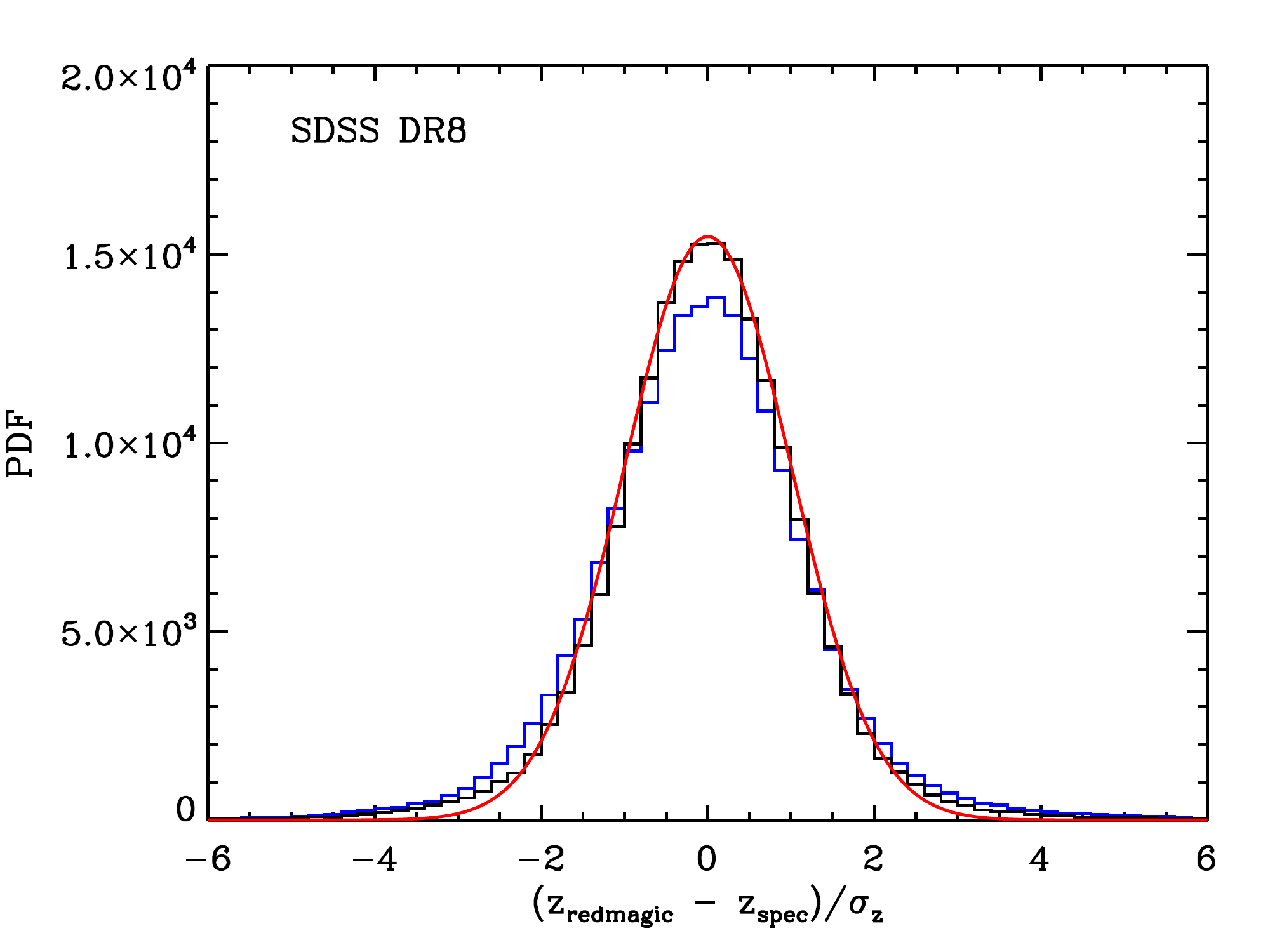}
\caption{{\bf Left:} \redmagic\ photometric redshift performance after training with
a spectroscopic sub-sample of galaxies.  Points with error bars show the
bias in the recovered redshifts.  The solid line shows the photometric redshift
scatter, while the dashed line shows the predicted redshift scatter. {\bf Right:}
A histogram of the quantity $\Delta = (\zspec-\zphoto)/\sigma_z$ where $\sigma_z$ is 
the reported photometric redshift uncertainty.  The blue histogram
is for our fiducial \redmagic\ sample, while the black histogram is for a spectroscopically
trained \redmagic\ sample.   The red curve is {\it not a fit.}
It is simply a Gaussian of zero mean and unit variance.  
}
\label{fig:ab}
\end{figure*}

For our spectroscopic afterburner, we apply the same procedure as outlined in
Section~\ref{sec:afterburner}, only now the initial redshift estimate is $\zrm$.
We label our final redshift estimate $\zsAB$ (for spectroscopic afterburner).
Similarly, we tweak the \photoz{} error using
median statistics as with our original afterburner.  
Having defined our new \redmagic\ spectroscopically-trained
\photoz{} estimates, we test the \redmagic\ \photoz{} performance using our
test sample. The results are shown in the left panel of Figure~\ref{fig:ab}.
The right panel of Figure~\ref{fig:ab} shows a histogram of the quantity
$\Delta_z=(\zspec-\zsAB)/\sigma_{\zsAB}$.  If all the \photozs{} were
Gaussian, unbiased, and we correctly estimated the \photoz{} error,
then a histogram of the quantity $\Delta_z$ would be well fit by a Gaussian of
zero mean and unit variance.  The right panel of Figure~\ref{fig:ab} shows the
$\Delta_z$ histogram for the \redmagic\ testing sample.  The red Gaussian is 
{\it not} a best fit: it is a Gaussian of zero mean and unit variance.

Given the improved performance for the spectroscopically trained \redmagic\ sample,
why do we not adopt 
this procedure as part of the \redmagic\ photometric redshift estimate by default?
As discussed in Section~\ref{sec:performance}, biased spectroscopic
sampling of our data set will introduce unknown and uncontrolled biases in the resulting 
photometric redshifts.  Consequently, we have opted not to apply this spectroscopic
afterburner until a fully representative spectroscopic galaxy sample becomes available,
or data augmentation techniques are advanced enough to extrapolate outside the
training sets.


\section{Understanding \redmagic\ Outliers}
\label{sec:outliers}


\begin{figure*}
\hspace{-12pt} \includegraphics[width=90mm]{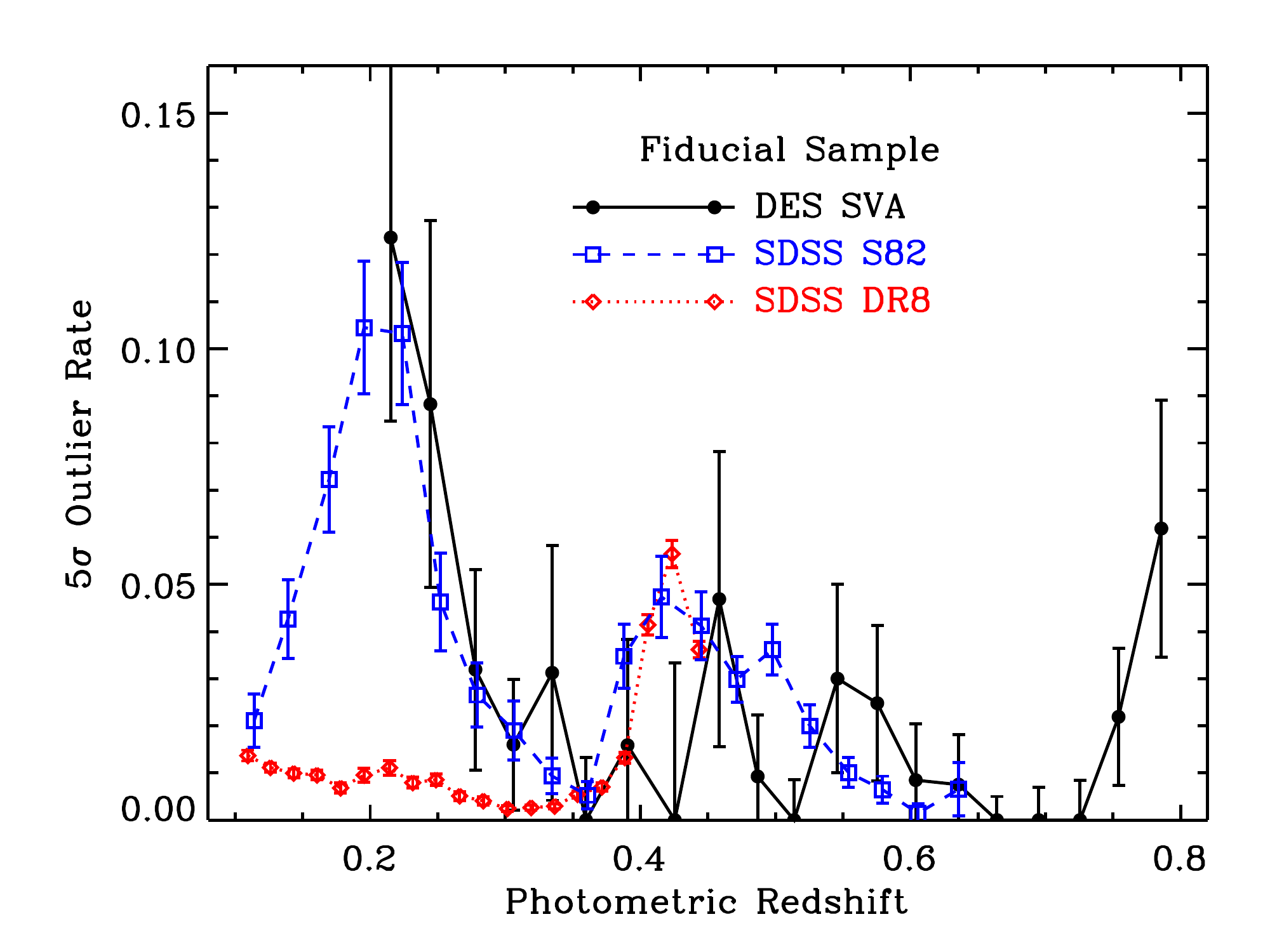}
\hspace{-12pt} \includegraphics[width=90mm]{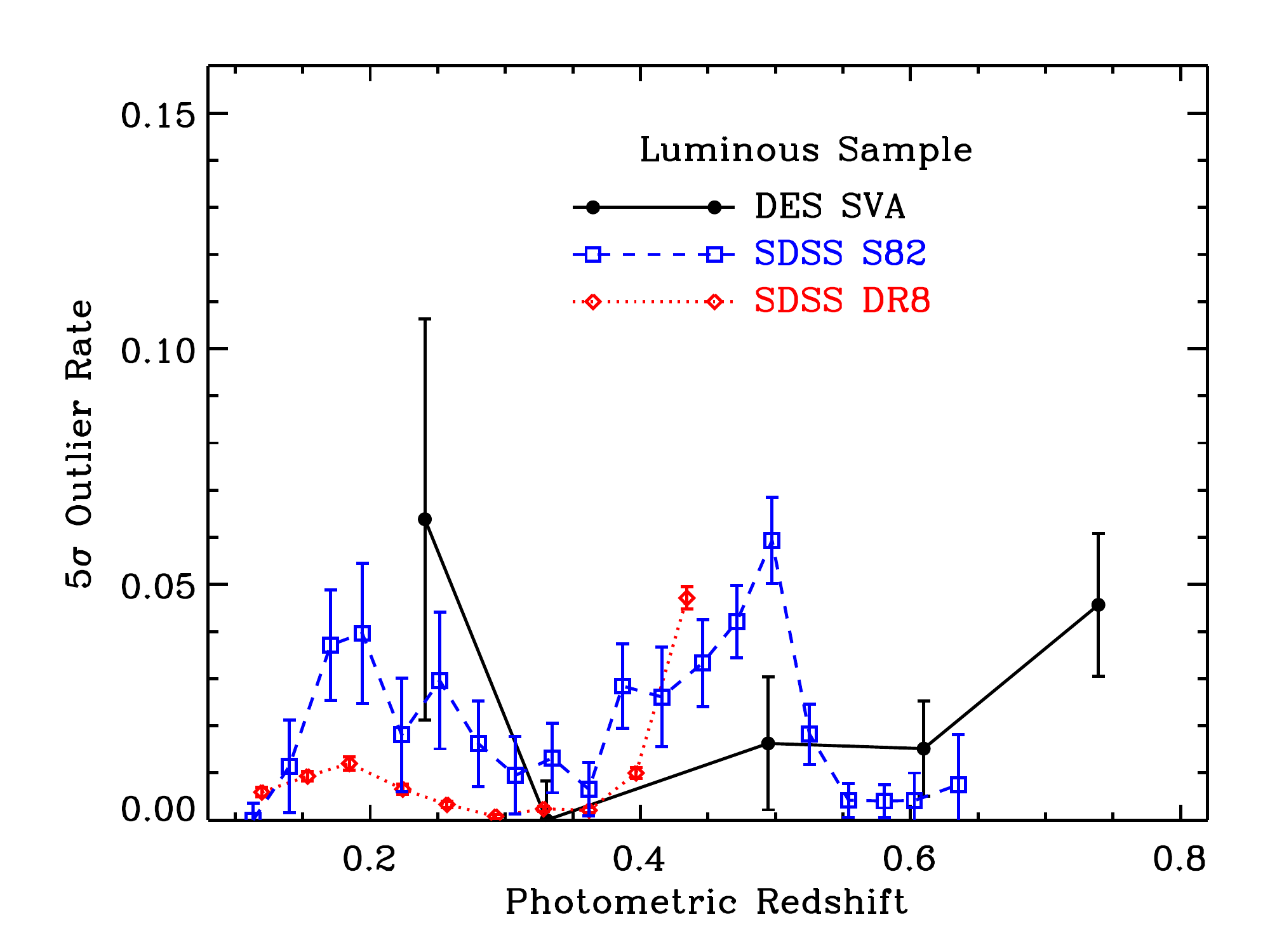}

\caption{$5\sigma$ outlier fraction for the fiducial and high luminosity DR8 (red), S82 (blue), and
DES (black) \redmagic\ samples as estimated using SDSS DR12 spectroscopy.
}
\label{fig:outliers}
\end{figure*}


We now investigate the \photoz{} outliers in the \redmagic\ galaxy sample.  We
consider a galaxy an outlier if its \photoz{} is more than $5\sigma$ away from
its spectroscopic redshift.  The outlier fraction of \redmagic\ galaxies as a
function of redshift is illustrated in Figure~\ref{fig:outliers} for both the
fiducial and high luminosity samples.  Perhaps the two most salient features in
this plot are: 1) the difference in the outlier fractions at low redshifts
between the SDSS DR8 and both the SDSS S82 and DES SV data sets; and 2) the
difference in the outlier fractions between the fiducial and high luminosity galaxy
samples.  The latter result is not surprising: the brighter the
galaxy, the easier it is to distinguish between red sequence and non
red sequence galaxies.  We will return to the difference between the DR8 and S82/DES SV
momentarily.

Consider first the DR8 outlier population.  The mean DR8 outlier fraction
is small, $\approx 0.7\%$, and is split among 3 sets of outlier clumps, as seen
in Figure~\ref{fig:photoz}.  
This last one is more readily apparent in the SDSS S82 data set.
We consider each of these in turn.

\subsection{Clump 1: Low Redshift Outliers}

\begin{figure}
\hspace{-12pt} \includegraphics[width=90mm]{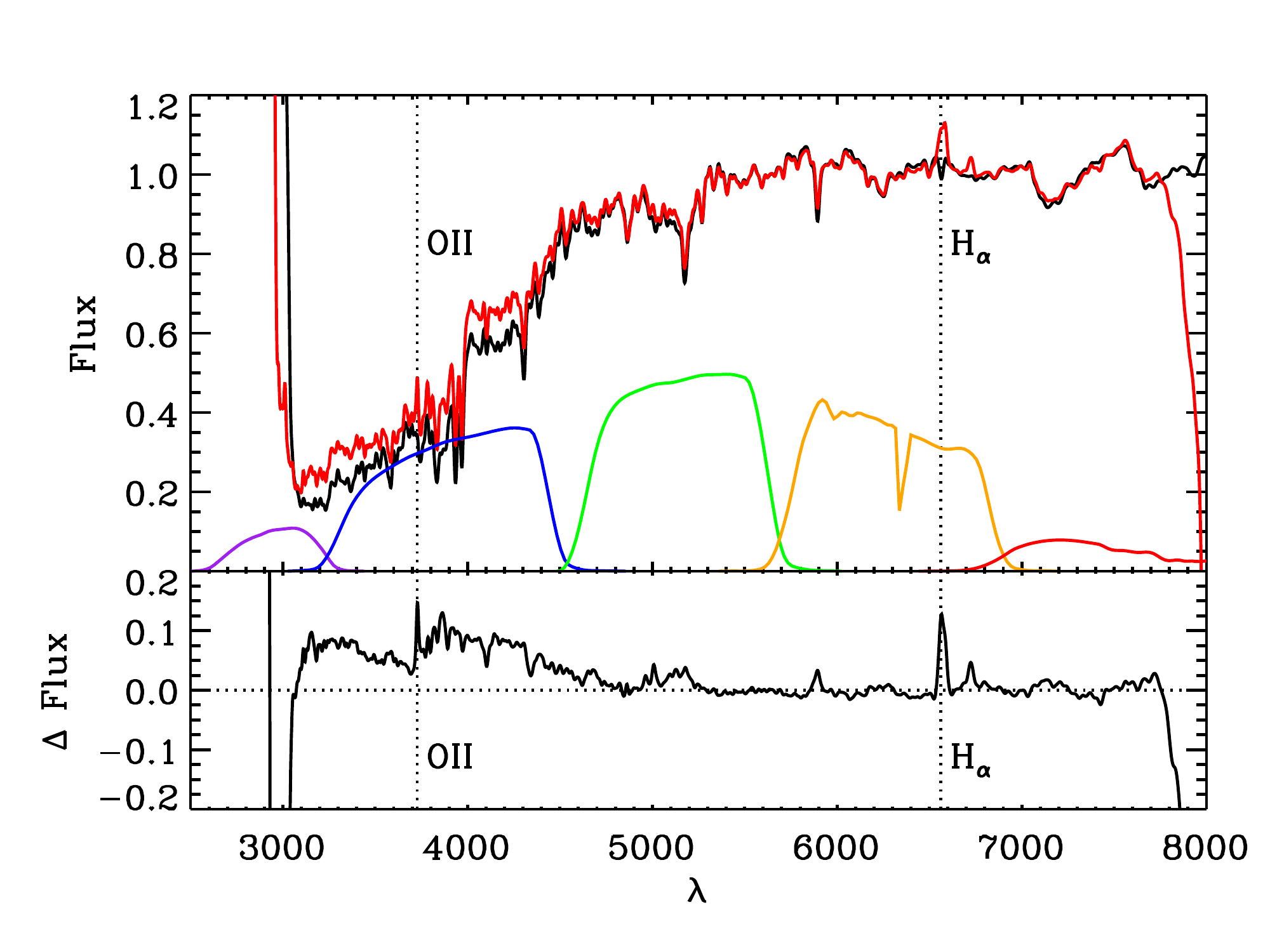}

\caption{{\bf Top panel:} Stacked rest-frame spectra for \redmagic\ galaxies with $\zphoto\in[0.18,0.22]$.
Outlier galaxies are shown in red (Clump 1 in Figure~\ref{fig:photoz}), and non-outliers in black.
Also shown are the SDSS $ugriz$ transmission curves for an extended source at $z=0.2$ assuming 1.3 air masses
(purple, blue, green, orange, red).
{\bf Bottom panel:} Difference between the two spectra in the top panel, showing the excess emission associated
with the outlier galaxy population.  The vertical dotted lines mark the \OII{} (left-most line)
and \Halpha{} (right-most line) emission lines.  Clump 1 galaxies have excess
blue light, as well as \OII{} and \Halpha{} emission indicative of a small
amount of residual star formation.}
\label{fig:stack1}
\end{figure}

We compare the rest-frame spectra of outliers in Clump 1 (the low redshift
outliers in Figure~\ref{fig:photoz}) to a control sample of 
non-outliers.  The control sample is comprised of galaxies with good \photozs\
(within $0.5\sigma$ of $\zspec=\zphoto$).
We randomly sample from the control sample so as to mirror the \photoz\
distribution of the outlier sample.  We median-stack the spectra of both sets
of galaxies, arbitrarily normalizing them to unity over the wavelength range
$\lambda=[5300\ \ang,\ 5800\ \ang]$.  We have further smoothed the spectra to
make the resulting stacks easier to interpret by eye.  The two stacked spectra
and their difference are shown in Figure~\ref{fig:stack1}.

We find that the two spectra are largely consistent with each other for
wavelengths $\lambda \gtrsim 5000\ \ang$.  At shorter wavelengths, however,
there is a clear excess of blue light in the photometric redshift outliers.  In
addition, the spectra of the outlier galaxies have obvious \Halpha{} and \OII{}
lines, demonstrating these galaxies have ongoing star formation.

\begin{figure}
\hspace{-12pt} \includegraphics[width=90mm]{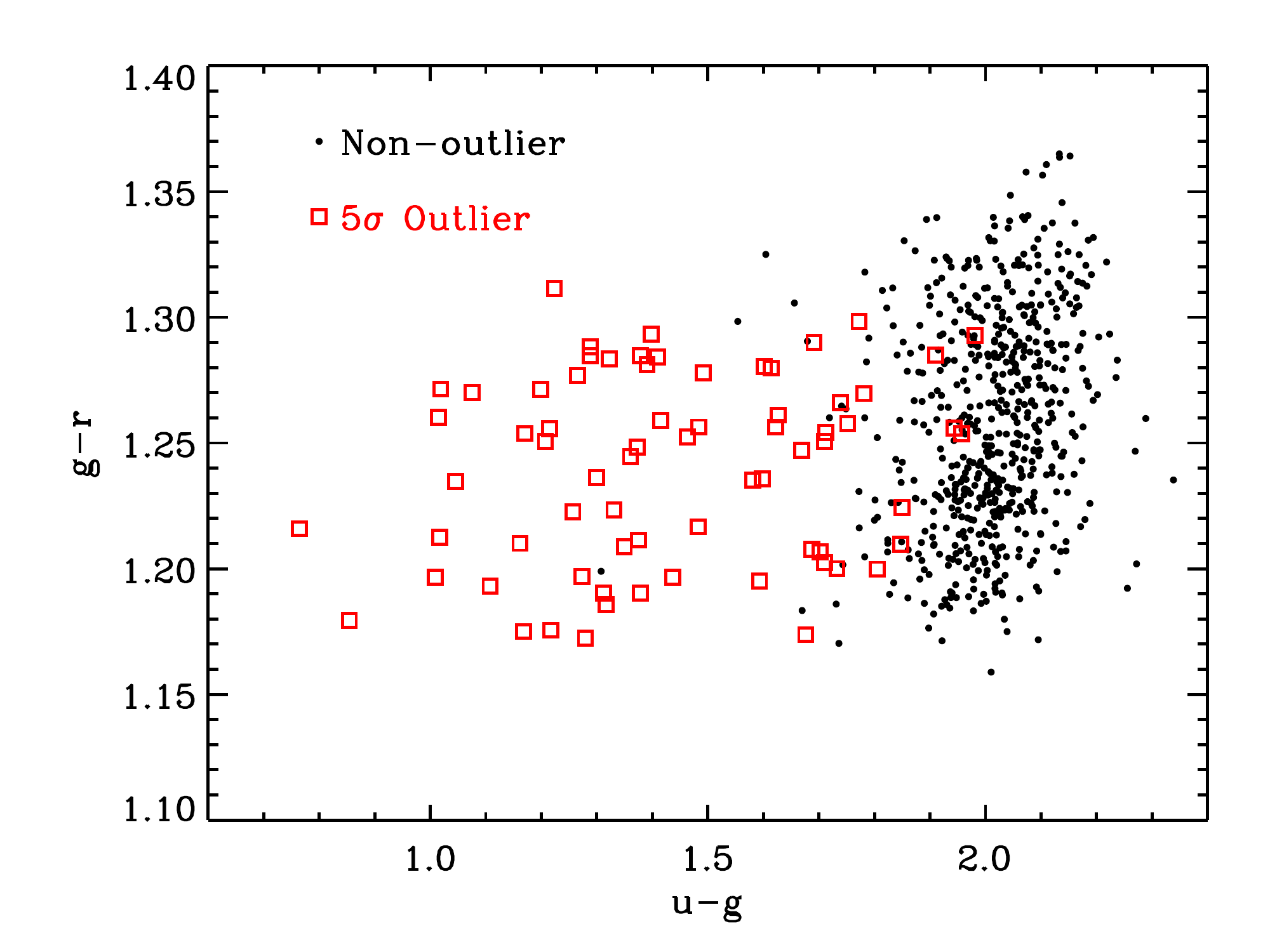}

\caption{Distribution of our fiducial S82 \redmagic\ galaxy sample in $u-g$
and $g-r$ space for galaxies in the photometric redshift bin $\zphoto \in [0.18,0.22]$.
Black points are galaxies where our photometric redshift estimate agrees with the spectroscopic
estimate within $2\sigma$, while red points correspond to $\geq 5\sigma$ redshift outliers.
}
\label{fig:s82_umg}
\end{figure}

Why is the fraction of outliers in Clump 1 is so much larger in
S82 and SV data sets relative to the DR8 sample?  This is because the S82 and
SV \redmagic{} selection was based solely on $griz$ photometry, while for DR8
we additionally included $u$-band photometry.  As the $u$-band is sensitive to
the enhanced star formation in Clump 1 galaxies, the relative contamination of
these outliers is dramatically decreased in DR8 relative to the S82 and SV
data sets.  While we did not use the S82 $u$-band in the construction of the
\redmapper{} and \redmagic{} catalogs --- its inclusion created problems with
the higher redshift ($z\sim0.5$) cluster calibration --- we do have the
data available for us to test our hypothesis.
Figure~\ref{fig:s82_umg} shows S82
\redmagic\ galaxies in the photometric redshift slice $\zphoto\in[0.18,0.22]$.
Black points are galaxies where the spectroscopic redshift of the galaxy is
within $2\sigma$ of our photometric estimate, while red points show $\geq
5\sigma$ outliers.  We see the vast majority of $5\sigma$ outliers are
unusually bright in $u$, as expected.


\subsection{Clump 2: Photo-$z$ Biased High}

\begin{figure}
\hspace{-12pt} \includegraphics[width=90mm]{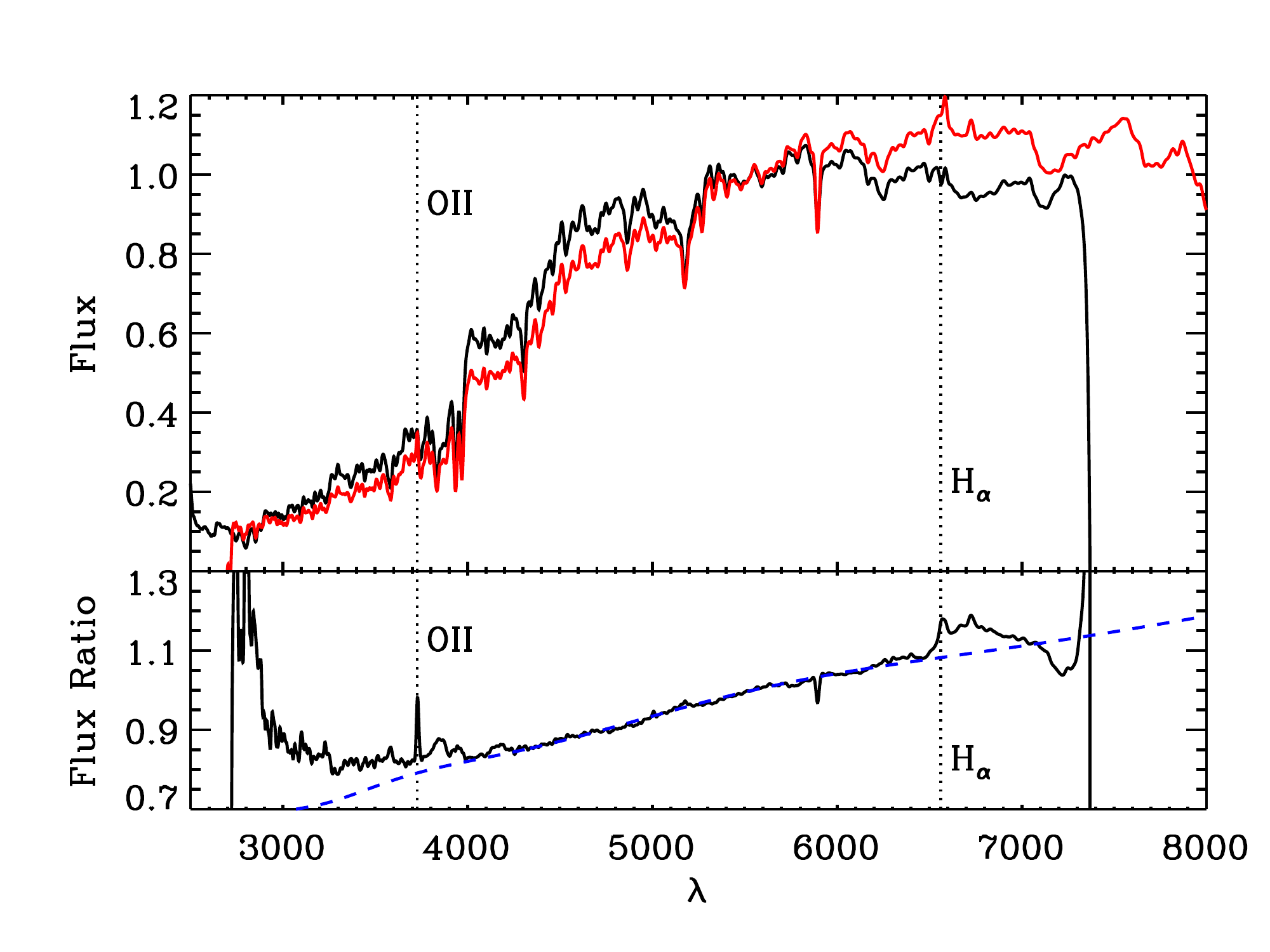}
\caption{{\bf Top panel:} Stacked rest-frame spectrum of outlier (red) and non-outlier (black) \redmagic\
galaxies for Clump 2 (see Figure~\ref{fig:photoz}).  
{\bf Bottom panel:} Ratio of the outlier to non-outlier spectra (black line). The dashed
blue line shows the effects of internal dust reddening with $E(B-V)=0.15$. 
The vertical dotted lines mark the \OII{}
and \Halpha{} emission lines, indicating a small amount of residual star
formation, as with the Clump 1 galaxies.
}
\label{fig:stack2}
\end{figure}

We repeat the spectra-stacking procedure above for Clump 2 galaxies (with
\photoz{} biased high in Figure~\ref{fig:photoz}).  For
reasons that will become apparent below, in Figure~\ref{fig:stack2} we plot not the
difference between the outlier and non-outlier spectra, but rather their ratios.
Both sets of spectra have been normalized as before.  A blue light
excess is immediately apparent, and we again see both \Halpha{} and \OII{}
emission.  However, the most salient feature is the slope of the flux ratio as
a function of wavelength, with the outlier spectra having a systematically
steeper continuum than the non-outlier galaxies.  This slope is consistent with
internal dust-reddening in the galaxy.  Specifically, the dashed blue line is the
predicted spectral ratio assuming an 
\citet{odonnell94} reddening law with $E(B-V)=0.15$. 

It is worth noting the reasons why these dusty galaxies show up in our
\redmagic{} selection only at this particular redshift range.  In particular,
at most redshifts the rest-frame reddening vector with broadband $griz$ photometry
is not parallel to the color evolution vector of the red sequence.  Consequently,
at most redshifts a galaxy that starts in the red sequence and is reddened simply
moves off the red sequence, and is not selected.  By contrast, 
at $z\approx 0.35$, the rest-frame
reddening vector is parallel to the color evolution vector of red sequence, so dust
reddening can move a galaxy from $z_{\mathrm{spec}}\sim0.3$ to $\zphoto\sim0.4$.  At the
same time, the internal reddening will suppress the excess blue emission, reducing excess blue
light as a discriminator for these galaxies.
It should also be noted that internal reddening also dims the galaxy, and thus
tends to increase photometric errors, making it even more difficult to
distinguish these galaxies from the expected template.


\subsection{Clump 3: Photo-$z$ Biased Low}

\begin{figure}
\hspace{-12pt} \includegraphics[width=90mm]{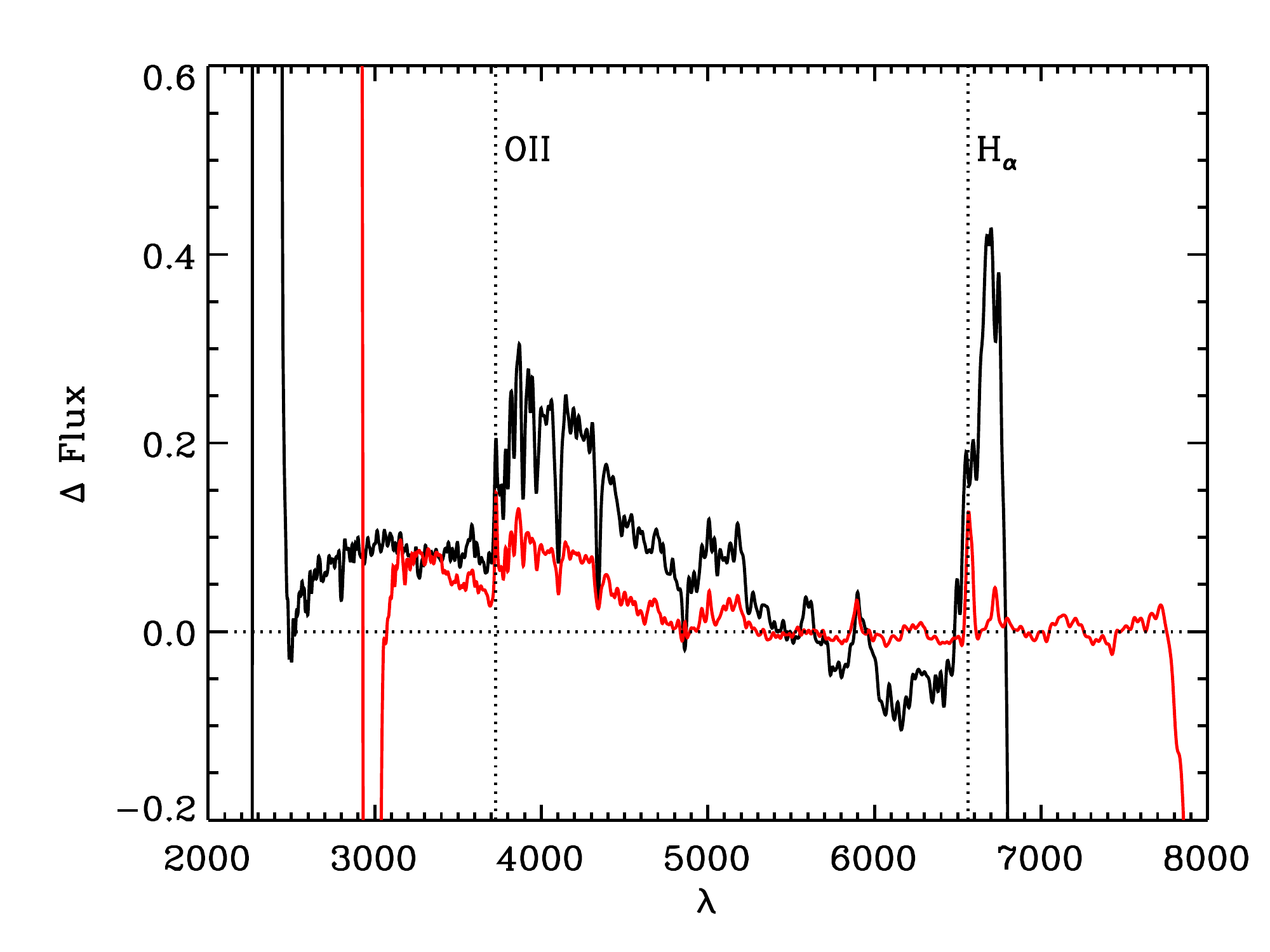}
\caption{Difference between the outlier and non-outlier stacked rest-frame spectra
for Clump 1 (red) and Clump 3 (black) galaxies (see Figure~\ref{fig:photoz}).  
The vertical dotted lines mark the \OII{} (left-most line)
and \Halpha{} (right-most line) emission lines.  Clump 3 galaxies are
qualtitatively similar to those in Clump 1, with residual star formation that
is not large enough to drive the galaxy from the photometric red sequence at
SDSS depths.
}
\label{fig:stack3}
\end{figure}

Finally, we repeat our spectral-stacking procedure for Clump 3 galaxies (with
\photoz{} biased low in Figure~\ref{fig:photoz}).  In
Figure~\ref{fig:stack3} we show the difference between the outlier and
non-outlier spectra (black line).  As a comparison, we show the difference
between outliers and non-outliers for Clump 1 (red dashed line), which are
similar in that they have $\zrm$ biased low.  We see that the differences are
qualitatively similar, but that the Clump 3 galaxies have excess emission that
is significantly larger than that of Clump 1.  This makes sense, as the SDSS
DR8 imaging is relatively shallow, and therefore the small photometric errors
for Clump 1 galaxies make the \redmagic{} selection more efficient.  In
contrast, at higher redshifts, the larger photometric errors allow for a larger
excess emission.

Having identified the physical origin of the various outlier populations of
\redmagic\ galaxies, it may be possible to construct observables that allow us
to reject such galaxies from the \redmapper\ sample.  We leave an exploration
of this possibility to future work.  Of course, it may be possible that some
of the outlier populations cannot be removed with the available photometry.
For instance, we expect Clump 1 outliers in the DES will be difficult to remove
without $u$-band.  If these outlier populations are irreducible, then they must
be adequately characterized and the corresponding $P(z)$ distributions for the
\redmagic\ galaxies must be correspondingly updated.  Alternatively, the corresponding
redshift regions ought to be excluded from high precision LSS studies.


\section{Summary and Conclusions}
\label{sec:conclusions}

Photometric redshift systematics are the primary challenge that must be
overcome for pursuing LSS studies with photometric data sets.  Based on the
fact that red sequence galaxies tend to have excellent photometric redshifts,
we have sought to address this challenge by refining red sequence selection
algorithms in the hope of creating a ``gold'' photometric galaxy sample for
photometric LSS studies.  A complementary goal is to develop a new photometric
redshift estimator for these galaxies.  The result is the \redmagic\ algorithm.

Conceptually, the algorithm is exceedingly simple: one specifies a desired comoving space density
and luminosity threshold.  The algorithm then fits {\it all}
galaxies with a red sequence template and assigns the galaxies a redshift.  Based on these redshifts,
we apply the desired luminosity threshold.  Finally, we then keep rank-order galaxies by the goodness-of-fit
statistic $\chi^2$, and keep the top $N$ galaxies that lead to the desired comoving space density.
In practice, the algorithm is necessarily more difficult to implement due to coupling
of the photometric redshift estimates to the galaxy density via the photometric redshift afterburner, but
the above description captures the spirit of the algorithm well.

As shown in Section~\ref{sec:performance}, we find that \redmagic{} is indeed
successful at identifying red sequence galaxies, and that the corresponding
photometric redshift estimates are of very high quality, with a low bias
$(\lesssim 0.5\%$), low scatter $\lesssim 1.6\%$, and low rate of catastrophic
outliers $\leq 2\%$, with the exact values depending on the precise sample
under consideration.  As demonstrated in Section~\ref{sec:selection}, the
\redmagic{} selection yields galaxies with superior \photoz{} performance 
to the standard color-cut selection method used to define the SDSS CMASS sample.  In addition,
the \photoz{} scatter is correctly estimated {\it a
priori}.  As detailed in Section~\ref{sec:spec}, this performance is
comparable to the best machine learning \photoz\ algorithms available today
when the same input data is used.  Machine learning algorithms can improve upon
the \photoz\ performance of \redmagic\ if additional information is provided,
though the improvement remains modest.

There are, however, two critical advantages of \redmagic\ \photozs\ relative to
machine learning based algorithms.  The first is that \redmagic\ has minimal
spectroscopic requirements: it is much easier to get the necessary cluster
redshifts that enable the \redmagic\ algorithm than it is to acquire
representative training samples for \redmagic.  The second important difference
is that, in the absence of representative spectroscopic sampling, machine
learning based algorithm are expected to be biased for galaxies that fall
outside the training data set, especially at the faint end as demonstrated in
Figure~\ref{fig:comparison}.  This failure mode is non-existent for \redmagic.

Of course, should representative spectroscopic training sets become available for \redmagic\ galaxies in the future,
one should pursue machine learning techniques to improve \redmagic\ \photozs.  Even with the context of
\redmagic, we explicitly demonstrated that representative spectroscopic sampling of \redmagic\ galaxies
enables \photoz\ estimation that is unbiased at the $0.1\%$ level, and with extremely well characterized
\photoz\ errors (Figure~\ref{fig:ab}, right panel).

Despite all of these successes, some additional work clearly remains.  First, the current photometric redshifts must
be extended into $P(z)$ distributions to properly capture skewness and kurtosis where it exists, for instance near
filter transitions.  Perhaps more importantly, however, the current samples clearly exhibit three distinct classes of redshift
outliers.  We have been able to identify the phyical origin of these outliers --- Clumps 1 and 3 in Figure~\ref{fig:photoz}
are ellipticals or S0 galaxies with residual star formation, while Clump 2 galaxies are very dusty ($E(B-V)\approx 0.15$)
elliptical/S0 galaxies.  These dusty galaxies also exhibit residual star formation, but the primary reason they are outliers
is their high dust content.  We defer the question of whether it is possible to photometrically identify these outliers
and remove them from the \redmagic\ sample to future work. 

\section*{Acknowledgements}

This paper has gone through internal review by the DES collaboration.
We are grateful for the extraordinary contributions of our CTIO colleagues and the DECam Construction, 
Commissioning and Science Verification
teams in achieving the excellent instrument and telescope conditions that have made this work possible.  
The success of this project also  relies critically on the expertise and dedication of the DES Data Management group.

This work was supported in part by the U.S. Department of Energy contract to
SLAC no. DE-AC02-76SF00515.

Funding for the DES Projects has been provided by the U.S. Department of Energy, the U.S. National Science Foundation, the Ministry of Science and Education of Spain, 
the Science and Technology Facilities Council of the United Kingdom, the Higher Education Funding Council for England, the National Center for Supercomputing 
Applications at the University of Illinois at Urbana-Champaign, the Kavli Institute of Cosmological Physics at the University of Chicago, 
the Center for Cosmology and Astro-Particle Physics at the Ohio State University,
the Mitchell Institute for Fundamental Physics and Astronomy at Texas A\&M University, Financiadora de Estudos e Projetos, 
Funda{\c c}{\~a}o Carlos Chagas Filho de Amparo {\`a} Pesquisa do Estado do Rio de Janeiro, Conselho Nacional de Desenvolvimento Cient{\'i}fico e Tecnol{\'o}gico and 
the Minist{\'e}rio da Ci{\^e}ncia, Tecnologia e Inova{\c c}{\~a}o, the Deutsche Forschungsgemeinschaft and the Collaborating Institutions in the Dark Energy Survey. 
The DES data management system is supported by the National Science Foundation under Grant Number AST-1138766.

The Collaborating Institutions are Argonne National Laboratory, the University of California at Santa Cruz, the University of Cambridge, Centro de Investigaciones En{\'e}rgeticas, 
Medioambientales y Tecnol{\'o}gicas-Madrid, the University of Chicago, University College London, the DES-Brazil Consortium, the University of Edinburgh, 
the Eidgen{\"o}ssische Technische Hochschule (ETH) Z{\"u}rich, 
Fermi National Accelerator Laboratory, the University of Illinois at Urbana-Champaign, the Institut de Ci{\`e}ncies de l'Espai (IEEC/CSIC), 
the Institut de F{\'i}sica d'Altes Energies, Lawrence Berkeley National Laboratory, the Ludwig-Maximilians Universit{\"a}t M{\"u}nchen and the associated Excellence Cluster Universe, 
the University of Michigan, the National Optical Astronomy Observatory, the University of Nottingham, The Ohio State University, the University of Pennsylvania, the University of Portsmouth, 
SLAC National Accelerator Laboratory, Stanford University, the University of Sussex, and Texas A\&M University.

The DES participants from Spanish institutions are partially supported by MINECO under grants AYA2012-39559, ESP2013-48274, FPA2013-47986, and Centro de Excelencia Severo Ochoa SEV-2012-0234.
Research leading to these results has received funding from the European Research Council under the European UnionÕs Seventh Framework Programme (FP7/2007-2013) including ERC grant agreements 
 240672, 291329, and 306478.

Based in part on observations taken at the Australian Astronomical Observatory under program A/2013B/012.

Funding for SDSS-III has been provided by the Alfred P. Sloan Foundation, the
Participating Institutions, the National Science Foundation, and the
U.S. Department of Energy Office of Science. The SDSS-III web site is
http://www.sdss3.org/.

SDSS-III is managed by the Astrophysical Research Consortium for the
Participating Institutions of the SDSS-III Collaboration including the
University of Arizona, the Brazilian Participation Group, Brookhaven National
Laboratory, University of Cambridge, Carnegie Mellon University, University of
Florida, the French Participation Group, the German Participation Group,
Harvard University, the Instituto de Astrofisica de Canarias, the Michigan
State/Notre Dame/JINA Participation Group, Johns Hopkins University, Lawrence
Berkeley National Laboratory, Max Planck Institute for Astrophysics, Max Planck
Institute for Extraterrestrial Physics, New Mexico State University, New York
University, Ohio State University, Pennsylvania State University, University of
Portsmouth, Princeton University, the Spanish Participation Group, University
of Tokyo, University of Utah, Vanderbilt University, University of Virginia,
University of Washington, and Yale University.

This publication makes use of data products from the Two Micron All Sky Survey,
which is a joint project of the University of Massachusetts and the Infrared
Processing and Analysis Center/California Institute of Technology, funded by
the National Aeronautics and Space Administration and the National Science
Foundation.

\newcommand\AAA{{A\& A}}
\newcommand\PhysRep{{Physics Reports}}
\newcommand\apj{{ApJ}}
\newcommand\PhysRevD{ {Phys. Rev. D}} 
\newcommand\prd{ {Phys. Rev. D}} 
\newcommand\PhysRevLet[3]{ {Phys. Rev. Letters} }
\newcommand\mnras{{MNRAS}}
\newcommand\PhysLet{{Physics Letters}}
\newcommand\AJ{{AJ}}
\newcommand\aj{{AJ}}
\newcommand\aap{ {A \& A}}
\newcommand\apjl{{ApJ Letters}}
\newcommand\apjs{{ApJ Supplement}}
\newcommand\aph{astro-ph/}
\newcommand\AREVAA{{Ann. Rev. A.\& A.}}
\newcommand{\pasj}{{PASJ}}
\newcommand{\pasp}{Pub. of the Astro. Soc. of the Pacific}
\newcommand{\aaps}{ {A \& A Supplement}}

\bibliographystyle{mn2e}
\bibliography{mybib}

\begin{thebibliography}{}

\bibitem[\protect\citeauthoryear{{Abazajian} et~al.,}{{Abazajian}
  et~al.}{2009}]{abazajianetal09}
{Abazajian} K.~N.,  et~al., 2009, \apjs, 182, 543

\bibitem[\protect\citeauthoryear{{Abdalla}, {Banerji}, {Lahav} \&
  {Rashkov}}{{Abdalla} et~al.}{2011}]{abdallaetal11}
{Abdalla} F.~B.,  {Banerji} M.,  {Lahav} O.,    {Rashkov} V.,  2011, \mnras,
  417, 1891

\bibitem[\protect\citeauthoryear{{Ahn} et~al.,}{{Ahn}  et~al.}{2013}]{dr10}
{Ahn} C.~P.,  et~al., 2013, ArXiv: 1307.7735.

\bibitem[\protect\citeauthoryear{{Aihara} et~al.,}{{Aihara}
  et~al.}{2011}]{dr8}
{Aihara} H.,  et~al., 2011, \apjs, 193, 29

\bibitem[\protect\citeauthoryear{{Annis} et~al.,}{{Annis}
  et~al.}{2011}]{annisetal11}
{Annis} J.,  et~al., 2011, ArXiv: 1111.6619

\bibitem[\protect\citeauthoryear{{Ben{\'{\i}}tez}}{{Ben{\'{\i}}tez}}{2000}]{bpz00}
{Ben{\'{\i}}tez} N.,  2000, \apj, 536, 571

\bibitem[\protect\citeauthoryear{{Bertin}}{{Bertin}}{2010}]{bertin10}
{Bertin} E., , 2010, {SWarp: Resampling and Co-adding FITS Images Together},
  Astrophysics Source Code Library

\bibitem[\protect\citeauthoryear{{Bertin}}{{Bertin}}{2011}]{bertin11}
{Bertin} E.,  2011, in {Evans} I.~N.,  {Accomazzi} A.,  {Mink} D.~J.,   {Rots}
  A.~H.,  eds, Astronomical Data Analysis Software and Systems XX Vol.~442 of
  Astronomical Society of the Pacific Conference Series, {Automated Morphometry
  with SExtractor and PSFEx}.
p.~435

\bibitem[\protect\citeauthoryear{{Bertin} \& {Arnouts}}{{Bertin} \&
  {Arnouts}}{1996}]{bertinarnouts96}
{Bertin} E.,  {Arnouts} S.,  1996, \aaps, 117, 393

\bibitem[\protect\citeauthoryear{{Blanton} \& {Roweis}}{{Blanton} \&
  {Roweis}}{2007}]{blanton07}
{Blanton} M.~R.,  {Roweis} S.,  2007, \aj, 133, 734

\bibitem[\protect\citeauthoryear{{Bleem} et~al.,}{{Bleem}
  et~al.}{2015}]{bleemetal15}
{Bleem} L.~E.,  et~al., 2015, \apjs, 216, 27

\bibitem[\protect\citeauthoryear{{Bonnett}}{{Bonnett}}{2015}]{bonnet15}
{Bonnett} C.,  2015, \mnras, 449, 1043

\bibitem[\protect\citeauthoryear{{Bradshaw} et~al.,}{{Bradshaw}
  et~al.}{2013}]{bradshawetal13}
{Bradshaw} E.~J.,  et~al., 2013, \mnras, 433, 194

\bibitem[\protect\citeauthoryear{{Bruzual} \& {Charlot}}{{Bruzual} \&
  {Charlot}}{2003}]{bruzualcharlot03}
{Bruzual} G.,  {Charlot} S.,  2003, \mnras, 344, 1000

\bibitem[\protect\citeauthoryear{{Budav{\'a}ri} et~al.,}{{Budav{\'a}ri}
  et~al.}{2000}]{budavarietal00}
{Budav{\'a}ri} T.,  et~al., 2000, \aj, 120, 1588

\bibitem[\protect\citeauthoryear{{Carlstrom} et~al.,}{{Carlstrom}
  et~al.}{2011}]{carlstrometal11}
{Carlstrom} J.~E.,  et~al., 2011, \pasp, 123, 568

\bibitem[\protect\citeauthoryear{{Colless}, {Dalton}, {Maddox}, {Sutherland} \&
  {the 2dF collaboration}}{{Colless} et~al.}{2001}]{collessetal01}
{Colless} M.,  {Dalton} G.,  {Maddox} S.,  {Sutherland} W.,    {the 2dF
  collaboration} 2001, \mnras, 328, 1039

\bibitem[\protect\citeauthoryear{{Collister} \& {Lahav}}{{Collister} \&
  {Lahav}}{2004}]{collisteretal04}
{Collister} A.~A.,  {Lahav} O.,  2004, \pasp, 116, 345

\bibitem[\protect\citeauthoryear{{Cooper}, {Yan}, {Dickinson}, {Juneau},
  {Lotz}, {Newman}, {Papovich}, {Salim}, {Walth}, {Weiner} \&
  {Willmer}}{{Cooper} et~al.}{2012}]{cooperetal12}
{Cooper} M.~C.,  {Yan} R.,  {Dickinson} M.,  {Juneau} S.,  {Lotz} J.~M.,
  {Newman} J.~A.,  {Papovich} C.,  {Salim} S.,  {Walth} G.,  {Weiner} B.~J.,
  {Willmer} C.~N.~A.,  2012, \mnras, 425, 2116

\bibitem[\protect\citeauthoryear{{Csabai} et~al.,}{{Csabai}
  et~al.}{2007}]{csabaietal07}
{Csabai} I.,  et~al., 2007, Astronomische Nachrichten, 328, 852

\bibitem[\protect\citeauthoryear{{Dawson} et~al.,}{{Dawson}
  et~al.}{2013}]{dawsonetal13}
{Dawson} K.~S.,  et~al., 2013, \AJ, 145, 10

\bibitem[\protect\citeauthoryear{{Desai}, {Armstrong}, {Mohr}, {Semler}, {Liu},
  {Bertin}, {Allam}, {Barkhouse}, {Bazin}, {Buckley-Geer}, {Cooper}, {Hansen},
  {High}, {Lin}, {Lin}, {Ngeow}, {Rest}, {Song}, {Tucker} \& {Zenteno}}{{Desai}
  et~al.}{2012}]{desaietal12}
{Desai} S.,  {Armstrong} R.,  {Mohr} J.~J.,  {Semler} D.~R.,  {Liu} J.,
  {Bertin} E.,  {Allam} S.~S.,  {Barkhouse} W.~A.,  {Bazin} G.,  {Buckley-Geer}
  E.~J.,  {Cooper} M.~C.,  {Hansen} S.~M.,  {High} F.~W.,  {Lin} H.,  {Lin}
  Y.-T.,  {Ngeow} C.-C.,  {Rest} A.,  {Song} J.,  {Tucker} D.,    {Zenteno} A.,
   2012, \apj, 757, 83

\bibitem[\protect\citeauthoryear{{Diehl} et~al.,}{{Diehl}
  et~al.}{2012}]{diehletal12}
{Diehl} T.,  et~al., 2012, Physics Procedia, 37, 1332

\bibitem[\protect\citeauthoryear{{Driver} et~al.,}{{Driver}
  et~al.}{2011}]{gamadr1}
{Driver} S.~P.,  et~al., 2011, \mnras, 413, 971

\bibitem[\protect\citeauthoryear{{Eisenstein} et~al.,}{{Eisenstein}
  et~al.}{2001}]{eisensteinetal01}
{Eisenstein} D.~J.,  et~al., 2001, \AJ, 122, 2267

\bibitem[\protect\citeauthoryear{{Eisenstein} et~al.,}{{Eisenstein}
  et~al.}{2005}]{eisensteinetal05}
{Eisenstein} D.~J.,  et~al., 2005, \apj, 633, 560

\bibitem[\protect\citeauthoryear{{Flaugher} et~al.,}{{Flaugher}
  et~al.}{2015}]{flaugheretal15}
{Flaugher} B.,  et~al., 2015, ArXiv e-prints

\bibitem[\protect\citeauthoryear{{Franzetti}, {Garilli}, {Guzzo}, {Marchetti}
  \& {Scodeggio}}{{Franzetti} et~al.}{2014}]{franzettietal14}
{Franzetti} P.,  {Garilli} B.,  {Guzzo} L.,  {Marchetti} A.,    {Scodeggio} M.,
   2014, ArXiv e-prints

\bibitem[\protect\citeauthoryear{{Garilli} et~al.,}{{Garilli}
  et~al.}{2008}]{garillietal08}
{Garilli} B.,  et~al., 2008, \aap, 486, 683

\bibitem[\protect\citeauthoryear{{Garilli} et~al.,}{{Garilli}
  et~al.}{2014}]{garillietal14}
{Garilli} B.,  et~al., 2014, \aap, 562, A23

\bibitem[\protect\citeauthoryear{{G{\'o}rski}, {Hivon}, {Banday}, {Wandelt},
  {Hansen}, {Reinecke} \& {Bartelmann}}{{G{\'o}rski}
  et~al.}{2005}]{gorskietal05}
{G{\'o}rski} K.~M.,  {Hivon} E.,  {Banday} A.~J.,  {Wandelt} B.~D.,  {Hansen}
  F.~K.,  {Reinecke} M.,    {Bartelmann} M.,  2005, \apj, 622, 759

\bibitem[\protect\citeauthoryear{{Graff} et~al.,}{{Graff}
  et~al.}{2014}]{graffetal14}
{Graff} P.,  et~al., 2014, \mnras, 441, 1741

\bibitem[\protect\citeauthoryear{{Hoffleit} \& {Jaschek}}{{Hoffleit} \&
  {Jaschek}}{1991}]{hoffleitjaschek91}
{Hoffleit} D.,  {Jaschek} C.~.,  1991, {The Bright star catalogue}

\bibitem[\protect\citeauthoryear{{Hoyle} et~al.,}{{Hoyle}
  et~al.}{2015}]{hoyleetal15}
{Hoyle} B.,  et~al., 2015, \mnras, 449, 1275

\bibitem[\protect\citeauthoryear{{Hoyle}, {Rau}, {Bonnett}, {Seitz} \&
  {Weller}}{{Hoyle} et~al.}{2015}]{hoyleetal15b}
{Hoyle} B.,  {Rau} M.~M.,  {Bonnett} C.,  {Seitz} S.,    {Weller} J.,  2015,
  \mnras, 450, 305

\bibitem[\protect\citeauthoryear{{Jarvis}, {Bernstein} \& {Jain}}{{Jarvis}
  et~al.}{2004}]{jarvisetal04}
{Jarvis} M.,  {Bernstein} G.,    {Jain} B.,  2004, \mnras, 352, 338

\bibitem[\protect\citeauthoryear{{Kelly}, {von der Linden}, {Applegate},
  {Allen}, {Allen}, {Burchat}, {Burke}, {Ebeling}, {Capak}, {Czoske},
  {Donovan}, {Mantz} \& {Morris}}{{Kelly} et~al.}{2014}]{kellyetal14}
{Kelly} P.~L.,  {von der Linden} A.,  {Applegate} D.~E.,  {Allen} M.~T.,
  {Allen} S.~W.,  {Burchat} P.~R.,  {Burke} D.~L.,  {Ebeling} H.,  {Capak} P.,
  {Czoske} O.,  {Donovan} D.,  {Mantz} A.,    {Morris} R.~G.,  2014, \mnras,
  439, 28

\bibitem[\protect\citeauthoryear{{LSST Science Collaboration.}}{{LSST Science
  Collaboration.}}{2009}]{lsst09}
{LSST Science Collaboration.} 2009, ArXiv: 0912.0201

\bibitem[\protect\citeauthoryear{{McLure} et~al.,}{{McLure}
  et~al.}{2013}]{mclureetal13}
{McLure} R.~J.,  et~al., 2013, \mnras, 428, 1088

\bibitem[\protect\citeauthoryear{Nelder \& Mead}{Nelder \&
  Mead}{1965}]{nelder65}
Nelder J.~A.,  Mead R.,  1965, Computer Journal, pp 308--313

\bibitem[\protect\citeauthoryear{{O'Donnell}}{{O'Donnell}}{1994}]{odonnell94}
{O'Donnell} J.~E.,  1994, \apj, 422, 158

\bibitem[\protect\citeauthoryear{{Ross} et~al.,}{{Ross}
  et~al.}{2011}]{rossetal11}
{Ross} A.~J.,  et~al., 2011, \mnras, 417, 1350

\bibitem[\protect\citeauthoryear{{Rozo} et~al.,}{{Rozo}
  et~al.}{2014}]{rozoetal14e}
{Rozo} E.,  et~al., 2014, ArXiv: 1410.1193

\bibitem[\protect\citeauthoryear{{Rykoff} et~al.,}{{Rykoff}
  et~al.}{2014}]{rykoffetal14}
{Rykoff} E.~S.,  et~al., 2014, \apj, 785, 104

\bibitem[\protect\citeauthoryear{{S{\'a}nchez} et~al.,}{{S{\'a}nchez}
  et~al.}{2014}]{sanchezetal14}
{S{\'a}nchez} C.,  et~al., 2014, \mnras, 445, 1482

\bibitem[\protect\citeauthoryear{{Sevilla} et~al.,}{{Sevilla}
  et~al.}{2011}]{sevillaetal11}
{Sevilla} I.,  et~al., 2011, ArXiv:1109.6741

\bibitem[\protect\citeauthoryear{{Sinnott}}{{Sinnott}}{1988}]{sinnott88}
{Sinnott} R.~W.,  1988, {The complete new general catalogue and index
  catalogues of nebulae and star clusters by J. L. E. Dreyer}

\bibitem[\protect\citeauthoryear{{Skrutskie} et~al.,}{{Skrutskie}
  et~al.}{2006}]{skrutskieetal06}
{Skrutskie} M.~F.,  et~al., 2006, \aj, 131, 1163

\bibitem[\protect\citeauthoryear{{Stoughton} et~al.,}{{Stoughton}
  et~al.}{2002}]{stoughtonetal02}
{Stoughton} C.,  et~al., 2002, \AJ, 123, 485

\bibitem[\protect\citeauthoryear{{The Dark Energy Survey Collaboration}}{{The
  Dark Energy Survey Collaboration}}{2005}]{desetal05}
{The Dark Energy Survey Collaboration} 2005, ArXiv Astrophysics e-prints

\bibitem[\protect\citeauthoryear{{Tojeiro} et~al.,}{{Tojeiro}
  et~al.}{2014}]{tojeiroetal14}
{Tojeiro} R.,  et~al., 2014, \mnras, 440, 2222

\bibitem[\protect\citeauthoryear{{York} et~al.,}{{York}
  et~al.}{2000}]{yorketal00}
{York} D.~G.,  et~al., 2000, \AJ, 120, 1579

\bibitem[\protect\citeauthoryear{{Yuan} et~al.,}{{Yuan}
  et~al.}{2015}]{ozdes15}
{Yuan} F.,  et~al., 2015, ArXiv: 1504.03039

\end{thebibliography}
\vspace{20pt}

\noindent {\bf Affiliations:}\\
$^{1}$Department of Physics, University of Arizona, Tucson, AZ 85721, USA\\
$^{2}$Kavli Institute for Particle Astrophysics \& Cosmology, P. O. Box 2450, Stanford University, Stanford, CA 94305, USA\\
$^{3}$SLAC National Accelerator Laboratory, Menlo Park, CA 94025, USA\\
$^{4}$Institut de F\'{\i}sica d'Altes Energies, Universitat Aut\`onoma de Barcelona, E-08193 Bellaterra, Barcelona, Spain\\
$^{5}$Institut de Ci\`encies de l'Espai, IEEC-CSIC, Campus UAB, Carrer de Can Magrans, s/n,  08193 Bellaterra, Barcelona, Spain\\
$^{6}$ARC Centre of Excellence for All-sky Astrophysics (CAAS- TRO)\\
$^{7}$School of Mathematics and Physics, University of Queensland, QLD 4072, Australia\\
$^{8}$Universit\"ats-Sternwarte, Fakult\"at f\"ur Physik, Ludwig-Maximilians Universit\"at M\"unchen, Scheinerstr. 1, 81679 M\"unchen, Germany\\
$^{9}$Department of Physics \& Astronomy, University College London, Gower Street, London, WC1E 6BT, UK\\
$^{10}$Department of Physics, Stanford University, 382 Via Pueblo Mall, Stanford, CA 94305, USA\\
$^{11}$Cerro Tololo Inter-American Observatory, National Optical Astronomy Observatory, Casilla 603, La Serena, Chile\\
$^{12}$Institute of Astronomy, University of Cambridge, Madingley Road, Cambridge CB3 0HA, UK\\
$^{13}$Kavli Institute for Cosmology, University of Cambridge, Madingley Road, Cambridge CB3 0HA, UK\\
$^{14}$Department of Physics and Astronomy, University of Pennsylvania, Philadelphia, PA 19104, USA\\
$^{15}$CNRS, UMR 7095, Institut d'Astrophysique de Paris, F-75014, Paris, France\\
$^{16}$Sorbonne Universit\'es, UPMC Univ Paris 06, UMR 7095, Institut d'Astrophysique de Paris, F-75014, Paris, France\\
$^{17}$Fermi National Accelerator Laboratory, P. O. Box 500, Batavia, IL 60510, USA\\
$^{18}$Institute of Cosmology \& Gravitation, University of Portsmouth, Portsmouth, PO1 3FX, UK\\
$^{19}$Laborat\'orio Interinstitucional de e-Astronomia - LIneA, Rua Gal. Jos\'e Cristino 77, Rio de Janeiro, RJ - 20921-400, Brazil\\
$^{20}$Observat\'orio Nacional, Rua Gal. Jos\'e Cristino 77, Rio de Janeiro, RJ - 20921-400, Brazil\\
$^{21}$€University of Notre Dame, Notre Dame, IN 46556, USA\\
$^{22}$Department of Astronomy, University of Illinois, 1002 W. Green Street, Urbana, IL 61801, USA\\
$^{23}$National Center for Supercomputing Applications, 1205 West Clark St., Urbana, IL 61801, USA\\
$^{24}$Research School of Astronomy and Astrophysics, Australian National University,Canberra, ACT 2611, Australia\\
$^{25}$George P. and Cynthia Woods Mitchell Institute for Fundamental Physics and Astronomy, and Department of Physics and Astronomy, Texas A\&M University, College Station, TX 77843,  USA\\
$^{26}$Excellence Cluster Universe, Boltzmannstr.\ 2, 85748 Garching, Germany\\
$^{27}$Faculty of Physics, Ludwig-Maximilians University, Scheinerstr. 1, 81679 Munich, Germany\\
$^{28}$Jet Propulsion Laboratory, California Institute of Technology, 4800 Oak Grove Dr., Pasadena, CA 91109, USA\\
$^{29}$Department of Astronomy, University of Michigan, Ann Arbor, MI 48109, USA\\
$^{30}$Department of Physics, University of Michigan, Ann Arbor, MI 48109, USA\\
$^{31}$Kavli Institute for Cosmological Physics, University of Chicago, Chicago, IL 60637, USA\\
$^{32}$€Centre for Astrophysics \& Supercomputing, Swinburne University of Technology, Victoria 3122, Australia\\
$^{33}$Max Planck Institute for Extraterrestrial Physics, Giessenbachstrasse, 85748 Garching, Germany\\
$^{34}$Center for Cosmology and Astro-Particle Physics, The Ohio State University, Columbus, OH 43210, USA\\
$^{35}$Department of Physics, The Ohio State University, Columbus, OH 43210, USA\\
$^{36}$Lawrence Berkeley National Laboratory, 1 Cyclotron Road, Berkeley, CA 94720, USA\\
$^{37}$Australian Astronomical Observatory, North Ryde, NSW 2113, Australia\\
$^{38}$Departamento de F\'{\i}sica Matem\'atica,  Instituto de F\'{\i}sica, Universidade de S\~ao Paulo,  CP 66318, CEP 05314-970, S\~ao Paulo, SP,  Brazil\\
$^{39}$Department of Astronomy, The Ohio State University, Columbus, OH 43210, USA\\
$^{40}$Instituci\'o Catalana de Recerca i Estudis Avan\c{c}ats, E-08010 Barcelona, Spain\\
$^{41}$Department of Physics and Astronomy, Pevensey Building, University of Sussex, Brighton, BN1 9QH, UK\\
$^{42}$Centro de Investigaciones Energ\'eticas, Medioambientales y Tecnol\'ogicas (CIEMAT), Madrid, Spain\\
$^{43}$Instituto de F\'\i sica, UFRGS, Caixa Postal 15051, Porto Alegre, RS - 91501-970, Brazil\\
$^{44}$Department of Physics, University of Illinois, 1110 W. Green St., Urbana, IL 61801, USA\\
$^{45}$Argonne National Laboratory, 9700 South Cass Avenue, Lemont, IL 60439, USA

\appendix

\section{Star-Galaxy Separation}
\label{app:stargal}

\begin{figure}
  \hspace{-12pt} \includegraphics[width=90mm]{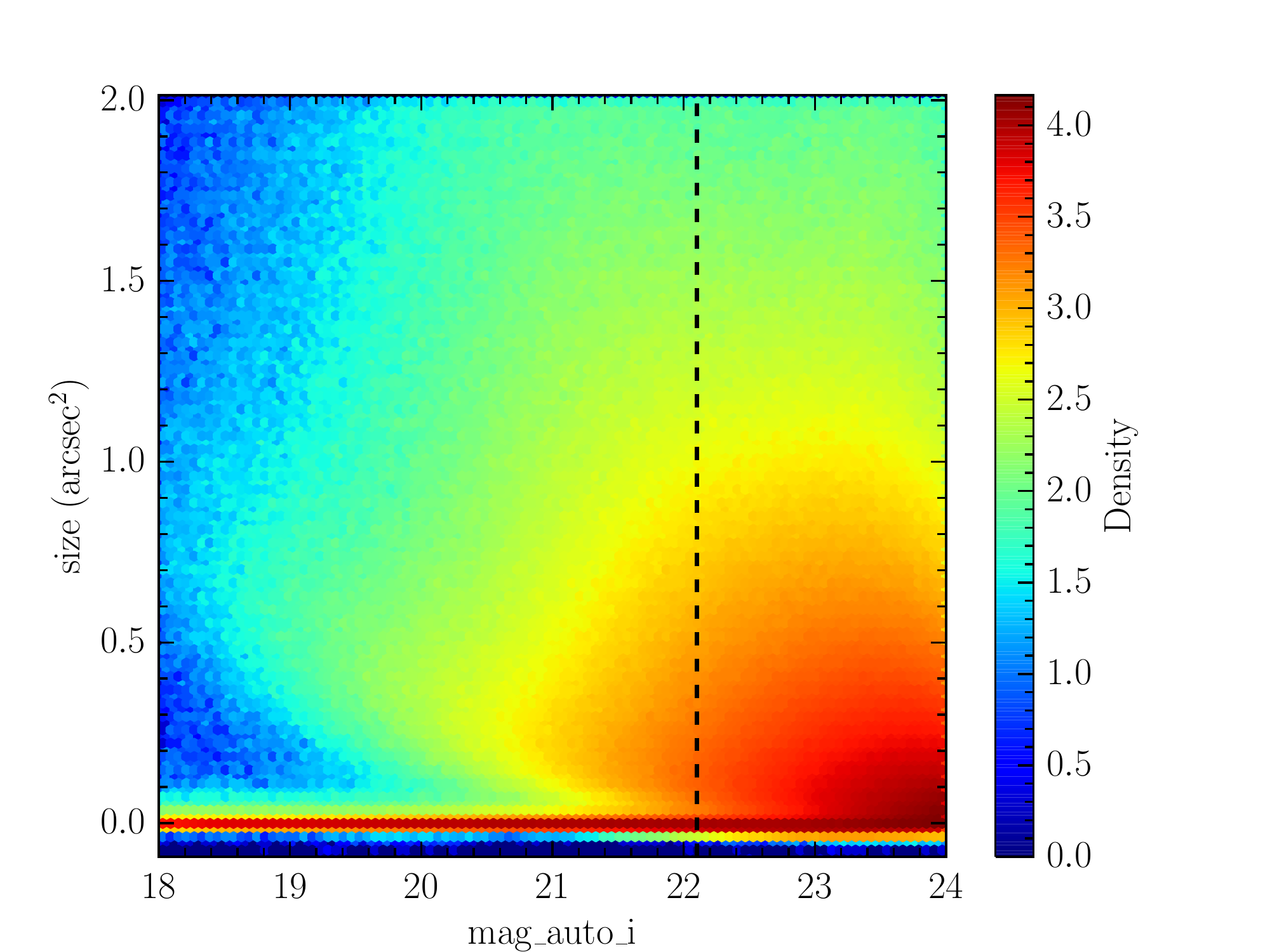}
  \caption{Intrinsic object size, {\tt exp\_t}, as a function of $\mi$ (as
    estimated with {\tt MAG\_AUTO}.  At the bright end, the stars are clearly
    separated from the galaxies, while the confusion is apparent at
    $\mi\sim23$.  The magnitude of the galaxies in the \redmagic{} sample described
    here, with $z\lesssim0.8$ and $L/L_* > 0.5$, is shown with a dashed black line.}
  \label{fig:ngmix_stargal}
\end{figure}

To perform star/galaxy separation, we use object size estimates from the
\ngmix{} multi-epoch shape fitting catalog~(Jarvis et al, in prep).  The
\ngmix{} algorithm fits an exponential disc profile to each object (in all
individual observations of each $griz$ band), and
estimates an intrinsic (psf-deconvolved) size ({\tt exp\_t}), as well as an
error on that size ({\tt exp\_t\_err}).  Figure~\ref{fig:ngmix_stargal} shows a
distribution of object sizes as a function of magnitude in the SPTE footprint.
The stellar locus at zero-size is obviously separated from the galaxy locus at the
bright end.  At the faint end, where the intrinsic size of the galaxies is
close to the typical seeing, it is harder to distinguish between the two loci.  
Our goal here is to select as complete a galaxy sample as possible while
minimizing stellar contamination.  Our task is made a little easier by the fact
that we are limiting ourselves to red galaxies with $z\leq 0.8$ and $L/L_* >
0.5$, the magnitude limit of which is denoted with a dashed red line in the
figure.


Until we develop a full probabilistic star/galaxy separator from the \ngmix{}
size estimator, we have decided to make use of simple cuts based on the
intrinsic size and error on the size.  At the bright end, we see that true
stars do not have intrinsic size $\mathtt{exp\_t} > 0.002$.  At the faint end,
we wish to make a selection that has as high a galaxy completeness as possible,
minimizing stellar contamination.  We make the ansatz that such a cut will
take the form 
\be
\mathrm{exp\_t} + n\times \mathrm{exp\_t\_err} > 0.002,
\label{eq:cut}
\ee
where $n$ is some number to be determined, and we expect $n\approx 2$.
That is, we keep all objects that are consistent with being extended sources within
observational errors.  

In order to choose a value of $n$, we have decided to make use of
cross-correlation tests.  Specifically, stars and galaxies should be uncorrelated
with each others.  Consequently, a non-zero cross correlation between a galaxy
sample and a known stellar sample is indicative of stellar contamination in
the galaxy sample.

Consier a sample of $n$ total objects that contains $n_g$ galaxies and $n_*$ stars.
One has then
\begin{equation}
  n = \bar{n}_g(1+\delta_g) + \bar{n}_*(1+\delta_*),
\end{equation}
and therefore
\begin{equation}
  1+\delta = \frac{\bar{n}_g}{\bar{n}} (1+\delta_g) +
  \frac{\bar{n}_*}{\bar{n}}(1+\delta_*).
\end{equation}
Defining the stellar fraction of the sample $f_*=\bar n_*/\bar n_g$, 
we arrive at
\begin{equation}
  \delta = (1-f_*)\delta_g + f_*\delta_*.
\end{equation}
Now, if we cross-correlate this sample (subscript ``$\obs$'') with a known sample of stars, then we have:
\begin{equation}
  w_{\obs,s} = < \delta_s\delta > = f_* < \delta_s \delta_* > = f_* < \delta_s \delta_s > = f_* w_{ss},
\end{equation}
where $\delta_s$ is the fluctuation of a known stellar population,
and we have assumed $\delta_s=\delta_*$.
It follows from this assumption that the cross correlation $w_{\obs,s}$
is proportional to the stellar auto-correlation $w_{s,s}$.
Consequently, we can estimate the stellar contamination via
\be
f_* = \frac{w_{s,s}}{w_{\obs,s}}.
\ee
By computing the above ratio for a galaxy selected using a cut $n$
as per equation~\ref{eq:cut}, we seek to optimize our sample selection.
To measure the cross-correlations, we make use of the {\tt TreeCorr}
code \citep{jarvisetal04}.  $f_*$ is obtained by computing the median value of the above
ratio on scales of 1 to 10 arcmin. 

We can use a similar method to estimate the completeness associated with our stellar--galaxy separation
cut.  Specifically, consider again 
equation~\ref{eq:cut}.  For large $n$, the selected sample should be
highly complete.  Suppose that at a large $n$, call it $\nmax$, the sample has 
$N(\nmax)$ objects, and a stellar fraction $f_*(\nmax)$ estimated via cross correlations.
It follows that the number of galaxies is $N(\nmax)f_*(\nmax)$.  At a lower $n$,
the number of galaxies $N(n)f_*(n)$ will have decreased, and the relative completeness is simply
\be
C(n) = \frac{N(n)f_*(n)}{N(\nmax)f_*(\nmax)}
\ee
We set $\nmax=5$ to define the relative completeness, and look
for the value of $n$ which results in the best compromise between purity
and completeness.


\begin{figure}
  \hspace{-12pt} \includegraphics[width=90mm]{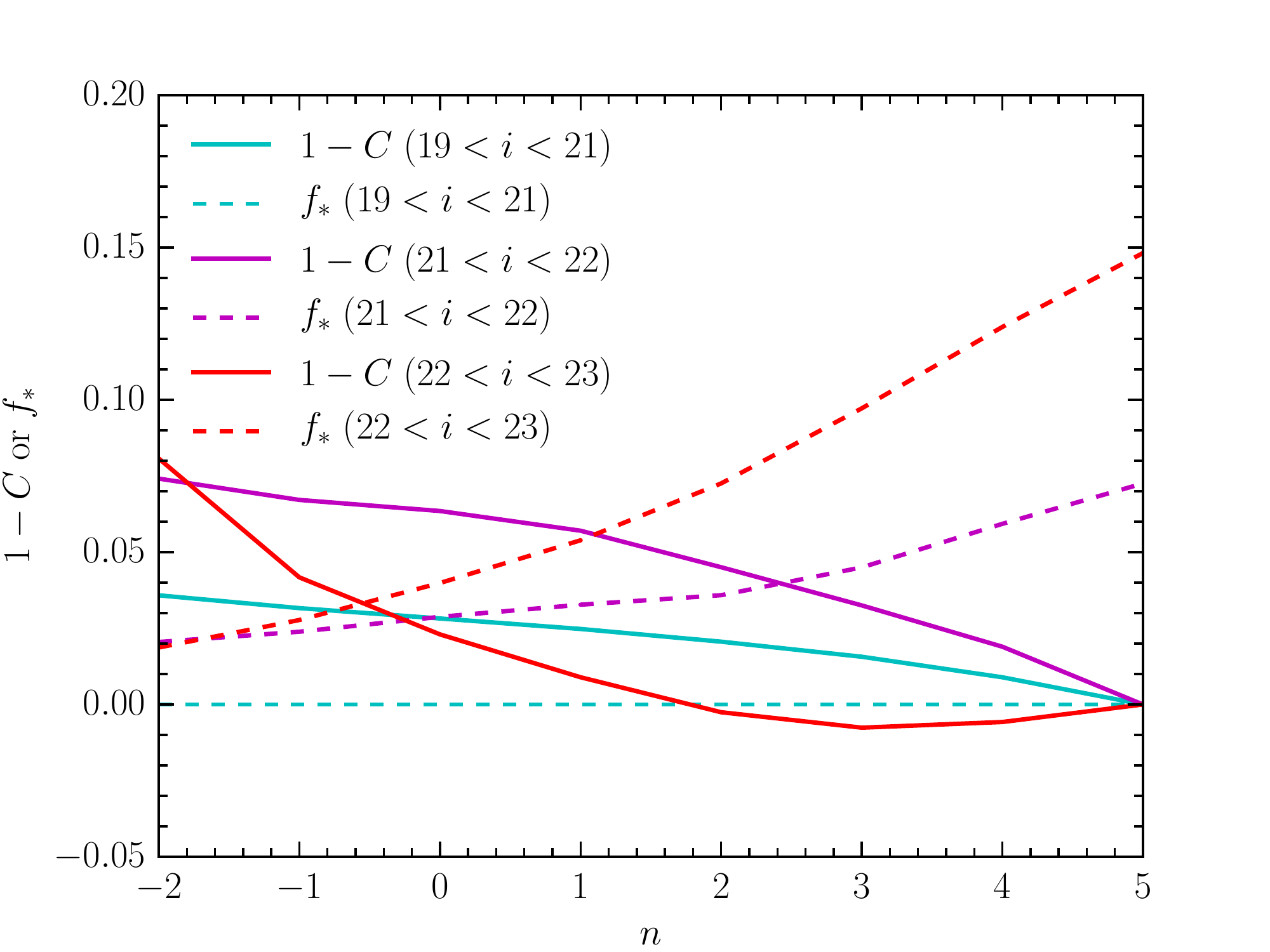}
  \caption{Incompleteness ($1-C$, dashed lines) and stellar contamination ($f_*$,
    solid lines) for four different magnitude bins, as a function of the selection
    parameter $n$.  The fainter galaxies tend to have lower completeness and
    larger stellar contamination. }
  \label{fig:stargal_results}
\end{figure}


We have implemented the above method with two stellar selections, a
bright sample ($19.0<i<21.5$), and a faint sample ($21.5<i<22.5$). 
Figure~\ref{fig:stargal_results} shows the results for the faint sample.  
Results for the bright sample are difficult to interpret (see below for further details).
The solid lines in Figure~\ref{fig:stargal_results} 
show the incompleteness ($1-C$) as a function of $n$ for three
different magnitude bins.  The dashed lines show the $f_*$ value for the same bins.
The faintest galaxies in the fiducial \redmagic\ sample have $i\approx 22$, and thus
lie in-between the red and purple lines.  The point $f_*=1-C$ for these two
lines is $n\approx -0.5$ and $n\approx 2.5$ respectively.  We adopt as our
fiducial cut the mean of the these two values, $n=1$.  From the figure, we 
expect $\approx 4\%$ stellar contamination and 4\% galaxy incompleteness.

Results from the bright stellar reference sample are difficult if not impossible
to interpret.  For instance, the completeness $C$ estimated as above using
the bright sample is larger than unity.  The estimated stellar fraction using
the bright stellar reference sample is $\approx 10\%$.  
The difference between the bright and faint stellar reference samples
suggests that the assumption $\delta_s=\delta_*$ is in fact incorrect,
and that a more reasonable model might be $\delta_s=k\delta_*$ for some $k$.
Since all we seek here is an optimal star--galaxy separation criterion, we 
adopt the proposed cut with $n=1$ here, and leave the problem of 
a more accurate estimate of the stellar contamination for the \redmagic\
galaxy sample to future work.

We emphasize the stellar contamination fractions quoted above are those relevant
for the full galaxy catalog given the star--galaxy separation criterion we have adopted.
The stellar fraction of the \redmagic\ catalog is much suppressed, since an object must
also have red sequence colors in order to make it into the \redmagic\ catalog.  The only
redshift at which the stellar locus crosses the red sequence is $z\approx 0.7$, so we
expect $\approx 5\%$ stellar contamination at $z\approx 0.7$, but essentially no contamination
at other redshifts.

\section{Data Catalog Formats}
\label{app:catalogs}

The full \redmagic{} SDSS DR8 and DES SV catalogs will be available at {\tt
  http://risa.stanford.edu/redmapper/} in FITS format, and from the online
journal in machine-readable formats. A summary of the DR8 catalog is given in
Table~\ref{tab:dr8catformat} and the SV catalog is given in
Table~\ref{tab:sva1catformat}.  Absolute magnitudes in the tables are computed
using {\tt kcorrect} v4.2~\citep{blanton07}.  $k$-corrections are applied
assuming an LRG template, band shifted to $z=0.1$.

The SDSS catalogs will be made publicly available upon publication of this article
in a journal.  
We plan to release the DES \redmagic\ catalogs publicly by January, 2016.  
See the Dark Energy Survey website\footnote{http://www.darkenergysurvey.org/}
for instructions on how to download the catalogs.

\begin{center}
\begin{table*}
\centering
\caption{\redmagic{} SDSS DR8 \redmagic{} Catalog Format}
\begin{tabular}{llcl}
Column & Name & Format & Description\\
\hline
1 & OBJID & I18 & SDSS DR8 CAS object identifier\\
2 & RA & F12.7 & Right ascension in decimal degrees (J2000)\\
3 & DEC & F12.7 & Declination in decimal degrees (J2000)\\
4 & IMAG & F6.3 & SDSS $i$ CMODEL magnitude (dereddened)\\
5 & IMAG\_ERR & F6.3 & error on $i$ CMODEL magnitude\\
6 & MODEL\_MAG\_U & F6.3 & SDSS $u$ model magnitude (dereddened)\\
7 & MODEL\_MAGERR\_U & F6.3 & error on $u$ model magnitude\\
8 & MODEL\_MAG\_R & F6.3 & SDSS $g$ model magnitude (dereddened)\\
9 & MODEL\_MAGERR\_R & F6.3 & error on $g$ model magnitude\\
10 & MODEL\_MAG\_I & F6.3 & SDSS $r$ model magnitude (dereddened)\\
11 & MODEL\_MAGERR\_I & F6.3 & error on $r$ model magnitude\\
12 & MODEL\_MAG\_Z & F6.3 & SDSS $i$ model magnitude (dereddened)\\
13 & MODEL\_MAGERR\_Z & F6.3 & error on $i$ model magnitude\\
14 & MODEL\_MAG\_Y & F6.3 & SDSS $z$ model magnitude (dereddened)\\
15 & MODEL\_MAGERR\_Y & F6.3 & error on $z$ model magnitude\\
16 & MABS\_U & F6.3 & Absolute magnitude in $u$\\
17 & MABS\_ERR\_U & F6.3 & Error on absolute magnitude in $u$\\
18 & MABS\_G & F6.3 & Absolute magnitude in $g$\\
19 & MABS\_ERR\_G & F6.3 & Error on absolute magnitude in $g$\\
20 & MABS\_R & F6.3 & Absolute magnitude in $r$\\
21 & MABS\_ERR\_R & F6.3 & Error on absolute magnitude in $r$\\
22 & MABS\_I & F6.3 & Absolute magnitude in $i$\\
23 & MABS\_ERR\_I & F6.3 & Error on absolute magnitude in $i$\\
24 & MABS\_Z & F6.3 & Absolute magnitude in $z$\\
25 & MABS\_ERR\_Z & F6.3 & Error on absolute magnitude in $z$\\
26 & ILUM & F6.3 & $i$ band luminosity, units of $L_*$\\
26 & ZREDMAGIC & F6.3 & \redmagic{} photometric redshift\\
27 & ZREDMAGIC\_E & F6.3 & error on \redmagic{} photometric redshift\\
28 & CHISQ & F6.3 & $\chi^2$ of fit to \redmagic{} template\\
29 & Z\_SPEC & F8.5 & SDSS spectroscopic redshift (-1.0 if not available)\\
\end{tabular}
\label{tab:dr8catformat}
\end{table*}
\end{center}

\begin{center}
\begin{table*}
\centering
\caption{\redmagic{} DES SV \redmagic{} Catalog Format}
\begin{tabular}{llcl}
Column & Name & Format & Description\\
\hline
1 & COADD\_OBJECT\_ID & I18 & DES SVA1 object identifier\\
2 & RA & F12.7 & Right ascension in decimal degrees (J2000)\\
3 & DEC & F12.7 & Declination in decimal degrees (J2000)\\
4 & MAG\_AUTO\_G & F6.3 & $g$ MAG\_AUTO magnitude (SLR corrected)\\
5 & MAGERR\_AUTO\_G & F6.3 & error on $g$ MAG\_AUTO magnitude\\
6 & MAG\_AUTO\_R & F6.3 & $r$ MAG\_AUTO magnitude (SLR corrected)\\
7 & MAGERR\_AUTO\_R & F6.3 & error on $r$ MAG\_AUTO magnitude\\
8 & MAG\_AUTO\_I & F6.3 & $i$ MAG\_AUTO magnitude (SLR corrected)\\
9 & MAGERR\_AUTO\_I & F6.3 & error on $i$ MAG\_AUTO magnitude\\
10 & MAG\_AUTO\_Z & F6.3 & $z$ MAG\_AUTO magnitude (SLR corrected)\\
11 & MAGERR\_AUTO\_Z & F6.3 & error on $z$ MAG\_AUTO magnitude\\
12 & MABS\_G & F6.3 & Absolute magnitude in $g$\\
13 & MABS\_ERR\_G & F6.3 & Error on absolute magnitude in $g$\\
14 & MABS\_R & F6.3 & Absolute magnitude in $r$\\
15 & MABS\_ERR\_R & F6.3 & Error on absolute magnitude in $r$\\
16 & MABS\_I & F6.3 & Absolute magnitude in $i$\\
17 & MABS\_ERR\_I & F6.3 & Error on absolute magnitude in $i$\\
18 & MABS\_Z & F6.3 & Absolute magnitude in $z$\\
19 & MABS\_ERR\_Z & F6.3 & Error on absolute magnitude in $z$\\
20 & ZLUM & F6.3 & $z$ band luminosity, units of $L_*$\\
21 & ZREDMAGIC & F6.3 & \redmagic{} photometric redshift\\
22 & ZREDMAGIC\_E & F6.3 & error on \redmagic{} photometric redshift\\
23 & CHISQ & F6.3 & $\chi^2$ of fit to \redmagic{} template\\
24 & Z\_SPEC & F8.5 & spectroscopic redshift (-1.0 if not available)\\

\end{tabular}
\label{tab:sva1catformat}
\end{table*}
\end{center}

\label{lastpage}

\end{document}